%% file: 000_main_paper.tex
\documentclass{scrartcl}
\usepackage{amsthm}
\usepackage{amsfonts}
\usepackage{amsmath}

\usepackage{bm}
\usepackage{natbib}
\usepackage[dvipsnames]{xcolor}
\usepackage[colorlinks,allcolors=MidnightBlue ]{hyperref}
\usepackage{graphicx}
\usepackage{subcaption}
\usepackage{dutchcal} 
\usepackage[utf8]{inputenc}
\usepackage{amssymb}
\usepackage{mathtools}
\usepackage{bbm}

\usepackage{graphicx}
\usepackage[ruled,vlined]{algorithm2e}
\usepackage{booktabs}
\usepackage{placeins}
\usepackage{tikz}
\usepackage{algpseudocode}

\usetikzlibrary{shapes,arrows,chains}

\newcommand{\bedit}{}
\newcommand{\eedit}{}
\newcommand{\by}{\boldsymbol{y}}
\newcommand{\bM}{\boldsymbol{M}}
\newcommand{\bS}{\boldsymbol{S}}
\newcommand{\bmu}{\boldsymbol{\mu}}

\newcommand{\bLambda}{\boldsymbol{\Lambda}}

\newcommand{\bomega}{\boldsymbol{\omega}}
\newcommand{\bpi}{\boldsymbol{\pi}}

\newcommand{\bv}{\boldsymbol{v}}
\newcommand{\ind}[1]{\mathbbm{1}_{\{#1\}}}
\newcommand{\prob}[1]{\mathbb{P}\left[#1\right]}

\newcommand{\Beta}{\mathrm{Beta}}

\newcommand{\E}[1]{\mathbb{E}\left[#1\right]}
\newcommand{\pr}{\mathrm{Pr}}
\newcommand{\Dirichlet}{\mathrm{Dirichlet}}

\newcommand{\eppf}{\mathrm{EPPF}}
\newcommand{\fiSAN}{\mathrm{fiSAN}}
\newcommand{\fSAN}{\mathrm{fSAN}}

\DeclareMathOperator*{\argmax}{arg\,max}

\usepackage{natbib}
\newtheorem{theorem}{Theorem}[section]
\newtheorem{proposition}{Proposition}[section]

\usepackage{booktabs}

\usepackage{authblk}

\title{\LARGE{A finite-infinite shared atoms nested model for the Bayesian analysis of large grouped data}}
\author[1]{Laura D'Angelo\thanks{Department of Economics, Management and Statistics; University of Milano-Bicocca; Milan, Italy. \url{laura.dangelo@unimib.it}}~ and Francesco Denti\thanks{Department of Statistics; University of Padua, Italy.\\ \url{francesco.denti@unipd.it} } 
}

\date{}

\begin{document}

\maketitle

\begin{abstract}
\input{01_abstract}

\end{abstract}
\addtocontents{toc}{\protect\setcounter{tocdepth}{-1}}
\input{text4arxiv}

\bibliographystyle{apalike-comma}
\bibliography{04_ref_overcam}

\setcounter{page}{0} 

\addtocontents{toc}{\protect\setcounter{tocdepth}{2}}
\appendix
\setcounter{figure}{0}
\setcounter{table}{0}
\renewcommand{\thetable}{S\arabic{table}}
\renewcommand{\thefigure}{S\arabic{figure}}
\clearpage

\begin{center}
{
\huge
\textbf{Supplementary Material}
}    
\end{center}

\input{03_text_suppl}

\end{document}

%% file: 01_abstract.tex
The use of hierarchical mixture priors with shared atoms has recently flourished in the Bayesian literature for partially exchangeable data. Leveraging on nested levels of mixtures, these models allow the estimation of a two-layered data partition: across groups and across observations. This paper discusses and compares the properties of such modeling strategies when the mixing weights are assigned either a finite-dimensional Dirichlet distribution or a Dirichlet process prior.
Based on these considerations, we introduce a novel hierarchical nonparametric prior based on a finite set of shared atoms, a specification that enhances the flexibility of the induced random measures and the availability of fast posterior inference. 
To support these findings, we analytically derive the induced prior correlation structure and partially exchangeable partition probability function. Additionally, we develop a novel mean-field variational algorithm for posterior inference to boost the applicability of our nested model to large multivariate data. We then assess and compare the performance of the different shared-atom specifications via simulation. 
We also show that our variational proposal is highly scalable and that the accuracy of the posterior density estimate and the estimated partition is comparable with state-of-the-art Gibbs sampler algorithms.
Finally, we apply our model to a real dataset of Spotify's song features, simultaneously segmenting artists and songs with similar characteristics.

%% file: text4arxiv.tex
\section{Introduction}
\label{sec::introduction}
Inference with grouped data is a well-established statistical challenge and a common scenario in real-data analyses. Hierarchical models provide a standard framework for working with this type of data and are particularly well-suited for a Bayesian treatment. 
They allow having a large enough number of parameters to fit the data well without incurring the risk of overfitting; they enable borrowing information between samples while admitting the presence of heterogeneity. These strengths are fundamental in several application areas and lines of research. 
In multicenter clinical trials, for example, it is relevant to study the cross-group variability for assessing the performance of a specific treatment~\citep{gray1994_hier_multicenters}; in the early stages of an epidemic, integrating data from different outbreaks allows timely interventions~\citep{lee2020_hier_covid}; innovative cancer radiomics techniques use computational approaches to improve diagnosis of tumors by borrowing information across subjects~\citep{Xiao2021_hier_radiomics}; causal inference models can use data from randomized trials to study related observational scenarios~\citep{wang2019_hier_causal}. 

Meanwhile, nonparametric Bayesian methods have become increasingly popular thanks to their non-restrictive assumptions on the parametric form of the data and general algorithms for posterior inference.
In the nonparametric context, the need to develop hierarchical modeling strategies has led to several formulations of layered prior distributions based on multi-level Dirichlet processes (DP) specifications. 
The hierarchical Dirichlet process (HDP) of~\citet{Teh2006} provides a remarkable contribution in this direction. The HDP formulation relies on modeling groups of observations using distinct DPs with a common base measure, which, in turn, is itself a realization from another DP. Since the draws from this DP are almost surely discrete, all group-specific distributions are based on a shared set of atoms. In the framework of model-based clustering, this formulation leads to the nice property of allowing cross-group clusters of observations (observational clusters, OC), favoring the interpretation of the latent structure of the data. 
A different perspective on multicenter studies was considered by~\citet{Rodriguez2008} with the nested Dirichlet process (nDP). 
\bedit 
Instead of focusing on partitioning observations, they considered the problem of investigating clusters of groups, that is, samples with similar distributional characteristics (distributional clusters, DC).
\eedit
This result is achieved by replacing the random atoms of a DP with random probability measures, which are themselves sampled from a DP, thus specifying a mixture over \emph{distributional atoms}. The discreteness of this mixing measure at the distributional level leads to a positive probability of modeling two distinct samples with the same distribution. 
\bedit
Additionally, the discreteness of the distributional atoms also results in the clustering of observations within the same distributional cluster. 
\eedit
Models based on the nDP have been widely employed in various contexts: see, for example,~\citet{Graziani2015,Rodriguez2014,Zuanetti2018}. 
\bedit
However, despite the attractiveness of this construction,~\citet{Camerlenghi2019} revealed a critical drawback, which occurs whenever an observational atom is shared by two distinct groups. In this case, the nDP imposes the full exchangeability of the two samples, forcing their complete homogeneity (this behavior is also referred to as \textit{degeneracy}). 
%
%
In response to this drawback, several authors have proposed alternatives to the nDP that admit the presence of cross-group observational clusters and avoid the degeneracy issue: notable examples are the semi-hierarchical DP of~\citet{beraha2021} and the hidden HDP (HHDP) of~\citet{lijoi2023}. 
\eedit
In particular, the latter work investigates the theoretical properties of \emph{admixture models}~\citep[see, e.g.,][]{Agrawal2013,Balocchi2022} where a nDP structure is placed over the realizations of an HDP, obtaining a discrete distribution over the space of random measures with shared atoms.
Another formulation that conveys both the cross-DC observational clustering of the HDP and the distributional clustering of the nDP has been recently proposed by~\citet{Denti2021} with the common atoms model (CAM). 
Their specification closely resembles that of the nDP, but has a crucial difference: the observational atoms are assumed to be the same across all distributional atoms. 
On the one hand, the shared atoms are essential for avoiding degeneracy while maintaining a clustering between observations in different groups. On the other hand, they pose a constraint on the random measures constituting the distributional atoms, which, by construction, are bound to have a correlation above $0.5$.
Therefore, this approach could lead to biased posterior inference in scenarios where, for example, the distributional clusters are well-separated. 

Somehow surprisingly, the literature on the parametric counterpart of nested mixture models is very limited. A finite version of the CAM was proposed by~\citet{DAngelo2022} for analyzing neuroimaging data. The proposed model made use of the telescoping sampler of~\citet{fs2011} to define nested levels of finite mixtures with a random number of components. Such a specification exhibited promising results both on simulated data as well as in real settings, and indeed their work suggested even improved performances compared to the nonparametric formulation. However, despite the empirical results, they did not provide theoretical justifications on the rationale of this behavior. The modeling framework we introduce and analyze also provides a formal justification for their findings. 
\bedit
For the interested reader, Section A.1 of the Supplementary Material contains a brief discussion in which we outline and compare the nDP, the CAM, and the fCAM. 
\eedit

In view of these considerations, we investigate nested priors based on a finite set of shared observational atoms. In particular, we explore a new class of models that we call the Shared Atoms Nested (SAN) priors.
Our formulation adopts the flexible two-level structure of the CAM, but it departs from it for the use of a set of observational atoms of finite dimension.
This modification preserves the simple structure of the CAM and, at the same time, has a considerably positive effect on the prior properties.
The proposed structure can be combined with different specifications of the distributional weights to enhance its flexibility. Moreover, it allows the derivation of a coordinate ascent variational inference (CAVI) algorithm to improve scalability. 
Thanks to the availability of this fast and efficient algorithm for posterior inference, we can indeed develop this modeling framework in the context of large multivariate data to widen the applicability of nested Bayesian nonparametric models to datasets with thousands of observations and hundreds of groups.

The paper is organized as follows. In Section~\ref{sec::nested_priors}, we introduce the general setting of SAN models, present several prior properties, and compare them with other nested priors. Section~\ref{sec::posterior_inference} details the CAVI algorithm. In Section~\ref{sec::simulation_study}, we perform a simulation study to compare the adequacy of the proposed prior for clustering and density estimation with other state-of-the-art models. In particular, we analyze how the proposed framework scales with multivariate data and compare the performances and accuracy of the proposed variational algorithm against a standard Gibbs sampler approach.
Finally, in Section~\ref{sec::application}, we apply our model to a dataset provided by Spotify that contains numerical features of thousands of songs authored by hundreds of artists. Our model is able to identify a sensible two-level clustering of similar artists and songs, which can be used to provide listening suggestions.

\section{Shared atoms nested priors}
\label{sec::nested_priors}
Consider a nested design, where the data $\by = (\by_1,\dots,\by_J)$ are divided into $J$ different groups. Within each group $j=1,\ldots,J$, $N_j$ observations are measured: $\by_j=(\by_{1,j},\dots,\by_{N_j,j})$, $\by_{i,j}\in\mathcal{Y}$, with $\mathcal{Y}$ the common support of dimension $d\geq 1$. 
\bedit
We assume a partially exchangeable framework, in the sense that observations are assumed exchangeable \textit{within} each sample, and we adopt the following mixture model specification: 
\eedit
\begin{equation}
    \by_{1,j},\dots,\by_{N_j,j} \mid f_j \overset{ind.}{\sim} f_j, \quad \quad
    f_j(\cdot) = \int_{\Theta} p(\cdot\mid \theta) G_j(d\theta),
    \label{eq::intro:mixture_model}
\end{equation}
for $j=1,\dots,J$, where $p(\cdot\mid\theta)$ denotes a parametric kernel on $\mathcal{Y}\times \Theta$ and $G_j$ is a group-specific mixing measure. Let $\Theta$ be the space of the mixing parameter $\theta$, endowed with the respective Borel $\sigma$-field $\mathcal{X}$. We assume that the mixing measures $G_j$'s are
sampled from an almost surely discrete distribution $Q$, defined over the space of probability distributions
on $\mathcal{X}$. In particular, to induce a two-layer clustering structure, we assume $Q$ to have the following form, as originally proposed by~\citet{Rodriguez2008} with the nDP: 
\begin{equation}
    G_1,\dots,G_J \mid Q \overset{iid}{\sim} Q, \quad \quad
    Q = \sum_{k= 1}^{K} \pi_k \delta_{G^*_k}.
    \label{eq::intro:nested_prior}
\end{equation}
Furthermore, we assume that each distributional atom $G^*_k$ is given by
\begin{equation}
    G^*_k = \sum_{l= 1}^{L}\omega_{l,k}\delta_{\theta^*_l}, \label{eq::intro:common_atoms}
\end{equation} 
with $\{\theta^*_l\}_{l= 1}^L$ randomly sampled from a non-atomic base measure $H$ defined on $(\Theta,\mathcal{X})$. 
In the following, we will refer to the sequence $\{\theta^*_l\}_{l= 1}^L$ as \emph{observational} atoms, and to $\{G^*_k\}_{k= 1}^K$ as \emph{distributional} atoms.
The parameters $K$ and $L$ play a crucial role in the prior distributions we introduce. They indicate the number of mixture components, with $K\in \mathbb{N}\cup \{\infty\}$ and $L\in \mathbb{N}\cup \{\infty\}$. 
It is crucial to note that according to~\eqref{eq::intro:common_atoms}, the atoms $\theta_l^*$ are \emph{shared} across all distributions $G^*_k$, following the idea introduced by~\citet{Denti2021}.
This shared set of atoms is essential for allowing cross-DC observational clusters. 

The distributional properties of such priors are intrinsically defined by the law of the observational weights $\bomega_k = \{\omega_{l,k}\}_{l=1}^L$ for $k=1,\dots, K$, and the distributional weights $\bpi = \{\pi_k\}_{k=1}^K$. The CAM resorted to a stick-breaking construction for the random weights of $Q$ and $G^*_k$'s. Specifically, they considered $K=L=\infty$, $\bpi\sim\mathrm{GEM}(\alpha)$ and $\bomega_k\sim\mathrm{GEM}(\beta)$ for $k\geq 1$, i.e., $\pi_k = v_k \prod_{j=1}^{k-1}(1-v_j)$ and $\omega_{l,k} = u_{l,k} \prod_{r=1}^{l-1}(1-u_{r,k})$ with $v_k \sim \mathrm{Beta}(1,\alpha)$ and $u_{l,k} \sim \mathrm{Beta}(1,\beta)$ for all $k\geq 1 $, $l\geq 1 $~\citep{Sethuraman1994}. This construction, however, leads to a well-known stochastically decreasing ordering of the weights: hence the atoms that appear earlier in the sequence will have larger associated mass, \emph{for all distributions}. This feature stands at the basis of the rigid correlation structure of the CAM mentioned in the Introduction.
We propose to modify the distribution of the mixing weights to ``break'' the correspondence between ordered weights and atoms.
In this sense, we distinguish our proposals based on \emph{shared} atoms, where no 
\bedit
(implicit) 
\eedit
order is assumed, from the model of~\citet{Denti2021} based on \emph{common} atoms, where the $\theta^*_l$'s have stochastically decreasing weight in all distributional atoms.
Specifically, we place a symmetric Dirichlet distribution on the observational weights instead of a Dirichlet process. Although it could appear as a simplification, this choice has a notable positive impact on the model's performance. Symmetric Dirichlet distributions indeed imply that, \textit{a priori}, all atoms are equally likely to be resampled across random measures. This characteristic grants a more flexible correlation structure, improving both prior and posterior properties. 

The finite-infinite Shared Atoms Nested (fiSAN) prior is defined by Equations~\eqref{eq::intro:nested_prior}-\eqref{eq::intro:common_atoms} and assumes $K=\infty$, a finite $L\in\mathbb{N}$, and that the mixing weights are assigned the following prior distributions
\begin{equation}
    \begin{aligned}
    \bpi = \{\pi_k\}_{k=1}^{\infty} &\sim \mathrm{GEM}(\alpha),\\
    \bomega_k = (\omega_{1,k},\ldots,\omega_{L,k}) &\sim \Dirichlet_L(b_k, \ldots, b_k).
    \label{eq::intro:nested_dirichlet_distrib}
    \end{aligned}
\end{equation}
For simplicity, from here on, we assume $b_k \equiv b$ for $k\geq 1$. This prior is based on a ``hybrid'' formulation, where a Dirichlet process drives the distributional partition, and finite Dirichlet distributions control the observational one. 
The crucial aspect of this prior, which will be extensively analyzed in the following sections, is that the finiteness of the Dirichlet distribution allows spreading the mass homogeneously over all the atoms $\theta^*_l$, without favoring a small subset of them.
Indeed, we will show that this prior has several interesting properties and that this mixed approach can have advantages over a purely nonparametric specification.

Since many key properties of the prior are driven by the prior on $\bomega_k$, one could specify different distributions at the distributional level. In particular, we will also study the fully finite case, i.e., $K\in\mathbb{N}$ and $L\in\mathbb{N}$ finite, and
\begin{equation}
	\bpi = (\pi_1,\dots,\pi_K) \sim \Dirichlet_K(a, \ldots, a),
	\label{eq::finite_pi}
\end{equation}
and we will call this specification a finite SAN (fSAN) prior. 

The use of finite mixtures has long caused concerns about the choice of the dimension of the Dirichlet distribution. However, in the recent literature, several effective strategies have been proposed to overcome this issue. 
These works leverage the distinction between \emph{components} and \emph{clusters}, with the latter indicating the ``filled'' mixture components used to generate the data~\citep{Richardson1997,nobile}.  
\citet{Rousseau2011} demonstrated that in the context of sparse ``overfitted'' mixtures (i.e., mixtures where the number of components is deliberately set larger than the number of clusters present in the data), the posterior distribution asymptotically concentrates on the true mixture if appropriate concentration parameters are specified. In other words, the posterior will empty the extra components, allowing the model to automatically estimate the dimension of the partition of the observed sample. The practical effectiveness of this strategy has also been showcased by~\citet{MalsinerWalli2016}. 
\bedit
Another strategy that has received renewed interest is the specification of a prior over the dimension of the Dirichlet distribution: this leads to a mixture with a finite, random number of components. 
This class comprises several notable instances. 
Using a symmetric Dirichlet distribution with random dimension and a fixed parameter, as discussed in~\citet{miller2018mixture}, results in a special case of Gibbs-type prior~\citep{GTP2, GTP1, GTP3}. An even more flexible approach was studied by~\citet{fs2011}, where they allowed the Dirichlet parameter to vary as a function of the dimension.
\eedit
In general, one could use the preferred approach to fix or estimate $L$ and $K$ in~\eqref{eq::intro:nested_dirichlet_distrib} and~\eqref{eq::finite_pi} without leading to restrictive assumptions. 
The model of~\citet{DAngelo2022} is precisely an instance of this kind of approach, where both $K$ and $L$ are finite and random, following the mixture of finite mixtures framework of~\citet{fs2011}. 
However, the study of the properties induced by different prior specifications of $p(K)$ and $p(L)$ is beyond the scope of this paper. Indeed, in the following, we will study the properties of finite nested mixture models for fixed $L$ and $K$ (or conditionally on their values).

\subsection{Correlation structure}
\label{subsec::corr}
The discreteness of the nested priors on the two levels, together with the commonality of the atoms, allows for the presence of observational and distributional ties. To compare the different priors, we analyze the dependence between pairs of distributions and observations. 
\bedit
The following properties hold for fixed concentration parameters, 
\eedit
all proofs are deferred to Section~B of the Supplementary Material. 
The first important quantity to investigate is the probability of observational and distributional co-clustering. Consider two distributions $G_j$ and $G_{j'}$ defined on $(\Theta,\mathcal{X})$, with $G_j,G_{j'}\mid Q \sim Q$ and $Q$ defined as in~\eqref{eq::intro:nested_prior}. Moreover, consider two observations $\theta_{i,j}\mid G_j\sim G_j$ and $\theta_{i',j'}\mid G_{j'}\sim G_{j'}$. 

Under the finite-infinite SAN prior, the co-clustering probabilities are
\begin{equation*}
	\prob{G_j = G_{j'}} = \frac{1}{1+\alpha}
	\quad 
	\text{and}  \quad
	\prob{\theta_{ij} = \theta_{i'j'}} = \frac{L+\alpha+L(b+\alpha b)}{L(\alpha+1)(Lb+1)};
\end{equation*}
while for the finite SAN prior, it is immediate to show that 
\begin{equation*}
    \prob{G_j = G_{j'}} = \frac{1+a}{1+Ka}
\quad 
\text{and} \quad 
    \prob{\theta_{ij} = \theta_{i'j'} } = \frac{a(L+K-1)+L(b+Kab+1)}{L(Ka+1)(Lb+1)}.
\end{equation*}

These expressions show that the probability of observational co-clustering between two observations belonging to different groups is non-null. Moreover, noticing that $\prob{\theta_{ij} = \theta_{i'j'} \mid G_j \neq G_{j'}}>0$, we have a confirmation of the ability of these priors to convey a cross-DC observational clustering.
Finally, the distributional co-clustering probability of the fiSAN is equivalent to that of the CAM and the nDP. 

The following proposition provides the expression for the correlation between random measures induced by the SAN priors.
\begin{proposition}
Consider two distributions $G_j,G_{j'}\mid Q \sim Q$ defined on $(\Theta,\mathcal{X})$, and $A \in \mathcal{X}$ a Borel set.
Under the finite-infinite SAN prior, the correlation between two random measures evaluated on the same set $A$ is:
\begin{equation}
	\rho^{(\text{fiSAN})}_{j,j\prime} := Corr(G_j(A), G_{j\prime}(A)) = 1-\frac{\alpha(L-1)}{L(\alpha+1)(b+1)}.
	\label{eq::corr_fisan}
\end{equation}
Under the finite SAN prior, the correlation is:
\begin{equation}
	\rho^{(\text{fSAN})}_{j,j\prime} := Corr(G_j(A), G_{j\prime}(A)) = 1-\frac{a(K-1)(L-1)}{L(Ka+1)(b+1)}.
	\label{eq::corr_fsan}
\end{equation}
\end{proposition}

\begin{figure}[t!]
    \centering
    \includegraphics[width = \linewidth]{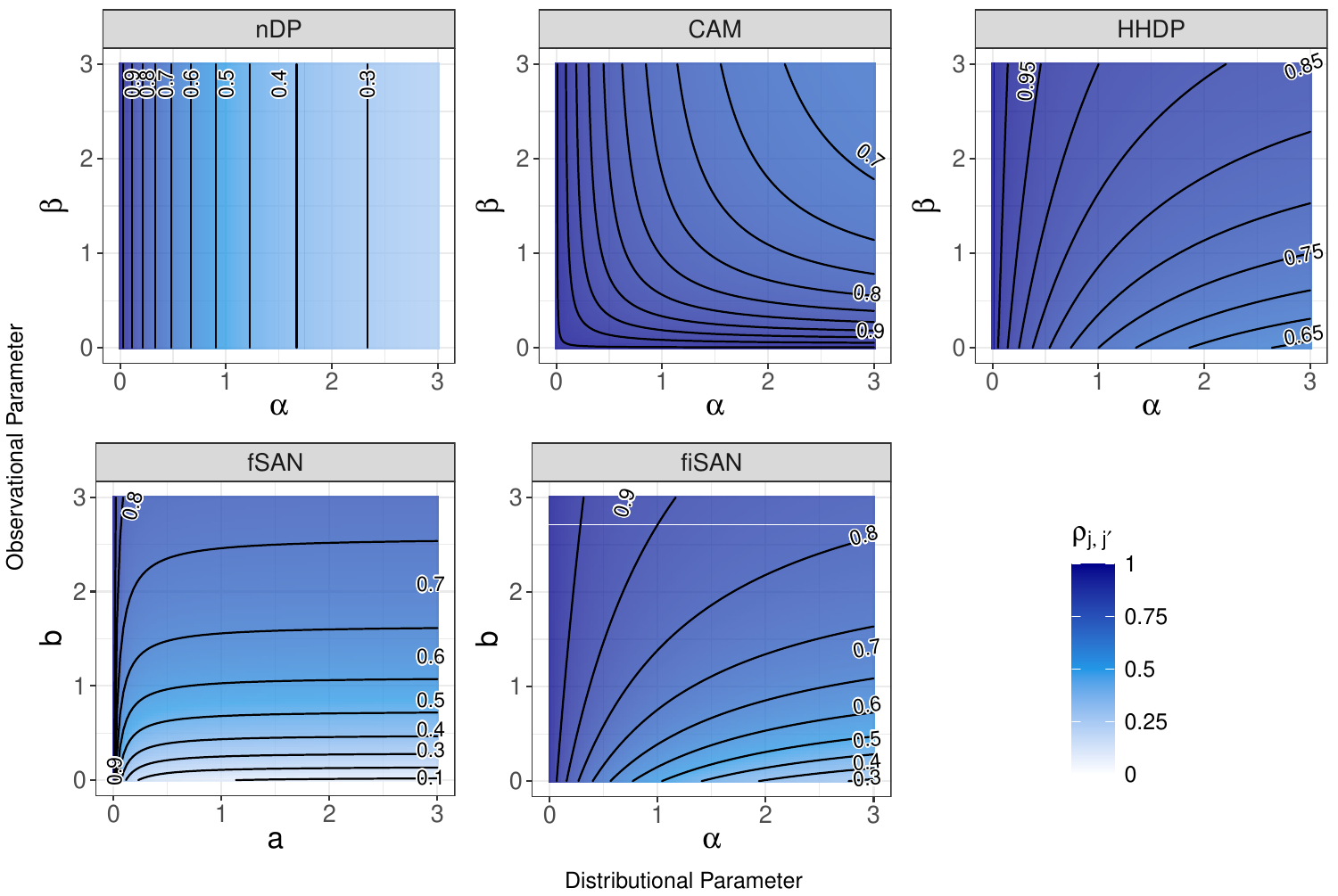}
    \caption{ Correlation of different nested models for varying observational and distributional concentration parameters. For the fiSAN we fixed $L=30$, and for the fSAN $L=K=30$. For the HHDP, we set $\beta_0=1$.}
    \label{fig:corrs_comparison}
\end{figure}

It is interesting to compare the prior correlation structure of SAN priors with the other nested nonparametric models; in particular, we consider the HHDP, the nDP, and the CAM (the correlation of these priors are derived in~\citealp{lijoi2023};~\citealp{Rodriguez2008}; and~\citealp{Denti2021}, respectively). 
For all these fully nonparametric models, we consider $\bomega_k \sim \mathrm{GEM}(\beta)$ for all $k$ at the observational level, and $\bpi \sim \mathrm{GEM}(\alpha)$ at the distributional level, 
\bedit
with both $\alpha$ and $\beta$ fixed. 
\eedit
Finally, under the HHDP, the $G_k^*$'s are realizations from an HDP, i.e., $G_k^*\sim DP(\beta,G_0)$ and $G_0\sim DP(\beta_0,H)$. 
The expressions for the correlations of these models are
\begin{equation}
\begin{gathered}
    	\rho_{j,j\prime}^{(nDP)} = \frac{1}{1+\alpha},\quad \quad 
    	\rho_{j,j\prime}^{(CAM)} = 1-\frac{\alpha}{1+\alpha}\cdot\frac{\beta}{1+2\beta }, \\ 	    	\rho_{j,j\prime}^{(HHDP)} = 1-\frac{\alpha \beta_0}{(\alpha+1)\left(\beta+\beta_0+1\right)}.
    	\label{eq:corr_competitors}
\end{gathered}
\end{equation}
Figure~\ref{fig:corrs_comparison} shows the correlation between two random probability measures obtained for varying values of the concentration parameters in the different nested models we are considering.
Under the nDP formulation, the correlation does not depend on $\beta$ since the model assumes independence between observational atoms of measures $G_j$ and $G_{j'}$ assigned to different distributional clusters.
On the contrary, for both the HHDP and the CAM, $\alpha$ and $\beta$ have a joint impact on the induced correlation. 
There are several analogies between the two models: they are both based on nested levels of Dirichlet processes combined with a shared set of atoms. This similarity of the fundamental structure is reflected in the implied correlation: indeed, $\rho_{j,j\prime}^{(CAM)}$ and $\rho_{j,j\prime}^{(HHDP)}$ are considerably high for standard default values of the concentration parameters (e.g., $\alpha=\beta=\beta_0=1$).
The cause of this characteristic is rooted in the interaction between the weights and the (implicitly ordered) set of observational atoms. 
For the CAM, the stick-breaking weights explicitly assume the stochastic order of the atoms to be the same across the random measures. Indeed, it is not even possible to act on the parameters to relax this constrained dependence, and $\rho_{j,j\prime}^{(CAM)}>0.5$ by construction. 
This implicit assumption is more subtle in the HHDP: their formulation involves a ``resampling'' of the weights through an additional layer, which, in principle, allows eluding the issue. However, the atoms are initially sampled from a DP, which already suggests a common importance across distributional atoms.
To leverage this intermediate layer and relax the induced correlation, one should set $\beta_0$ to very large values; however, this aspect is often overlooked in practice. At the other extreme, it is remarkable that the limiting case of $\beta_0\rightarrow \infty$ would make the HHDP revert to the nDP.
Finally, we highlight how the limiting cases $\beta\to 0$ for the CAM and $\beta_0\rightarrow 0$ for the HHDP are particularly troublesome, as they would force all the distributional atoms to place all the probability mass on the same observational atom (in Section~A.2 of the Supplementary Material we report additional graphs that investigate this issue). 
We did not include the HDP in this discussion since it is not based on nested levels of DPs and therefore is not directly comparable.

SAN priors show instead a very flexible dependence structure. Moreover, unlike other models, there are no parameter combinations that lead to pathological behaviors. As we can see from Equations~\eqref{eq::corr_fsan} and~\eqref{eq::corr_fisan}, both $\rho^{(\text{fSAN})}_{j,j\prime}$ and $\rho^{(\text{fiSAN})}_{j,j\prime}$ lie in $(0,1)$, hence different combinations of the concentration parameters allow reaching a broad range of correlations.
Although fSAN already attains increased flexibility, Figure~\ref{fig:corrs_comparison} (first panel of the second row) shows that the correlation is adequately influenced by $a$ only when $b$ is very small. Indeed, its impact is almost negligible for most of its values, and all prior correlation is driven only by $b$. The fiSAN combines the best features of both the finite and infinite scenarios: both parameters sensibly affect the induced correlation without the need to push them close to the limit of their parameter space.

\subsection{Partially Exchangeable Partition Probability Functions}\label{subsec:peppf}
\bedit
Within the exchangeable framework, the probabilistic properties of the random partition can be characterized through the exchangeable partition probability function~\citep[EPPF, ][]{Pitman1995,pitman2006combinatorial}. 
This quantity is a fundamental tool for understanding the clustering structure induced by a process. When the data are organized into separate groups, an analogous notion to analyze the resulting partition is the \emph{partially exchangeable partition probability function}~\citep[pEPPF, ][]{lijoi2014, lijoi2014b}. 
Examples of pEPPFs in the Bayesian nonparametric literature for dependent models were provided by~\citet{lijoi2014} for additive processes and by~\citet{Federico2017} for hierarchical processes.
In the nested framework,~\citet{Camerlenghi2019} demonstrated the importance of the pEPPF for analyzing the clustering properties and assessing that no pathological behaviors arise, discovering, for example, the degeneracy issue of the nDP. Here, we provide explicit expressions for the pEFFPs of the fiSAN and fSAN models, discussing their connections with the nDP.
\eedit
For notational convenience, we introduce the cluster allocation variables that express the cluster membership for each group and observation. In particular, we consider two sets of auxiliary categorical variables: $\bS = \{S_j, j=1,\dots, J\}$ with $S_j \in \{1,\dots,K\}$ indicating the DC allocation, and $\bM = \{ M_{i,j}, i=1,\dots,N_j, j=1,\dots, J\}$ with $M_{i,j} \in \{1,\dots,L\}$, indicating the OC allocation. 
Notice that the cluster allocation variables can take any value between 1 and the number of mixture components. However, some components could be empty (not used to generate the data), hence $K$ and $L$ do not coincide with the number of clusters, in general.  
Clearly, $p(S_j\mid\boldsymbol{\pi})= \sum_{k=1}^{K} \pi_k \delta_k(\cdot)$, and $p(M_{i,j}\mid S_j,\boldsymbol{\omega}_{S_j})=\sum_{l=1}^L \omega_{l,S_j}\delta_l(\cdot)$.
The observations in each group can assume at most $L$ distinct values. Thus, ties among the observations can appear, inducing a clustering configuration. We denote the frequency of cluster $l\in\{1,\dots,L\}$ in the $j$-th group as $n_{l,j} = \sum_{i=1}^{N_j} \ind{M_{i,j}=l}$.

For simplicity, in the rest of this section, we consider the setting where we only have two samples $\bm{\theta}_1 = \{\theta_{1,1},\dots,\theta_{N_1,1}\}$ and $\bm{\theta}_2=\{\theta_{1,2},\dots,\theta_{N_2,2}\}$ of sizes $N_1$ and $N_2$, respectively. 
When only two groups are considered, the observed atoms can be either shared by the two samples or specific to only one group.
We denote with $s_0$ the number of atoms that appear in both groups (i.e., those for which $n_{l,1} n_{l,2} >0$); with $s_1$ the number of atoms specific to group 1 (i.e., $n_{l,1}>0$ and $n_{l,2} =0$); and, similarly, with $s_2$ the number of atoms specific to group 2. Hence, the number of empty clusters is $L-s$ with $s = s_0+s_1+s_2$. Our goal is to study the distribution of the partition of $(\bm{\theta}_1,\bm{\theta}_2)$.

Both the fSAN and the fiSAN model allow for a straightforward derivation and an analytical expression of the pEPPF, which is given in Theorem~\ref{th:peppf_san}. 

\begin{theorem}\label{th:peppf_san} 
Let $\bm{\theta}_1$ and $\bm{\theta}_2$ be two samples of sizes $N_1$ and $N_2$ from a SAN model. Let $s=s_0+s_1+s_2$ be the number of distinct values in $(\bm{\theta}_1,\bm{\theta}_2)$, and let $(\boldsymbol{n}_1,\boldsymbol{n}_2) = \big( (n_{1,1},\dots,n_{L,1}),\allowbreak (n_{1,2}, \dots,n_{L,2}) \big)$ be the frequencies of each OC, with $L$ the number of observational mixture components. 

The pEPPF of the finite-infinite SAN prior is expressed as
\begin{equation}
\Pi^{{(\text{fiSAN})}}_{N_1,N_2,s}\left(\boldsymbol{n}_1,\boldsymbol{n}_2\right) =  \frac{1}{\alpha+1} \Phi^{(D_L)}_{N_1+N_2,s}(\boldsymbol{n}_1+\boldsymbol{n}_2) + 
     \frac{\alpha}{\alpha+1}\:\mathcal{C}^L_{s_0,s_1,s_2}\prod_{j=1}^2 \Phi^{(D_L)}_{N_j,s_0+s_j}(\boldsymbol{n}_j),
\label{eq:peppf_fiSAN}
\end{equation}
%
where $\Phi^{(D_L)}_{N,s}(\boldsymbol{n})$ and $\mathcal{C}^L_{s_0,s_1,s_2}$ denote the EPPF of a $\Dirichlet_L$ distribution~\citep{Green2001} and a correction constant, respectively, with expressions
\begin{equation*}
    \Phi^{(D_L)}_{N,s}(\boldsymbol{n}) = \frac{L! \, \Gamma(Lb) \, \Gamma(b)^{-L}}{ (L-s)! \,\Gamma(Lb+ N)} \prod_{l=1}^L \Gamma(b + n_{l}), \quad \mathcal{C}^L_{s_0,s_1,s_2} = \frac{(L-s_0-s_1)!(L-s_0-s_2)!}{L!(L-s_0-s_1-s_2)!}.
\end{equation*} 

With the above definition of $\Phi^{(D_L)}_{N,s}(\boldsymbol{n})$, the pEPPF of the finite SAN is expressed as
\begin{equation}
\Pi^{{(\text{fSAN})}}_{N_1,N_2,s}\left(\boldsymbol{n}_1,\boldsymbol{n}_2\right) = \frac{(1+a)}{(1+Ka)} \:\Phi^{(D_L)}_{N_1+N_2,s}(\boldsymbol{n}_1+\boldsymbol{n}_2) +  \frac{(K-1)a}{(1+Ka)} \:\mathcal{C}^L_{s_0,s_1,s_2}\prod_{j=1}^2 \Phi^{(D_L)}_{N_j,s_0+s_j}(\boldsymbol{n}_j),
\label{eq:peppf_fSAN}
\end{equation}
where $K$ is the number of distributional mixture components. 
\end{theorem}

The pEPPFs in~\eqref{eq:peppf_fiSAN} and~\eqref{eq:peppf_fSAN} are convex combinations of two different scenarios, where the distributional cluster allocation probabilities, i.e., $\prob{S_1 = S_2}$ and $\prob{S_1 \neq S_2}$, play the role of mixing weights. The two scenarios correspond to the EPPFs of two extreme cases: the fully exchangeable case, represented by the $\eppf$ of the pooled sample, and the unconditional independence case. Notice that the latter needs to be corrected by a constant $\mathcal{C}^L_{s_0,s_1,s_2}$ to account for the presence of shared atoms. A pure convex combination between the fully exchangeable case and the unconditional independence case - a finite-dimensional version of the results in~\citet{Camerlenghi2019} - arises if, in our model, we replace $\theta^*_l\sim H$ with $\theta^*_{l,k}\sim H$, obtaining the finite-dimensional version of the nDP. However, this formulation would prevent the presence of any cross-DC observational cluster.

The pEPPFs derived in the previous paragraph are useful to understand the connection of SAN priors with other nonparametric models, in particular, the limiting behavior of the fSAN model when one sets $\alpha = a/K$ and studies $K\rightarrow \infty$, and that of the fiSAN model when one sets $b=\beta/L$ and $L\rightarrow \infty$.  

Consider the pEPPF of SAN priors introduced before. Then, the following relationship holds across the different nested priors:
\begin{equation}
\lim_{L,K\rightarrow\infty} \Pi_{N_1,N_2,s}^{(\fSAN)}(\boldsymbol{n}_1,\boldsymbol{n}_2) = 
\lim_{L\rightarrow\infty} \Pi_{N_1,N_2,s}^{(\fiSAN)}(\boldsymbol{n}_1,\boldsymbol{n}_2) = 
 \Pi_{N_1,N_2,s}^{(\mathrm{nDP})}(\boldsymbol{n}_1,\boldsymbol{n}_2),
\label{eq:eppf_links}
\end{equation}
where
\begin{align*}
\Pi_{N,s}^{(\mathrm{nDP})}(\boldsymbol{n}_1,\boldsymbol{n}_2)
=  \frac{1}{\alpha+1} \Phi^{(DP)}_{N_1+N_2,s}(\boldsymbol{n}_1+\boldsymbol{n}_2) + \frac{\alpha}{\alpha+1} \Phi^{(DP)}_{N_1,s}(\boldsymbol{n}_1) \Phi^{(DP)}_{N_2,s}(\boldsymbol{n}_2)\ind{s_0=0},
\label{eq:cam_eppf}
\end{align*}
and $\Phi^{(DP)}_{N,s}(\boldsymbol{n})$ denotes the EPPF of a DP. More details are provided in Section~B of the Supplementary Material.  

The relations in~\eqref{eq:eppf_links} can be used to formalize some intuitions given in Section~\ref{subsec::corr}: sharing an infinite number of atoms that are, a priori, \emph{all equally likely to be sampled} in each random measure $G^*_k$ is not sufficient to prevent the model from collapsing to the fully exchangeable case, as indeed the probability of cross-DC ties approaches zero. Differently, both the CAM and the HHDP models impose, in terms of prior expectation, larger mass to the same sets of atoms across different groups: this characteristic is at the same time the solution of the degeneracy issue and the leading cause of the high correlation.

\section{Posterior inference}\label{sec::posterior_inference}
In line with other nested mixture models, posterior inference for SAN models is not available in closed form, and we need to resort to computational approximations.
The standard approach is to rely on Markov chain Monte Carlo (MCMC) techniques: in Section~C.1 of the Supplementary Material, we outline a Gibbs sampler algorithm.
However, it is well-known that MCMC methods have limited scalability when dealing with large datasets.
As large amounts of data become ubiquitous, the computational burden of MCMC algorithms may constitute a problem and hinder the application of complex models.

\subsection{Mean-field variational inference}
Variational inference (VI) methods provide a viable solution to this problem, approaching posterior inference through optimization rather than simulation~\citep{Blei2017}. 
These methods rely on finding, among the elements of a set of simple distributions, the one that better resembles the actual posterior in terms of their Kullback-Leibler (KL) divergence. 
The price of this greater scalability is a less precise posterior approximation, especially regarding the estimated variability. 
\bedit
In the following, we extend mean-field VI strategies for mixtures models~\citep{Bishop2006} to the nested setting. In particular, we describe a coordinate ascent optimization algorithm to perform posterior inference 
\eedit
under the fiSAN model (in line with the application in Section~\ref{sec::application}), while we defer to Section~C.2 of the Supplementary Material for the corresponding derivation for the fSAN. To the best of our knowledge, no variational methods are yet available in the literature for nested Bayesian common atoms mixture models.

We consider $d$-variate observations $\by_{i,j}$, and we assume a mixture of multivariate Gaussian kernels $p(\cdot\mid \theta) = \phi_d(\cdot\mid \bmu,\bLambda^{-1})$, with $\bmu$ a $d$-dimensional mean vector and $\bLambda$ a $d\times d$ precision matrix. Also, we assume a conjugate normal-Wishart prior distribution on the model parameters, $(\bmu,\bLambda) \sim \mathrm{NW}(\bmu_0,\kappa_0,\tau_0,\boldsymbol{\Gamma}_0)$.   
To derive the proposed algorithm, we exploit the model formulation based on the data augmentation scheme introduced in Section~\ref{subsec:peppf}, which makes use of the cluster allocation variables $S_j\in\{1,2,\dots\}$, $j=1,\dots, J$, and $M_{i,j}\in\{1,\dots, L\}$, $i=1,\dots, N_j$, $j=1,\dots, J$.
Thus, we can write the model as 
\begin{equation*}
\begin{gathered}
    p(\by \mid \bM, \{\bmu_l,\bLambda_l\}_{l=1}^L) = \prod_{j=1}^J\prod_{i=1}^{N_j}\prod_{l=1}^{L} \phi_d(\by_{i,j}\mid \bmu_l,\bLambda^{-1}_l)^{\ind{M_{i,j}=l}},  \\
    p(\bM\mid \bS,\bomega) = \prod_{j=1}^J\prod_{i=1}^{N_j}\prod_{k=1}^{\infty}\prod_{l=1}^{L} {\omega_{l,k}}^{\ind{M_{i,j}=l \: \cap \: S_{j}=k} },\quad \quad \quad
    p(\bS\mid \bpi) = \prod_{j=1}^J\prod_{k=1}^{\infty} {\pi_{k}}^{\ind{S_{j}=k} }.
    \label{eq::memb_model}
\end{gathered}
\end{equation*}
Finally, a gamma hyperprior on the concentration parameter of the distributional DP is assumed: $\alpha \sim \mathrm{Gamma}(a_\alpha,b_\alpha)$~\citep{Escobar1995}. In contrast, the Dirichlet parameter $b$ is assumed to be known and set to a small value to ensure sparsity (i.e., $b<\zeta/2$, with $\zeta$ the dimension of the component-specific parameter, following~\citealp{Rousseau2011}). 

To set up a variational strategy for the fiSAN, we first need to define a suitable set of variational distributions. In the mean-field variational framework, this class is given by the set of densities where the latent variables are mutually independent.
Following~\citet{Blei2006}, we use a truncated variational family to deal with the nonparametric mixture at the distributional level, where the truncation level is denoted with $T$. 
The fully factorized family of distributions that we assume can be written as
\begin{equation*}
\begin{aligned}
    q(\bM, \bS, & \{\boldsymbol{\omega}_k\}_{k=1}^T, \bv, \{\bmu_l,\bLambda_l\}_{l=1}^L ; \boldsymbol{\lambda}) =
     \prod_{j=1}^J q(S_j; \{\rho_{j,k}\}_{k=1}^T )\:
     \prod_{j=1}^J \prod_{i=1}^{N_j} q(M_{i,j}; \{ \xi_{i,j,l} \}_{l=1}^{L} ) \times\\      & 
     \times \prod_{k=1}^{T} q(v_{k} ; \bar{a}_{k},\bar{b}_{k})\:
     q(\alpha ; s_1,s_2)\:
     \prod_{k=1}^{T} q(\bomega_k ; \{p_{l,k}\}_{l=1}^L)  
    \prod_{l=1}^L q(\bmu_l,\bLambda_l ; \bm{m}_l, t_l, c_l, \boldsymbol{D}_l), 
\end{aligned}
\end{equation*}
where $q(S_j; \{\rho_{j,k}\}_{k=1}^T )$ and $q(M_{i,j}; \{ \xi_{i,j,l} \}_{l=1}^{L} )$ are multinomial distributions; $q(v_{k} ; \bar{a}_{k},\bar{b}_{k})$ are beta distributions, and they are such that $q(v_{T}=1)=1$ and $q(v_{g}=0)=1$ for $g>T$;  $q(\alpha ; s_1,s_2)$ is a gamma distribution; $q(\bomega_k ; \{p_{l,k}\}_{l=1}^L)$ are Dirichlet distributions; and $q(\bmu_l,\bLambda_l ; \bm{m}_l, t_l, c_l, \boldsymbol{D}_l)$ are normal-Wishart distributions.  
Under this representation, the set of latent variables is $\boldsymbol{\Theta} = \left(\bS, \bM, \bv, \{\boldsymbol{\omega}_k\}_{k=1}^T, \{\bmu_l, \bLambda_l\}_{l=1}^L,\alpha \right)$ and the set of variational parameters is $\boldsymbol{\lambda} =\left(\boldsymbol{\rho}, \boldsymbol{\xi}, \bar{\boldsymbol{a}},\bar{\boldsymbol{b}},s_1,s_2, \boldsymbol{p},\boldsymbol{m},\boldsymbol{t},\boldsymbol{c},\boldsymbol{D}\right)$. 
Optimization is then carried out by looking for the combination of variational parameters $\boldsymbol{\lambda}^\star$ that maximizes the evidence lower bound (ELBO).
To this end, the most commonly used algorithm is the coordinate-ascent variational inference algorithm~\citep[CAVI - see, for example,][]{Bishop2006}.
We report the CAVI updating rules for fiSAN in Algorithm~\ref{algo::CAVIfiSAN}, while we defer to Section~C.3 in the Supplementary Material for additional details on the ELBO and its evaluation. We outline the algorithm for the specific case of multivariate Gaussian likelihood, however, this approach can be adapted in a straightforward manner whenever the data distribution is a member of the exponential family, as discussed in~\citet{Blei2006}.

\begin{algorithm}[H]
{\footnotesize
\hspace*{0em}\textbf{Input:} $t \gets 0$. Randomly initialize $\boldsymbol{\lambda}^{(0)}$. Define the threshold $\epsilon$ and randomly set $\Delta>\epsilon$. 

\While{$\Delta(t-1,t) > \varepsilon$}{
    Set $t = t+1$; Let $\boldsymbol{\lambda}^{(t-1)} = \boldsymbol{\lambda}^{(t)};$ \\
    
    Update the variational parameters according to the following CAVI steps:
    \begin{enumerate}
    \item For $j=1,\dots,J$, $q^\star(S_{j})$ is a $T$-dimensional multinomial, with $q^\star(S_{j}=k)=\rho_{j,k}$\\ for $k=1,\dots,T$,
      \begin{align*}
       \log\rho_{j,k} = 
            g(\bar{a}_k,\bar{b}_k) + \sum_{r=1}^{k-1}g(\bar{b}_r,\bar{a}_r)
       + \sum_{l=1}^L \left(\sum_{i=1}^{N_j} \xi_{i,j,l}\right) h_l(\boldsymbol{p}_k),
    \end{align*}
    where 
    $g(x,y) = \psi(x) - \psi(x+y)$ and $h_k(\boldsymbol{x}) = \psi(x_{k}) - \psi(\sum_{k=1}^K x_{k})$, with $\psi$ denoting \\ the digamma function.
    \item For $j=1,\dots,J$ and $i=1,\dots,N_j$, $q^\star(M_{i,j})$ is a $L$-dimensional multinomial, with $q^\star(M_{i,j}=l)=\xi_{i,j,l}$ for $l=1,\dots,L$,
    \begin{align*}
       \log\xi_{i,j,l} = \frac12\ell^{(1)}_{l} +\frac12\ell^{(2)}_{i,j} + \sum_{k=1}^T \rho_{j,k} h_{l}(\boldsymbol{p}_k).
    \end{align*}
    where $\ell^{(1)}_{l} =  \sum_{x=1}^d\psi\left( (c_l-x+1)/2\right) +d\log2 + \log|\boldsymbol{D}_l|$ and $\ell^{(2)}_{i,j,l}= - d/t_l- c_l  (\by_{i,j}-\bm{m}_l)^T \boldsymbol{D}_l(\by_{i,j}-\bm{m}_l)$.

    \item For $k=1,\dots,T$, {$q^\star(\boldsymbol{\omega}_k)$} is $\Dirichlet_L(\boldsymbol{p}_k)$ with
    $ p_{l,k} = b + \sum_{j=1}^J \sum_{i=1}^{N_j}\xi_{i,j,l}\rho_{j,k}.$
    \item For $k=1,\dots,T$, $q^\star(v_{k})$ is a $\Beta(\bar{a}_k,\bar{b}_k)$ distribution with $$
        \bar{a}_{k} = 1 + \sum_{j=1}^J\rho_{j,k}, \quad \bar{b}_{k} = s_1/s_2 + \sum_{j=1}^J\sum_{q=k+1}^{T-1}\rho_{j,q}.$$
 
    \item   For $l=1,\dots,L$ {$q^\star(\theta_l)$} is a $\mathrm{NW}(\bm{m}_l,t_l,c_l,\boldsymbol{D}_l)$ distribution with parameters
    \begin{align*}
        \bm{m}_l &= t_l^{-1}(\kappa_0\:\boldsymbol{\mu}_0+N_{\boldsymbol{\cdot} l}\bar{\by}_l),  \quad 
        t_l = \kappa_0 + N_{\boldsymbol{\cdot} l}, \quad 
        c_l = \tau_0 + N_{\boldsymbol{\cdot} l},\\
        \boldsymbol{D}_l^{-1} &= \boldsymbol{\Gamma}_0^{-1} +\frac{\kappa_0 N_{\boldsymbol{\cdot} l}}{\kappa_0 + N_{\boldsymbol{\cdot} l}} \left(\bar{\by}_l-\bmu_0\right) \left(\bar{\by}_l-\bmu_0\right)^T + \mathbcal{S}_{\boldsymbol{\cdot} l},
    \end{align*}
    where 
    $$\begin{gathered}
        N_{\boldsymbol{\cdot} l} = \sum_{j=1}^J\sum_{i=1}^{N_j} \xi_{i,j,l}, \qquad \bar{\by}_l = N_{\boldsymbol{\cdot} l}^{-1} (\sum_{j=1}^J\sum_{i=1}^{N_j} \xi_{i,j,l}\by_{i,j}), \\
        \mathbcal{S}_{\boldsymbol{\cdot} l}=\sum_{j=1}^J\sum_{i=1}^{N_j} \xi_{i,j,l}\left(\bm{y}_{i,j}-\bar{\by}_l\right)\left(\bm{y}_{i,j}-\bar{\by}_l\right)^T.
    \end{gathered}$$

    \item  $q^\star(\alpha)$ is a $\mathrm{Gamma}(s_1,s_2)$ distribution with parameters
    $$s_1 = a_\alpha + T-1, \quad s_2 = b_\alpha - \sum_{k=1}^{T-1} g(\bar{b}_{k},\bar{a}_{k}).$$
   \end{enumerate}
    Store the updated parameters in $\boldsymbol{\lambda}$ and let $\boldsymbol{\lambda}^{(t)}=\boldsymbol{\lambda}$;
    \\
    Compute $\Delta(t-1,t) = \mathrm{ELBO}(\boldsymbol{\lambda}^{(t)})- \mathrm{ELBO}(\boldsymbol{\lambda}^{(t-1)})$.\\ 
 }
 \Return{ $\boldsymbol{\lambda}^\star$, containing the optimized variational parameters. }
 }
 \caption{CAVI updates for the fiSAN model}
\label{algo::CAVIfiSAN}
\end{algorithm}

\section{Simulation Studies} 
\label{sec::simulation_study}
We illustrate two simulation studies to assess the performances of the proposed priors and computational approach on synthetic data. \bedit
The first study aims to compare the fSAN and fiSAN models introduced in Section~\ref{sec::nested_priors} with two methods based on the same common atoms structure, to ease the comparison (specifically, the CAM and fCAM). 
For this simulation study, we consider univariate Gaussian kernels. \eedit
This setting also provides a first assessment of the efficacy of the variational inference approach outlined in Section~\ref{sec::posterior_inference} in a simple univariate setting.
The second study focuses instead on the evaluation of how the proposed model and CAVI algorithm behave in a multivariate framework. Here, we especially focus on the clustering accuracy and scalability of MCMC and VI as the dimensionality and sample size increase.

\subsection{Univariate case}\label{subsec:simu1}
This simulation study aims to provide empirical evidence for the prior properties presented in the first part of the paper and, specifically, on the accuracy of different prior specifications. We compare both the ability to recover the true two-level partition of the data and to estimate the posterior densities of the different DCs.
Additionally, we test the accuracy and efficiency of variational inference compared to the well-established MCMC procedure. This way, we are able to simultaneously evaluate:
\begin{itemize}
    \item[i.] if the proposed fSAN and fiSAN models are competitive with state-of-the-art models (based on ``standard'' MCMC methods);
    \item[ii.] if the proposed CAVI algorithm for estimating SAN priors has accuracy comparable with a Gibbs sampler approach in a simple, univariate setting.
\end{itemize}
We compare the performances of the fSAN and fiSAN models based on overfitted finite mixtures with the CAM of~\citet{Denti2021} and the fCAM of~\citet{DAngelo2022}. In particular, under the latter model, both $L$ and $K$ are random and estimated using the telescoping sampler of~\citet{fs2011}. 
\bedit
While we have examined the similarities and differences of the correlation structure with other nested priors from a theoretical perspective in Section~\ref{sec::nested_priors}, the computational complexity of those models hinders a practical comparison in settings with many groups and large sample sizes.
\eedit

For this experiment, we considered a nested dataset where each group-specific distribution is a mixture of univariate Gaussian kernels, denoted as $\phi(y\mid \mu,\sigma^2)$. Specifically, the data-generating process is a nested mixture made of three distributional clusters with homogeneous probabilities $(\pi_1,\pi_2,\pi_3) = (1/3, 1/3, 1/3)$, where the density $f_k(y)$ characterizing the $k$-th cluster, $k=1,2,3$, is given by
\begin{gather*}
    f_1(y) = 0.5 \: \phi(y\mid -5, 0.6^2) + 0.5 \:\phi(y\mid -2, 0.6^2),\\
    f_2(y) = 0.5 \:\phi(y\mid 2, 0.6^2) + 0.5 \:\phi(y\mid 5, 0.6^2),\\
    f_3(y) = \phi(y\mid 0, 0.6^2).
\end{gather*}
We independently extracted $J=6$ groups with equal sample sizes from this distribution, obtaining two samples for each distributional cluster.
We considered four configurations corresponding to varying sample sizes of each group: $N_j\in\{10, 50, 500, 2500\}$, for $j=1,\ldots,6$. Therefore, the total sample size $N$ ranges from 60 to $15{\small,}000$. 
Considering small samples allows us to investigate if there are problematic situations when the information conveyed by the data is limited, and the posterior estimates are heavily influenced by the prior. On the contrary, the large-sample scenarios allow us to evaluate the computational burden of the algorithms.
For each configuration, we replicated the experiment over 50 independently simulated datasets. 

All priors comprise some relevant parameters that affect their distributional properties, as discussed in Section~\ref{subsec::corr}. Following~\citet{Denti2021}, for the CAM the concentration parameters $\alpha$ and $\beta$ are now assigned $\text{Gamma}(1,1)$ hyperpriors. Since the parameters are random, we do not have an analytical expression of the prior correlation between pairs of random measures. Instead, we computed a Monte Carlo estimate, which resulted in a mean correlation of $0.8914$. Similarly, for the fCAM, the Dirichlet parameters $a$ and $b$ are assigned $\text{Gamma}(10,10)$ distributions, while the parameters $L$ and $K$ are assigned beta-negative-binomial distributions $\text{BNB}(1,4,3)$, following the indications of~\citet{fs2011}. Interestingly, this specification leads to a Monte Carlo estimate of the correlation equal to the one of the CAM.

For models based on finite overfitted mixtures, choosing appropriate Dirichlet parameters (dimension and concentration parameter) is key for obtaining good posterior estimates. The Dirichlet concentration parameter should satisfy the condition derived in~\citet{Rousseau2011} to induce sparsity and empty superfluous components. 
The Dirichlet dimension should be set large enough to guarantee that it exceeds the number of clusters expected in the data. However, it is interesting that, as long as these conditions are satisfied, the posterior estimates are quite robust to the specific values of such parameters (a sensitivity analysis is reported in Section~D.3 of the Supplementary Material).
Specifically, for the fiSAN, the parameter $\alpha$ of the DP at the top level is assigned a $\text{Gamma}(1,1)$ hyperprior; at the observational level, the parameters are instead set to $L = 25$ and $b = 0.05$. This combination leads to a correlation of $0.6309$. For the fSAN, all parameters are fixed, and, specifically, $a = b = 0.05$, $K=20$, and $L=25$. The resulting prior correlation is equal to $0.5657$.
Notice that increasing the Dirichlet dimensions $K$ and $L$ requires allocating larger matrices, and this is particularly true for MCMC approaches. Therefore, especially with large data sets, the choice of these parameters should be made in a reasoned way, taking into account the computational cost. Finally, given the univariate nature of the problem, 
\bedit
we adopted a conjugate normal-inverse gamma base measure $(\mu_l,\sigma^2_l)\sim\mathrm{NIG}(\mu_0,\kappa_0,\tau_0,\Gamma_0)$, whose hyperparameters were fixed to $(\mu_0,\kappa_0,\tau_0,\Gamma_0) = (0,0.01,3,2)$. 
\eedit

Turning now to the parameters of the variational density, in the sensitivity analysis we evaluated the impact of the truncation parameter $T$ used in the variational version of the fiSAN (which uses a DP at the distributional level). For this parameter, a similar reasoning to that of $L$ and $K$ applies: as long as $T$ is fixed large enough, the algorithm can freely explore the space of reasonable partitions, and the estimates are unaffected by the specific truncation value. Specifically, we used $T = 20$. Finally, the variational distribution of the univariate kernel parameters is a normal-inverse gamma, denoted as $q(\mu_l,\sigma^2_l; m_l,t_l,c_l,d_l)$.

In the following, we discuss the relevant aspects of the posterior inference. The algorithms used for this simulation study are written in efficient Rcpp language, and they are available in the R packages \texttt{SANple}~\citep{sanple} and \texttt{SANvi}~\citep{sanvi}, both downloadable from the Comprehensive R Archive Network; the scripts to replicate the analyses are available at the GitHub repository \href{https://github.com/fradenti/san4ba/}{\texttt{Fradenti/SAN4ba}}. 
All analyses were performed on a Linux server running an AMD EPYC-Rome Processor with 405 GB of RAM.

\subsubsection*{Comparison between shared atoms nested priors}
\begin{figure}[t!]
    \centering
    \includegraphics[width=\linewidth]{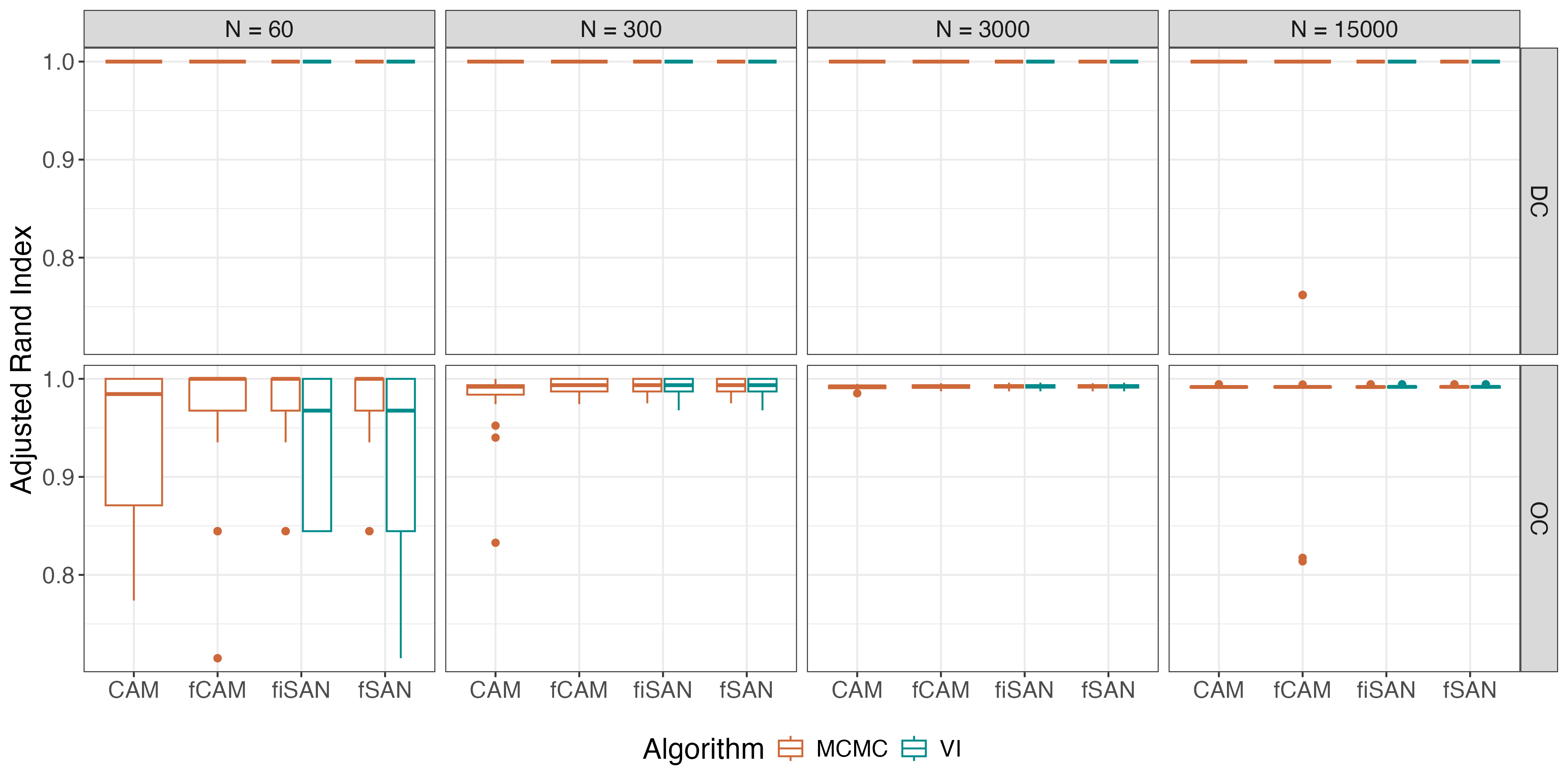}
    \caption{Accuracy of the estimated distributional (top panel) and observational (bottom panel) clustering for the CAM, fCAM, fiSAN, and fSAN. Each panel shows the distribution of the ARI obtained across the 50 replications, for each configuration. For the fiSAN and fSAN, colors correspond to the algorithm.}
    \label{fig:scenario1:clustering}
\end{figure}

\bedit
We start by analyzing the accuracy of the different nested priors in estimating the two-level partition. 
Depending on the computational strategy, the posterior point estimates of the clusters were obtained using different procedures. 
When dealing with the MCMC output, the chains of the cluster allocation variables $S_j$ and $M_{i,j}$ were used to compute the corresponding posterior similarity matrices. Then, the optimal partitions were estimated by minimizing the variation of information loss~\citep{Wade2018}, employing the algorithm developed by~\citet{Dahl2022}. 
When dealing with the VI approach, the algorithm returns instead the optimized variational parameters corresponding to the cluster assignment probabilities, i.e., $\hat{\rho}_{j,k}=q^*(S_j=k)$ and $\hat{\xi}_{i,j,l}=q^*(M_{i,j}=l)$. 
The former indicates the posterior point estimate of the probability of assigning the $j$-th group to the $k$-th distributional cluster; similarly, the latter represents the posterior estimate of the probability of assigning the $i$-th observation of the $j$-th group to the $l$-th observational cluster. 
Hence, in this case, the partitions were estimated as 
\begin{equation*}
\hat{S}_j= \argmax_{k=1,\ldots,T}\hat{\rho}_{j,k} \quad \text{and} \quad
\hat{M}_{i,j}=\argmax_{l=1,\ldots,L}\hat{\xi}_{i,j,l}
\end{equation*}
for $j=1,\dots,J$ and $i=1,\dots,N_j$. 
\eedit
Then, to measure the accuracy of the estimated partitions, we compared the adjusted Rand index (ARI,~\citealp{rand};~\citealp{hubert1985}) between the posterior point estimate and the true partition. An ARI equal to zero corresponds to random labeling, while an ARI equal to one indicates that the clusterings are identical. Figure~\ref{fig:scenario1:clustering} shows the distribution of the ARIs (over the 50 replications) obtained by the different models for each configuration. 
All models adequately estimate the distributional partition under each scenario. At the observational level, all models have remarkable performances, with an ARI over 0.8 for almost all replications; moreover, the posterior point estimates of the clustering consistently improve with increasing sample size.
For the fSAN and fiSAN, the variational approach produces slightly worse posterior estimates than the MCMC. Still, in general, the difference is small and is due to a very limited number of misclassified observations.

We now inspect the ability of the different models to estimate the density of the data. 
We first needed to obtain a posterior point estimate of the density of each group. When using an MCMC approach, we computed the pointwise mixture density on a grid of points at each iteration by substituting the current values of the chains into the theoretical data density; then, we obtained the posterior estimate by averaging them.
This strategy was necessary for dealing with the varying partition across iterations and the additional complications caused by the label-switching.
We used instead a slightly different procedure when using a CAVI algorithm since this approach does not produce a sample of replications but only a single estimated model. 
Having a unique point estimate of the model hyperparameters allows for overcoming the problem of label-switching; moreover, representing only one of the several modes (given by permutations of the indices) is sufficient to adequately perform posterior inference~\citep{Blei2017}.
Here, we substituted the estimated variational posterior expected values $\hat{\mu}_l$, $\hat{\sigma}^2_l$ and $\hat{\omega}_{l,k}$ into the parametric mixture density. In this context, $\hat{\mu}_l = m^\star_l$, $\hat{\sigma}^2_l = d^\star_l/(c_l^\star-1)$, while the weights $\hat{\omega}_{l,k}$'s are the posterior means of the corresponding Dirichlet distributions. With this estimate, we computed the density on the same grid of points.
\begin{figure}[t!]
    \centering
    \includegraphics[width=\linewidth]{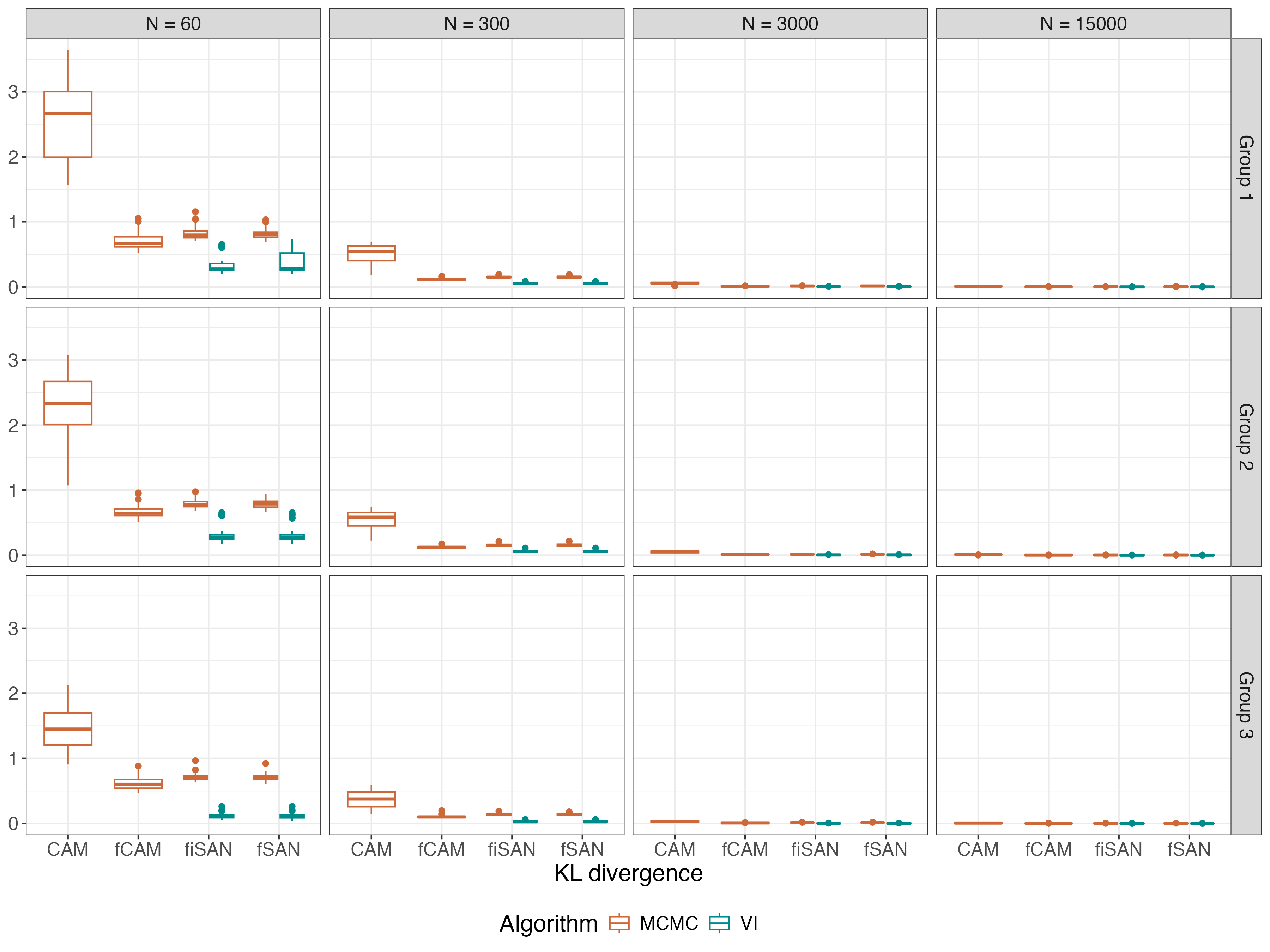}
    \caption{Accuracy of the density estimates for groups 1, 2, and 3, for the considered models. Each panel shows the distribution of the KL divergence (over the 50 replications), for each configuration. For the fiSAN and fSAN, colors correspond to the algorithm.}
    \label{fig:scenario1:KL}
\end{figure}
Figure~\ref{fig:scenario1:KL} shows the Kullback-Leibler divergence distribution between the true and the estimated mean density. In line with the previous analysis, we show the CAM, fCAM, fSAN, and fiSAN results, where the latter were estimated via Gibbs sampler and CAVI.
The plot clearly shows that the CAM has difficulties estimating the density when the sample size is small, as highlighted by the large KL divergence in the first configuration. 
Increasing the sample size alleviates this problem, as the data convey enough information to overcome the prior distribution. On the contrary, all models based on finite observational mixtures have good posterior estimates. This result is particularly interesting if one considers that the correlations of the CAM and fCAM were equal, \textit{a priori}. Hence, the improved posterior density estimates of the fCAM are purely a consequence of the Dirichlet distributions of the weights.

\begin{figure}[th!]
    \centering
    \includegraphics[width=\linewidth]{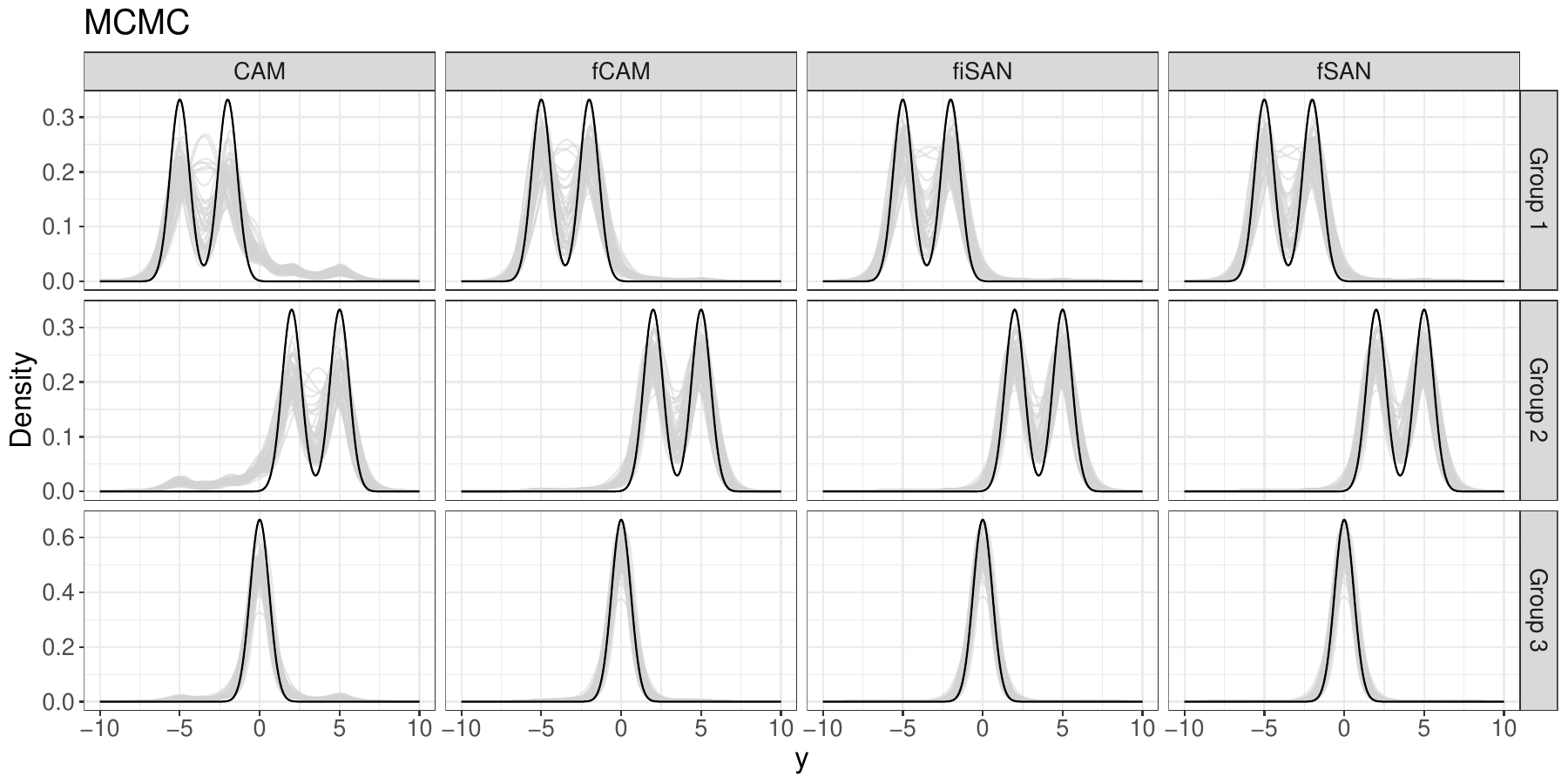}
    \caption{Posterior density estimate for groups 1, 2, and 3, for the CAM, fCAM, fiSAN, and fSAN. Each panel shows the true density (black line) and the posterior density estimates (grey lines) obtained over the 50 replications, for configuration 1 ($N = 60$). }
    \label{fig:scenario1:density}
\end{figure}

To gain more insight into the reason that led to this behavior, Figure~\ref{fig:scenario1:density} shows the posterior point estimate of the density of the first three groups 
\bedit
computed via Gibbs sampler under the first configuration, corresponding to $N=60$. Figure~S.6 in the Supplementary Material shows the posterior density estimate obtained via CAVI for the fSAN and fiSAN. 
We considered three groups whose distributions are representative of the three DCs. 
\eedit 
The first column shows the density estimates for CAM and highlights the pitfalls of using this model when the distributional atoms do not share observational atoms. The CAM indeed expects the different distributional clusters to have some atoms in common, and even if the data do not support this assumption, the model forces a similarity in the estimated density. In particular, the forced positive correlation between distributions causes the presence of ``ghost'' modes inherited from the other DCs. 
Conversely, models based on symmetric Dirichlet distributions can avoid this spurious borrowing of information, even for small sample sizes.

\subsubsection*{Comparison between computational approaches}

With the previous setting, we have already provided evidence that the proposed variational inference algorithm is a valid tool for estimating the data partition and density. That said, it is well-known that CAVI can severely underestimate the variability of the posterior distributions of the parameters. In this section, we further investigate and compare the performances of the Gibbs sampler and CAVI algorithms. Here, again, the goal is twofold: assessing the accuracy of the posterior distributions, and investigating the computational burden of the two approaches. Indeed, the actual need for a VI procedure for this type of model has not yet been discussed or corroborated by empirical evidence.
\begin{figure}[t]
    \centering
    \begin{subfigure}[b]{0.32\linewidth}
    \centering
    \includegraphics[width=\linewidth]{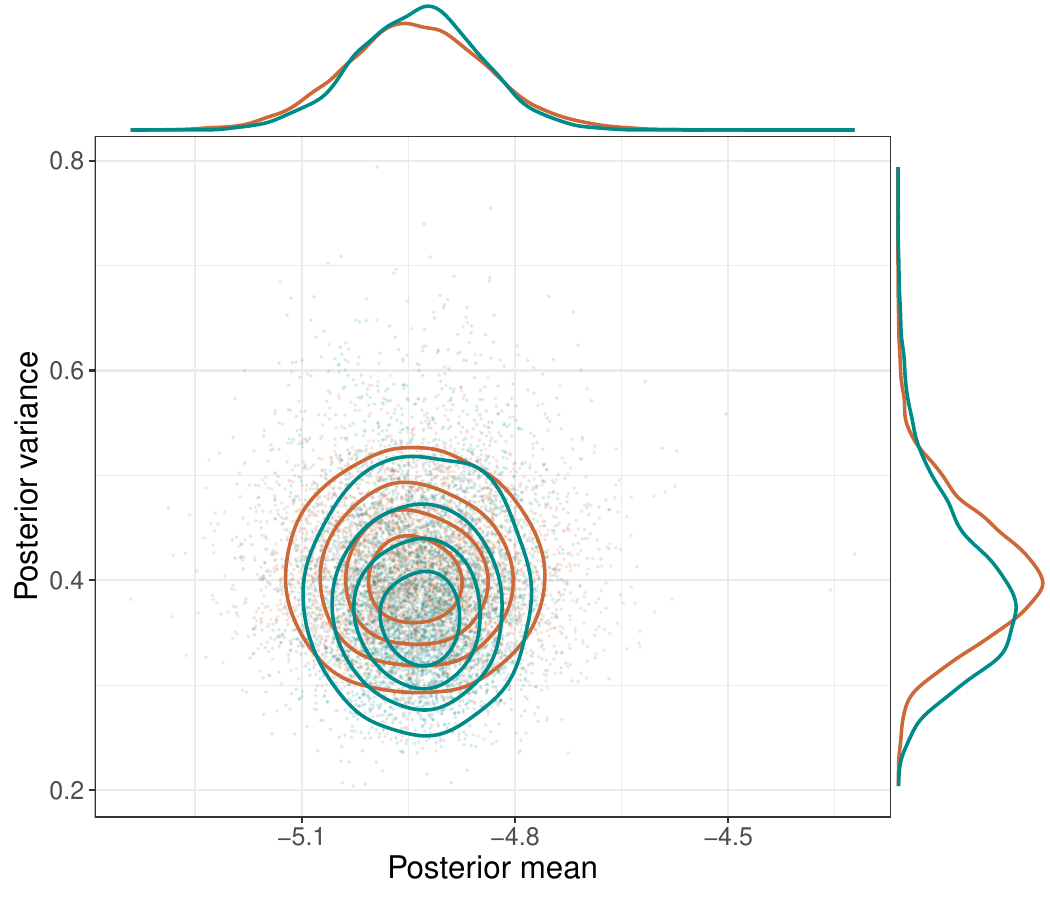} 
    \caption*{$l=1$}
    \end{subfigure}
    \begin{subfigure}[b]{0.32\linewidth}
    \centering
    \includegraphics[width=\linewidth]{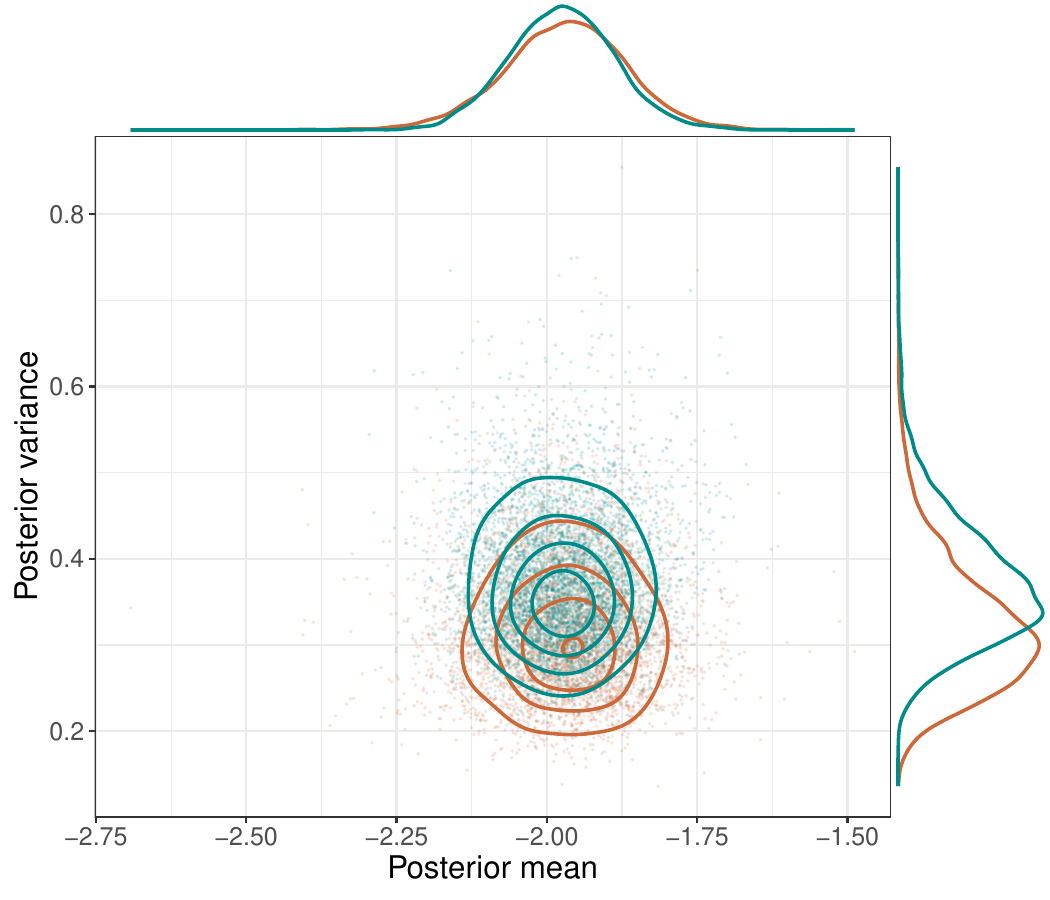} 
    \caption*{$l=2$}
    \end{subfigure}
    \begin{subfigure}[b]{0.32\linewidth}
    \centering
    \includegraphics[width=\linewidth]{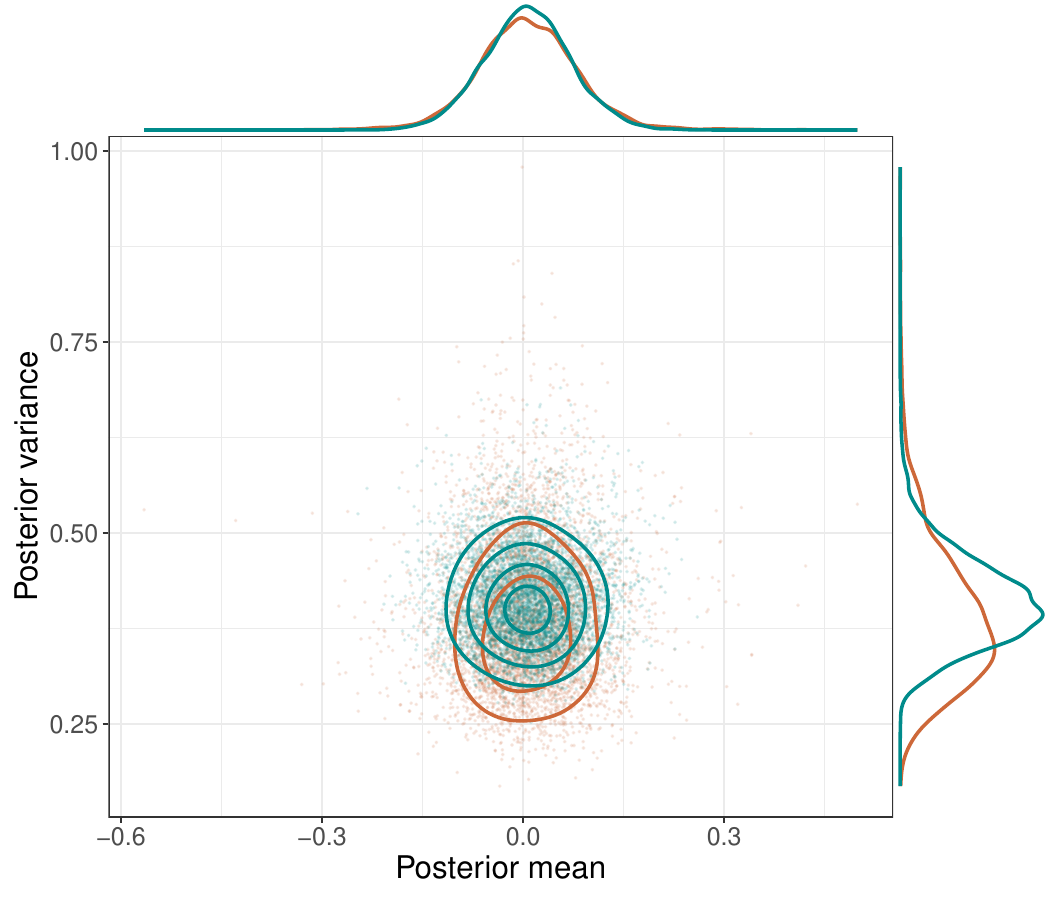} 
    \caption*{$l=3$}
    \end{subfigure} \\
    \begin{subfigure}[b]{0.32\linewidth}
    \centering
    \includegraphics[width=\linewidth]{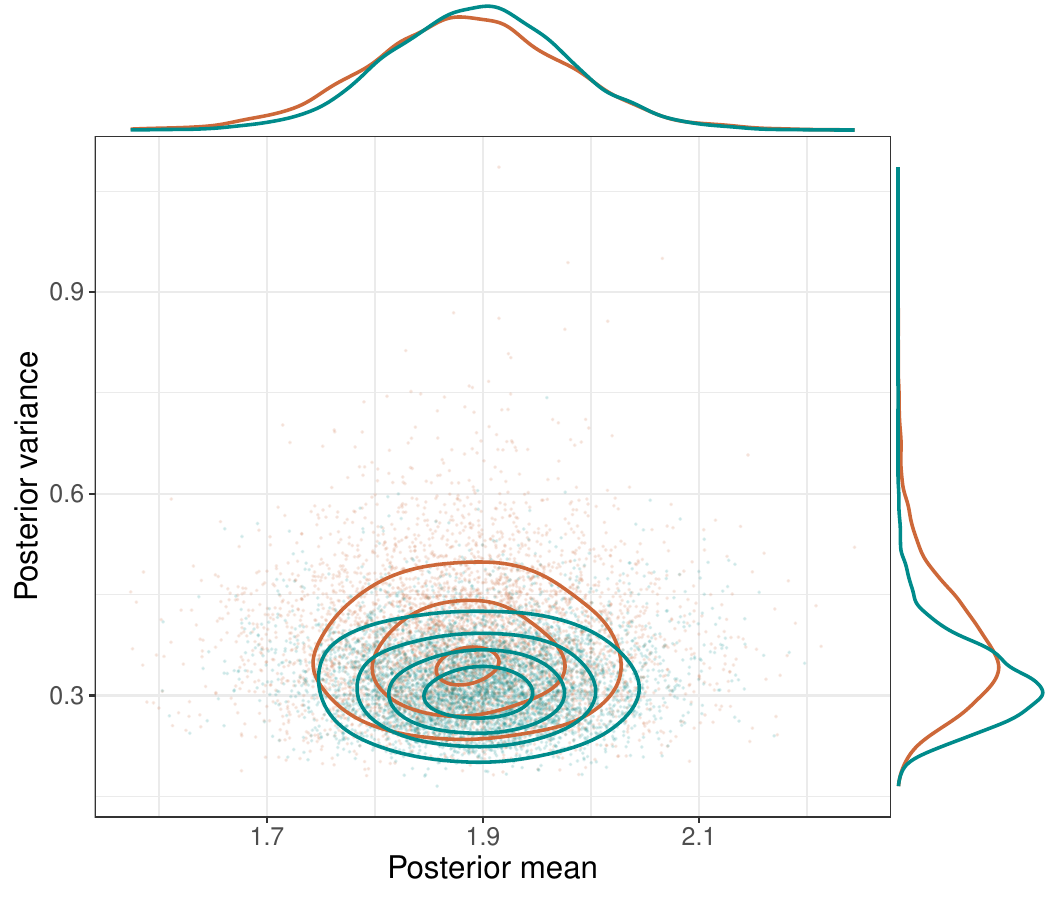} 
    \caption*{$l=4$}
    \end{subfigure}
    \begin{subfigure}[b]{0.32\linewidth}
    \centering
    \includegraphics[width=\linewidth]{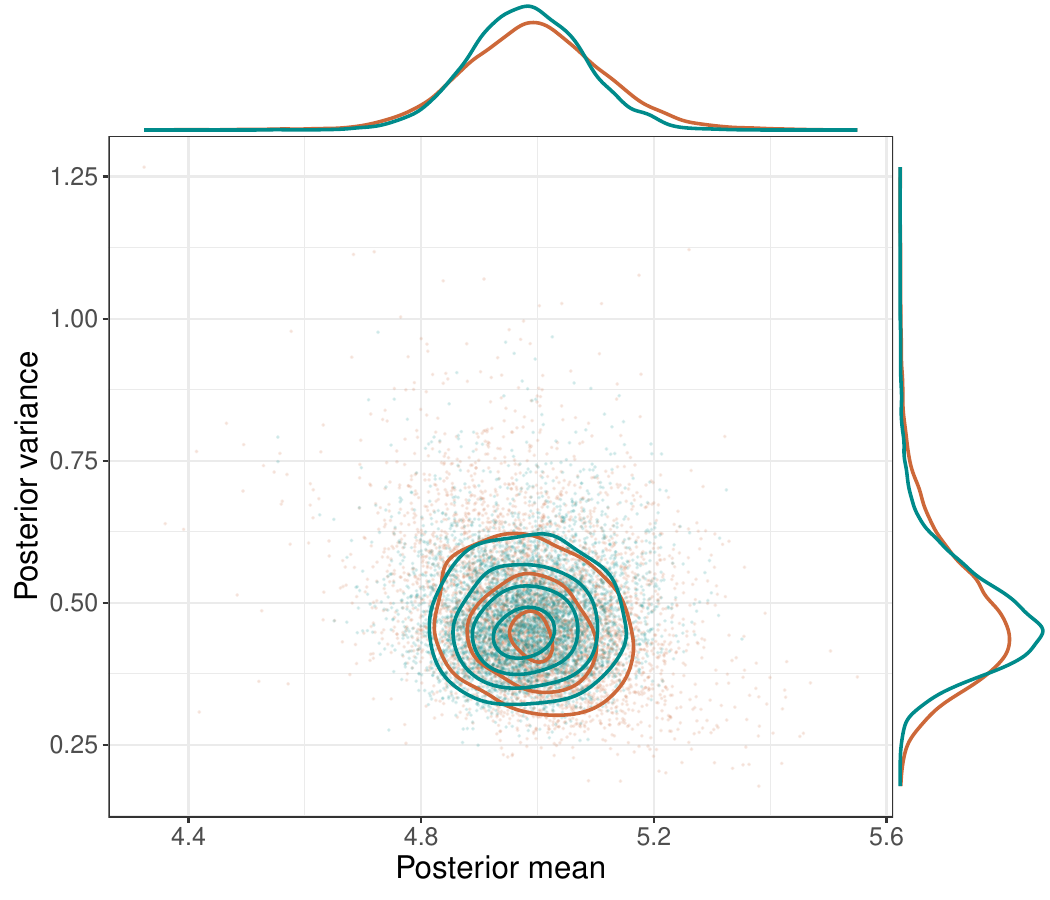} 
    \caption*{$l=5$}
    \end{subfigure}
    \caption{fiSAN prior - Configuration 2: posterior density estimate of $(\mu_l,\sigma^2_l)$, for $l=1,2,3,4,5$, obtained using a Gibbs sampler (orange line) and a CAVI algorithm (green line). Each panel shows the contour plot of the joint density, together with the two marginal densities.}
    \label{fig:scenario1:vi_vs_mcmc1}
\end{figure}

\bedit
Figure~\ref{fig:scenario1:vi_vs_mcmc1} shows the estimated posterior density of the cluster-specific parameters $(\mu_l,\sigma^2_l)$, for $l=1,2,3,4,5$ (i.e., the atoms actually used to generate the data), under the fiSAN prior estimated via MCMC and VI. 
\eedit
Each panel shows the contour plot of the joint density, together with the marginal densities. The results refer to the second configuration, corresponding to $N = 300$. 
\bedit
In Section~D.2 of the Supplementary Material, we provide additional graphs corresponding to different sample sizes and to the estimates under the fSAN prior. 
Inference on the parameters of the observational atoms is not immediate when employing an MCMC approach, as chains may be affected by label-switching, a common problem when dealing with mixture models~\citep{Stephens2000}. Moreover, a comparison between the two algorithms requires matching the cluster-specific parameters arising from two fundamentally different approaches. To solve the issue, we post-processed the chains using the relabelling Equivalence Classes Representatives algorithm~\citep{Rodriguezwalker2014}, as implemented in the R package \texttt{label.switching}~\citep{labswitchR}. Sections~D.1 and~D.2 of the Supplementary Material detail the post-processing to treat the label-switching and the procedure to derive these plots. 
\eedit
\begin{figure}[t]
    \centering
    \includegraphics[width=\linewidth]{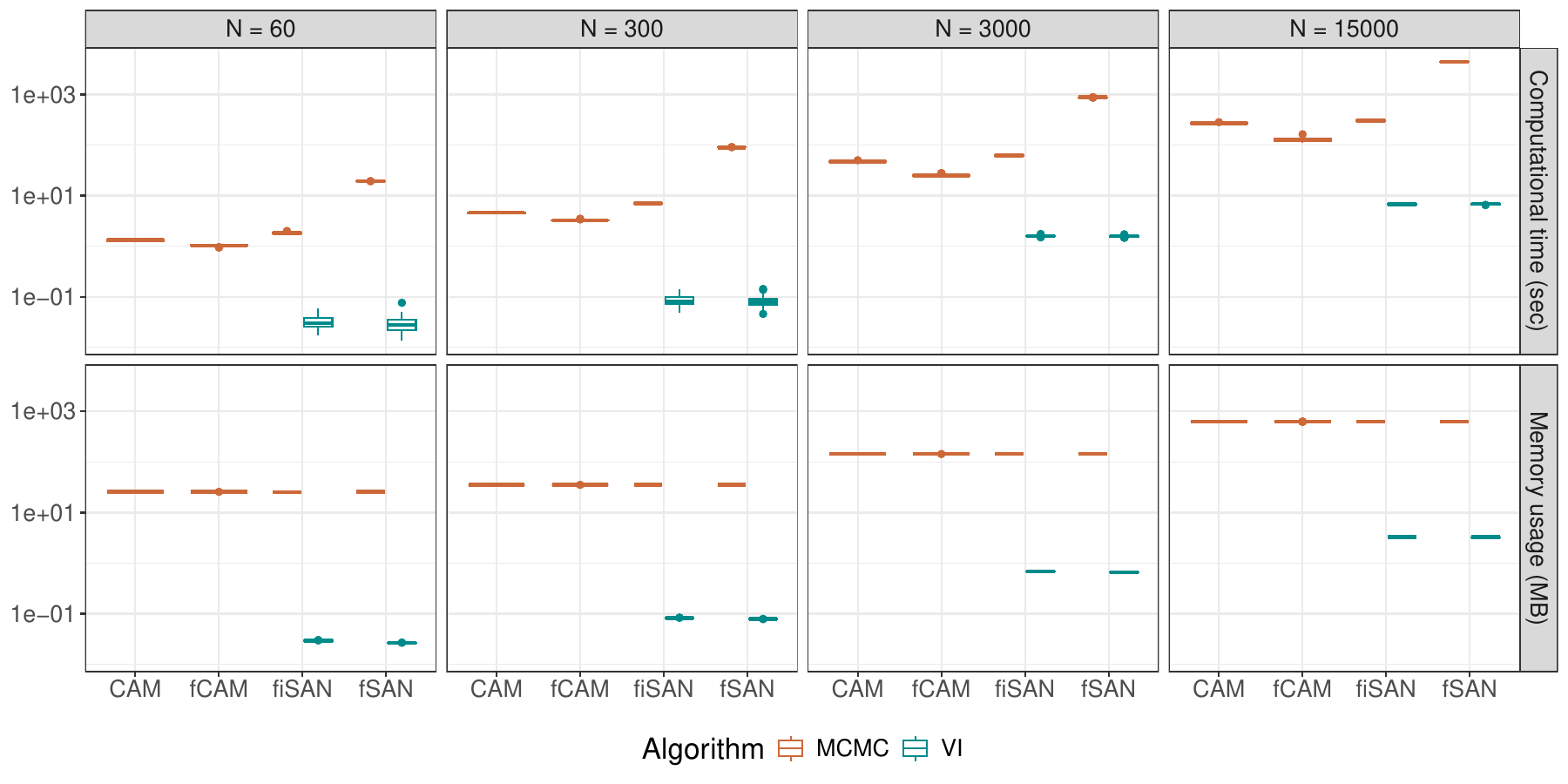}     
    \caption{Top row: distributions of the computing time (in seconds) over the 50 replications for the two algorithms, for each configuration. Bottom row: distributions of the memory usage (in MB) over the 50 replications for the two algorithms for each configuration. Values are displayed on $\log_{10}$ scale.}
    \label{fig:scenario1:memotime}
\end{figure}

Figure~\ref{fig:scenario1:vi_vs_mcmc1}, as well as Figures~S7, S8, and S9, do not highlight any systematic troubling behavior of the VI approach compared to the MCMC. 
\bedit
We acknowledge that, in some cases, the modes of the distributions do not strictly coincide, and the VI estimates have a larger bias compared to the MCMC. Moreover, the marginal variances of the posterior distributions are sometimes underestimated using the VI approach. \eedit
However, such discrepancies are not severe, and the CAVI estimates are overall satisfactory. Moreover, the computational advantages of VI largely compensate for the approximation, as displayed in the subsequent simulations, and, 
\bedit 
especially, in the multivariate setting discussed in Section~\ref{seubsec::multivariate}. 
\eedit

Turning now to the rationale for the need for a variational approach, we compare the computational cost of the two algorithms: here, we consider both the memory usage and the computing time. Indeed, for the MCMC methods, the need to store the entire Markov chains (at most, excluding the burn-in) can raise issues with memory allocation when dealing with big data. On the contrary, VI methods only require storing the optimized parameters. The panels in the top row of Figure~\ref{fig:scenario1:memotime} clearly show the advantage of a VI approach over the MCMC regarding memory usage for all sample sizes, even in this univariate case.

Obtaining a fair efficiency comparison is more complex since the two algorithms are fundamentally different. However, to guarantee convergence and adopt a procedure that would be reasonable in a real-data application, we proceeded as follows. The computational time of the Gibbs sampler algorithm was computed as the total run time to generate 10,000 iterations. 
\bedit
As for the VI approach, we considered $\Delta(t-1,t) < 10^{-4}$ as a stopping rule to define the convergence of the ELBO. 
Although the discrepancy between the variational distribution and the target posterior is reduced at each iteration, there is no guarantee that the CAVI algorithm will lead to a global optimum. 
On the contrary, depending on the initial configuration, it will likely obtain a local solution. 
Hence, we executed 50 distinct runs of the algorithm with different starting points, keeping the one with the highest ELBO to draw the inference. 
The ultimate advantage of VI is that these optimizations are easily parallelizable; thus, we report the maximum individual run time obtained over the 50 runs for each dataset, which can be seen as an indicator of the computational cost of the CAVI.
\eedit 
The panels in the top row of Figure~\ref{fig:scenario1:memotime}, displaying the elapsed seconds on the log scale, confirm that the VI approach is at least one order of magnitude faster than the MCMC, and this gap increases with increasing sample size.

\subsection{Multivariate case}\label{seubsec::multivariate}
The second simulation study is devised to investigate the performance of the fiSAN model when dealing with large multivariate data. \bedit
We focus solely on this model since the two competitors, CAM and fCAM, were only developed for univariate data. 
Moreover, from the findings in the previous paragraph, we see that all the formulations relying on a finite set of observational atoms appear to have overall the best performances. Since there is no clear evidence of superiority of a particular specification over the others, we decided to focus on the finite-infinite one. 
\eedit
Let $\by_{i,j}$ represent a vector of dimension $d\geq 1$, $\by_{i,j} = (y_{i,j,1},\dots,y_{i,j,d})^T$ for $i=1,\dots,N_j$, $j=1,\dots,J$. Here, we study and compare the MCMC and VI approaches in terms of classification accuracy and scalability in this multivariate framework. 

The data-generating process is now a nested mixture of multivariate Gaussian kernels. Specifically, it is an extension of the data-generating process of Section~\ref{subsec:simu1} to dimensions $d\in\{2,5,10\}$. Again, we have $J=6$ groups extracted from a nested mixture of three distributional atoms $f_k(\by)$, $k=1,2,3$, with homogeneous probabilities $1/3$, where each atom is a mixture of multivariate Gaussian kernels with different mean vectors and covariance matrices. The densities $f_k(\by)$ are defined as 
\begin{gather*}
    f_1(\by) = 0.5 \: \phi_d(\by\mid -5\cdot \bm{1}_d, 0.2\cdot \text{I}_d) + 0.5 \:\phi_d(\by\mid -2\cdot\bm{1}_d, 0.2\cdot \text{I}_d\cdot \text{R}_1)\\
    f_2(\by) = 0.5 \:\phi_d(\by\mid 2 \cdot\bm{1}_d, 0.2\cdot \text{I}_d ) + 0.5 \:\phi_d(\by\mid 5\cdot\bm{1}_d, 0.2\cdot \text{I}_d\cdot \text{R}_2 )\\
    f_3(\by) = \phi_d(\by\mid 0\cdot\bm{1}_d, 0.2\cdot \text{I}_d\cdot \text{R}_3).
\end{gather*}
where $\bm{1}_d$ denotes a $d$-dimensional vector of ones and $\text{I}_d$ a $d\times d$ identity matrix. The matrices $\text{R}_k$, $k=1,2,3$ are correlation matrices that induce different types of dependence across variables.
The correlation matrix $\text{R}_1$ is a band matrix, with entries equal to 0.25 for $\lvert h_1-h_2 \rvert<2$ and 0 otherwise ($h_1,h_2=1,\dots,d$); the matrix $\text{R}_2$ assumes a correlation equal to 0.5 between each pair of variables; finally, $\text{R}_3$ assumes a correlation equal to 0.85 between each pair of variables.
Similarly to the previous section, we considered homogeneous group sample sizes $N_j$ and studied the performances for varying $N_j\in\{50, 500, 1000\}$. Hence, the total sample size ranges from 300 to 6000. We replicated the experiment over 50 independently simulated datasets.
The Dirichlet parameters are equal to the ones in Section~\ref{subsec:simu1}. 
\bedit
The hyperparameters on the multivariate normal-Wishart base measures, $(\bmu_l,\bm{\Lambda}_l)\sim\mathrm{NW}(\bmu_0,\kappa_0,\tau_0,\boldsymbol{\Gamma}_0)$, are set to $(\bmu_0,\kappa_0,\tau_0,\boldsymbol{\Gamma}_0) = (\bm{0}_d,0.01,d+5, \text{I}_d)$. 
\eedit

\subsubsection*{Accuracy of posterior inference}
\begin{figure}[t!]
    \centering
    \includegraphics[width = \linewidth]{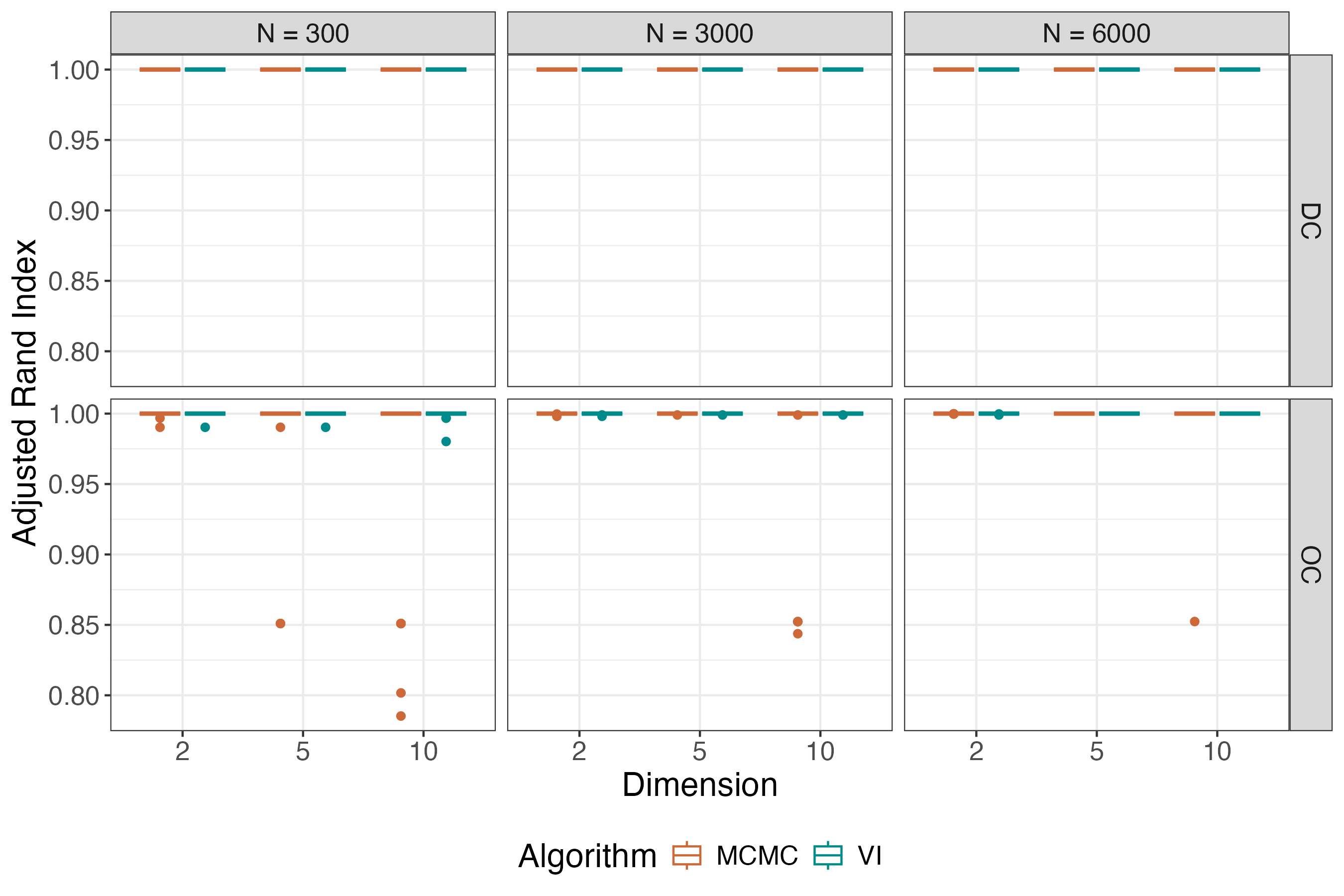}
    \caption{Distributions of the distributional (top row) and observational (bottom row) ARIs over the 50 replications for the fiSAN model estimated via VI and MCMC for each data dimension and sample size.}
    \label{fig:accuracy_multiv}
\end{figure}
We start by analyzing the accuracy of the fiSAN model in recovering the observational and distributional data partition. Figure~\ref{fig:accuracy_multiv} shows the observational and distributional ARI obtained in each scenario using both an MCMC and a VI approach. Overall, the model performs well in all cases, with larger sample sizes leading to better posterior point estimates. 
\bedit
Moreover, the two algorithms have comparable performances at clustering groups and observations, for all sample sizes and dimensions.
\eedit
\begin{figure}[t!]
    \centering
     \includegraphics[width = \linewidth]{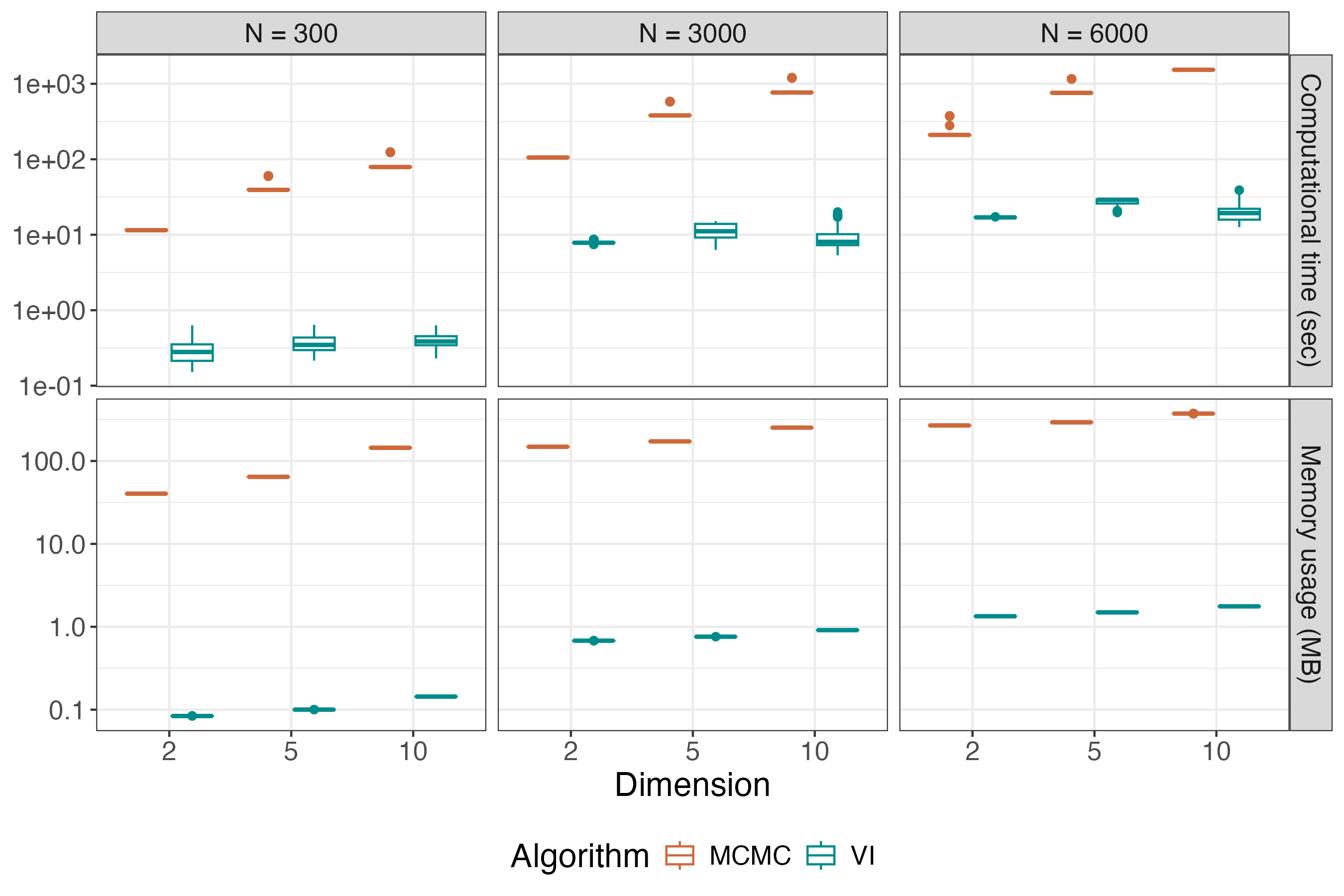}
   \caption{Distributions of the computing time (top row) and memory usage (bottom row) over the 50 replications for the fiSAN model estimated via VI and MCMC for each data dimension and sample size. Values are displayed on $\log_{10}$ scale.}
    \label{fig:mvt_comp_aspects}
\end{figure}

\subsubsection*{Computational aspects}
The computational advantages of variational inference algorithms compared to standard MCMC approaches are presumably the main reason for turning to these approximate methods. We already highlighted these advantages in the univariate case, but they become particularly evident in this multivariate framework. The bottom panels of Figure~\ref{fig:mvt_comp_aspects} show the distribution of the memory usage in MB for the two types of algorithm in each scenario.
The top panels show, instead, the total computing time. The computational time was obtained using the same definition as the previous simulation study. Both sets of plots showcase the need for a variational approach when the data are multivariate, and the sample size is moderately large. The memory usage and computational time of MCMC methods appear to grow with an exponential trend as $d$ and $N$ increase: both aspects can indeed hinder the application of complex Bayesian models to large multivariate data. Ultimately, variational inference appears to be a good compromise, balancing good posterior inference with a remarkably efficient implementation.

\section{Spotify data analysis for the discovery of music profiles} \label{sec::application}

Our study considers an open-source Spotify dataset available from the Kaggle platform\footnote{https://www.kaggle.com/ektanegi/spotifydata-19212020}. Spotify is one of the largest music streaming service providers. Part of its success is due to a personalized recommendation system, which analyzes the user's listening history and suggests new, potentially relevant tracks. Because of the success of these algorithms, increasing interest has gone into understanding what makes two songs ``similar'' from the user's enjoyment point of view. 

To this end, Spotify developed several scores to summarize various features of a song.
These features provide a description of a song's mood (e.g., \emph{danceability}, \emph{energy}), audio characteristics (e.g., \emph{loudness}, \emph{speechiness}, \emph{instrumentalness}), and context (e.g., \emph{liveness}, \emph{acousticness}). One can find more details about these features in the documentation available on the \emph{Spotify for developer} webpage\footnote{https://developer.spotify.com/discover/}.
The original dataset contains ten scores for over $160{\small,}000$ songs released between 1921 and 2020, authored by over $1500$ artists. We performed a preprocessing phase \bedit meant to discard outliers (e.g., tracks containing entire concerts, thus having exceptional duration; silent tracks, with extremely low energy, or, conversely, pure applause in live tracks, with extremely high energy levels). We then proceeded to select a large subset of artists for our analysis. We kept artists that authored more than 100 songs to ease the detection of DCs, and less than 200 songs (0.3\% of the total of artists), obtaining a dataset with $19{\small,}315$ songs partitioned into 154 artists. Additional details on the preprocessing phase are available as Supplementary Material (Sec.~E.1).\eedit

The goal of our analysis is to identify clusters of similar artists and songs based on their characteristics. This way, the system could rely on songs and artists within the same observational or distributional cluster, respectively, for the creation of playlists and listening suggestions. 
Specifically, we focus on three meaningful indicators: the \emph{duration} (D), the \emph{energy} (E), and the \emph{speechiness} (S) of each song. In this sense, our problem can be formalized as a two-level multivariate clustering, where the songs (i.e., the observations) are exchangeable data points ``within'' each artist (i.e., the groups). 
For example, during a workout, users could find it more enjoyable to listen to a brief, energetic rock song rather than a piano sonata, even if, in principle, they might like both. Hence, a segmentation driven by multiple features could convey sensible suggestions that go beyond simple genre similarities and personal taste. 

All the variables were properly transformed to fit a mixture of multivariate normal distributions: first, the three variables were marginally normalized in the $(0,1)$ interval; then, they were mapped onto the real line using a probit transformation. 
\begin{figure}
    \centering
    \includegraphics[width = \linewidth]{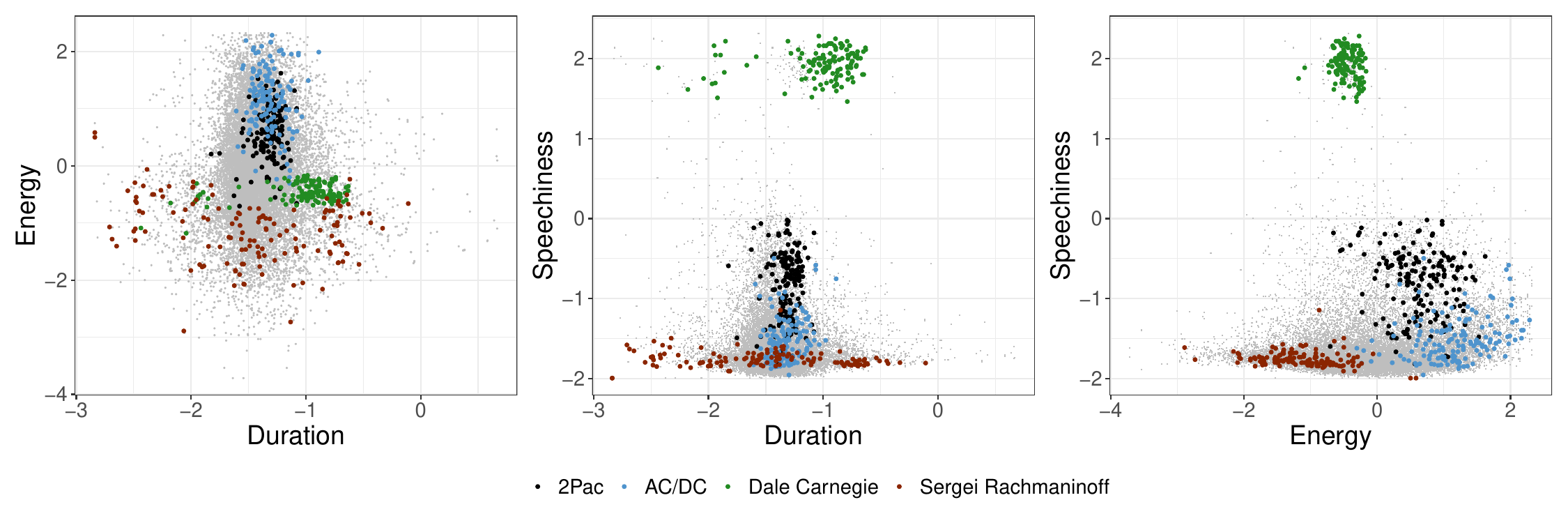}
    \caption{Distribution of the songs (points) across the three features. 
    Colored points correspond to songs by three artists: 2Pac (black), AC/DC (blue), \bedit{D. Carnegie (green)}, \eedit{and} S. Rachmaninoff (red).}
    \label{fig:spot-descr}
\end{figure}
Figure~\ref{fig:spot-descr} displays the three-dimensional data with the help of three pairwise scatterplots. Each point represents a song, and we highlighted the songs authored by 
\bedit
four different artists, i.e., 2Pac, AC/DC, D. Carnegie, and S. Rachmaninoff. The four artists belong to fundamentally different genres: 
\eedit
the scores of their tracks indeed show different distributional characteristics, although there is an overlap in some of the features. However, an adequate model should be able to recognize their differences and assign them to distinct distributional clusters.

We fit the proposed fiSAN model using the CAVI algorithm outlined in Section~\ref{sec::posterior_inference}. 
Indeed, in Section~\ref{sec::simulation_study}, we have seen how MCMC methods already pose major computational issues when the sample size is $N=6000$. Because of the fast increase of memory allocation and computing time of MCMC methods as $N$ increases, here, with almost $20{\small,}000$ observations, a Gibbs sampler approach would require considerable computational resources. Similarly to the procedure adopted in the simulation study, we ran our CAVI algorithm 1000 times using independent random initialization, and we kept the iteration with the highest ELBO to draw inference.
In line with the reasoning outlined in Section~\ref{sec::simulation_study}, we fixed the Dirichlet parameters $L = 35$ and $b = 0.05$; the truncation parameter $T$ of the Dirichlet process at the distributional level was set equal to $30$. As for the remaining hyperparameters, they were set as in the simulation studies. 
On this dataset, the fiSAN estimates 21 OCs and 20 DCs. 
In what follows, we discuss how we can exploit the estimated two-layer partition to obtain interesting insights.

\begin{table}[t]
\footnotesize
\centering
\begin{tabular}[t]{ll}
\toprule
DC & Artists\\
\midrule
\texttt{Audio lectures} & D. Carnegie - E. H. Gombrich\\
\midrule
\texttt{Classical} & A. Copland - A. Scriabin - B. Evans - B. Evans Trio - C. Mingus - C. Baker \\& E. Satie - F. Mendelssohn - F. J. Haydn - F. Liszt - F. Schubert - G. F. Handel \\& G. Mahler - M. Ravel - P. I. Tchaikovsky - R. Strauss - S. Rachmaninoff - S. Getz\\
\midrule
\texttt{Hard rock} & AC/DC - Aerosmith - blink-182 - Bob Seger - BTS - Def Leppard - Green Day \\& Iron Maiden - Journey - Judas Priest - KISS - Linkin Park - Nirvana - Ramones \\& Rush - The Smiths - Van Halen\\

\midrule
\texttt{Rap} & 2Pac - Beastie Boys - Beyoncé - JAY-Z - Kanye West - Lil Uzi Vert - Lil Wayne \\& Mac Miller -  Sublime - The Notorious B.I.G.\\
\bottomrule
\end{tabular}
\caption{Artists in the four analyzed distributional clusters.}
\label{tab:sumDC}
\end{table}
\begin{figure}
    \centering
    \includegraphics[width = \linewidth]{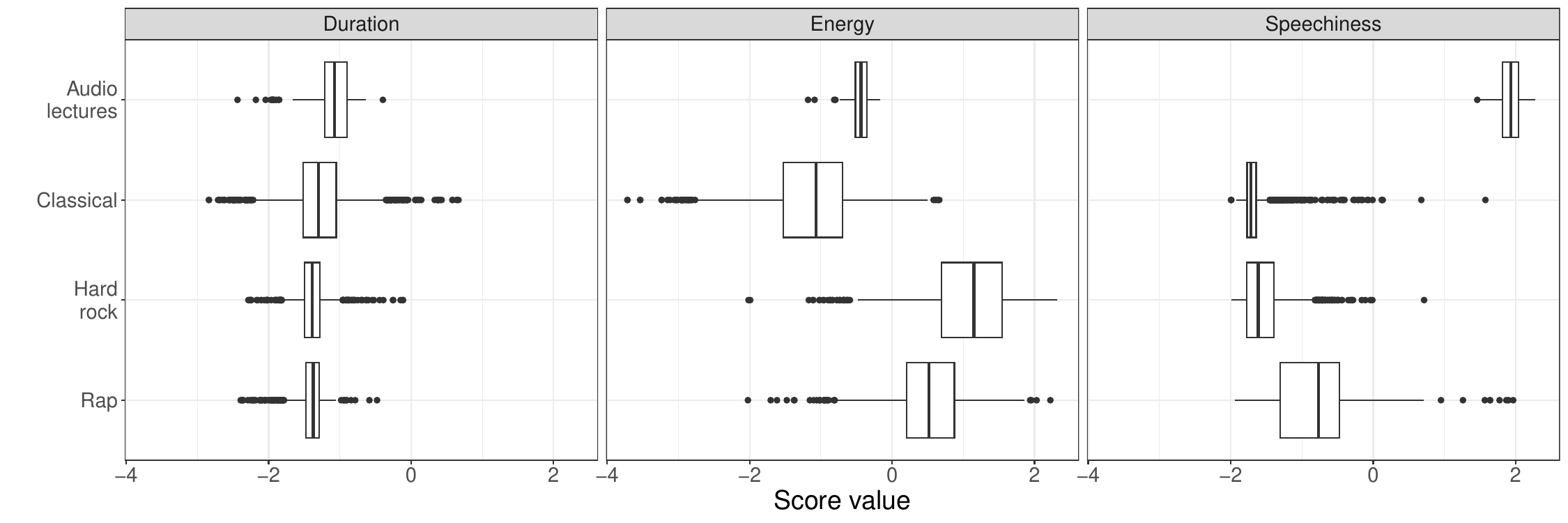}
    \caption{Distribution of the three features in the four analyzed distributional clusters.}
    \label{fig:boxplots}
\end{figure}

\textbf{Analysis of the clusters of artists (DC)}. Our algorithm groups the 154 artists into 20 clusters. In Table~S.2 of the Supplementary Material, we report the complete segmentation of the artists. Here, we analyze four notable DCs, whose members are reported in Table~\ref{tab:sumDC}. Looking at the members of these clusters, we are able to characterize them according to distinctive ``genres'', broadly interpreted as ``audio lectures'', ``classical'', ``hard rock'', and ``rap''. 
Figure~\ref{fig:boxplots} shows the distribution of the three features in these clusters. In terms of duration, all DCs are quite similar. However, tracks in the classical music cluster have a more variable duration than the other groups, and audio lectures last longer, on average. 
The energy and speechiness features are the ones with the greater heterogeneity across clusters.
Hard rock songs have the highest energy, followed by rap songs. However, rap songs have a higher speechiness score. Classical music has low energy and low speechiness, while the audio lectures are, as expected, the most verbose.

\begin{figure}[th!]
    \centering
    \includegraphics[width=.9\linewidth]{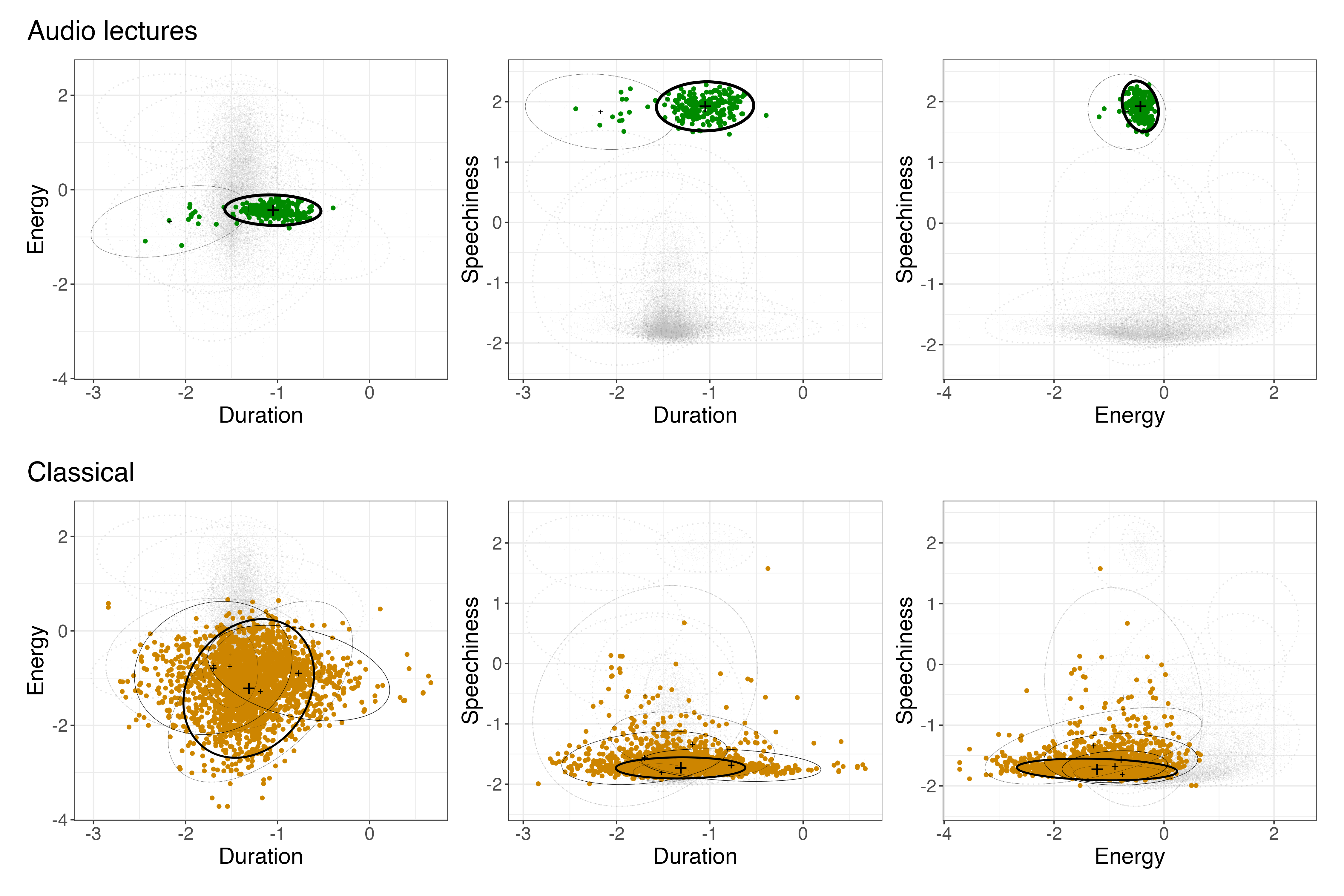}
    \includegraphics[width=.9\linewidth]{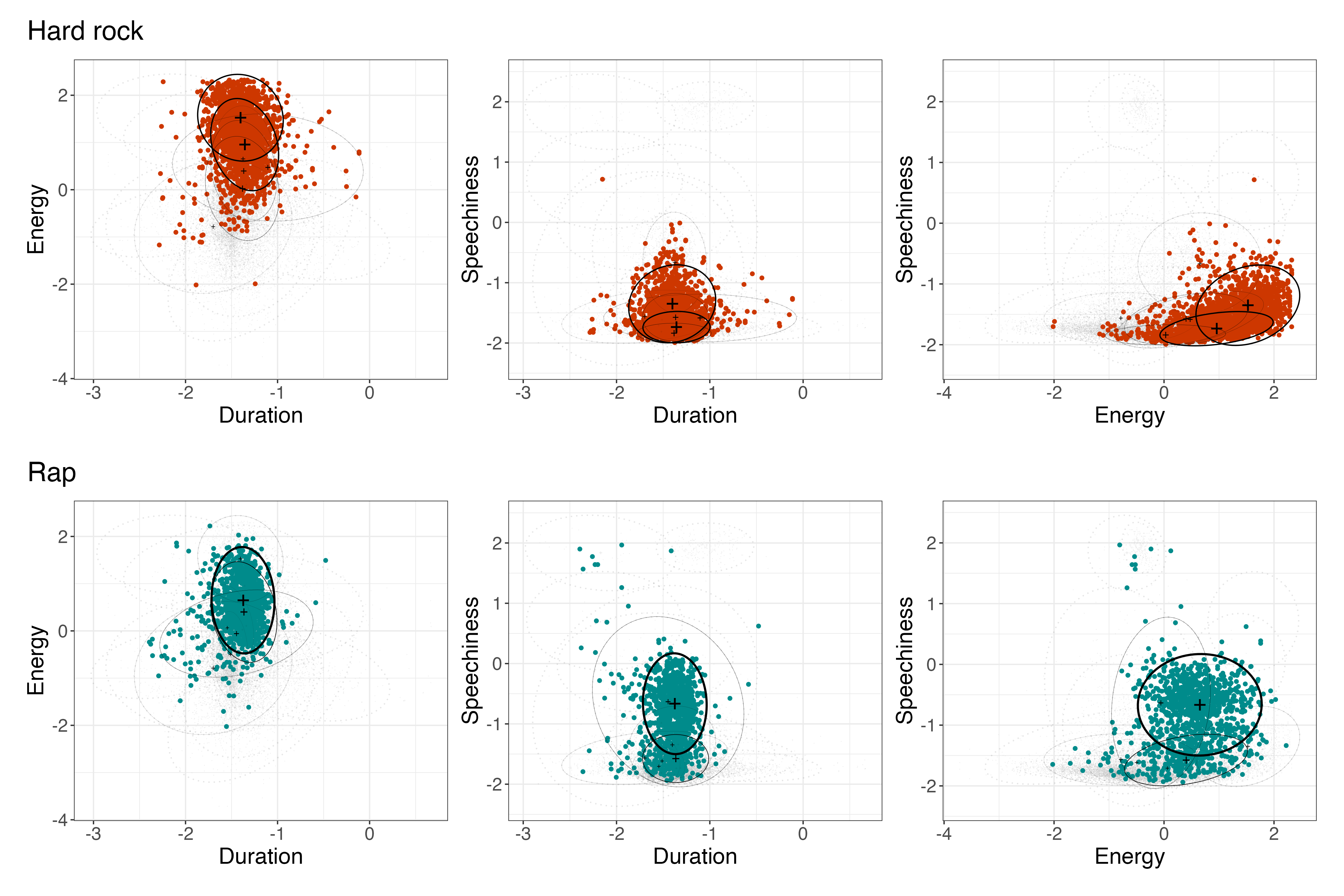}
    \caption{\bedit Pairwise scatterplots of songs (points). Each row corresponds to a different DC (``genre''), colored points indicate songs belonging to that DC. In each plot, crosses and ellipses correspond to the active observational atoms, and the intensity of the color is proportional to their posterior weight.\eedit}
    \label{fig:DCOC}
\end{figure}
Our model-based clustering approach, however, allows for much deeper insights into what drives the segmentation. 
\bedit
According to our model, the observational atoms are shared across all DCs; however, in each DC, some of these atoms may be assigned a negligible posterior probability. It is then interesting to examine the active components that ultimately characterize each DC. 
\eedit
In Figure~\ref{fig:DCOC}, we show again the pairwise scatterplots of the three song features. 
\bedit
Here, we highlighted the points belonging to the four analyzed DCs. Moreover, for every DC, we also represented all the ``active'' observational atoms, associated with non-empty observational clusters.
\eedit
Specifically, all observational atoms are represented by their estimated mean $\hat{\bmu}_l = (\hat{\mu}_{\text{D}}, \hat{\mu}_{\text{E}}, \hat{\mu}_{\text{S}})^T_l$ with crosses (2-dimensional subvectors) and covariances $\hat{\bm{\Sigma}}_l[r,s]$ $(r,s\in\{\text{D, E, S}\})$ with ellipses. 
\bedit
To favor the interpretation of each DC, we highlighted which atoms are the most relevant by drawing the intensity of the line color as proportional to the posterior weight $\hat{\omega}_{l,\hat{S}_j}$. 
Regarding the ``audio lecture'' cluster, we can appreciate that it is defined by a simple distribution made of two mixture components with very heterogeneous posterior weights. The leading OC is characterized by average duration and energy but very high speechiness, while the second one accounts for a few low-duration tracks. 
The ``rap'' DC comprises a larger number of active OCs; however, its distribution is fundamentally characterized by two predominant components. Looking at the posterior means of these two observational atoms, we see that they strongly overlap in the energy and duration variables (all rap songs have high energy and average duration). However, the speechiness feature distinguishes them and allows differentiating between more and less verbose tracks. 
\eedit
Finally, the ``classical'' and ``hard rock'' DCs are more complex and nuanced. \bedit In particular, the latter distribution can be described as a mixture of two predominant normal kernels that capture different traits in all three considered dimensions.\eedit

\begin{figure}[t!]
    \centering
    \includegraphics[width=\linewidth]{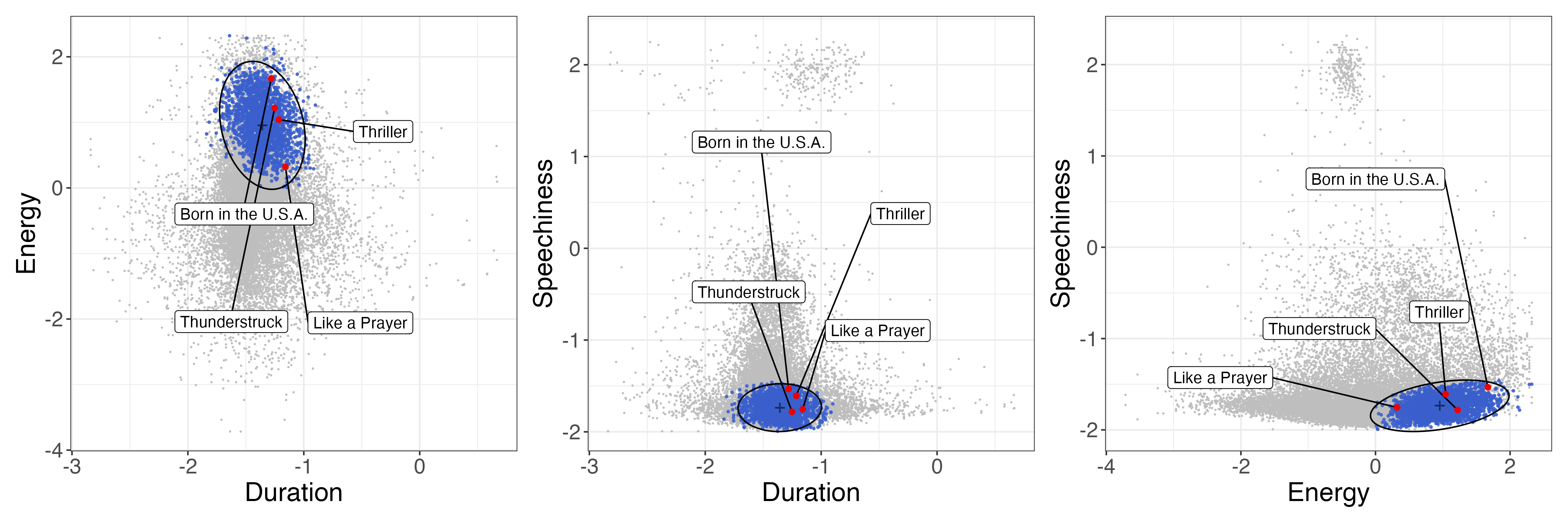}
    \caption{Pairwise scatterplots of song features. Colored (blue) points correspond to songs in one observational cluster, and red points highlight four famous tracks. The black crosses and ellipses indicate the estimated mean and variance/covariance.}
    \label{fig:OCanalysis}
\end{figure}

\textbf{Analysis of the clusters of songs (OC)}. Until now, we have only discussed the distributional clusters. Nonetheless, observational clusters allow for a refined analysis of the similarities between songs. Here, we show how we can take advantage of the shared-atom structure of our model to find songs that have similar characteristics, despite belonging to artists in different DCs.
We report an example in the scatterplots in Figure~\ref{fig:OCanalysis}. We highlighted songs assigned to the same OC: this cluster contains 2057 songs authored by 63 artists (belonging to 6 different DCs).
The red points identify four famous songs: ``Like a Prayer'' by Madonna, ``Thriller'' by Micheal Jackson, ``Thunderstruck'' by AC/DC, and ``Born in the U.S.A.'' by Bruce Springsteen. Despite belonging to different musical genres (a trait captured by the DC), the scores of these songs are similar, indicating that all these pieces have common characteristics. Specifically, they all have high energy, average duration, and low speechiness.

\section{Discussion}
In this paper, we developed a novel Bayesian nonparametric model for density estimation and clustering with grouped data, which conveys a two-level partition of groups and observations.
The advantages of the proposed modeling framework are threefold. First, it allows shared observational clusters across different random measures; second, it provides a more flexible correlation structure, which induces improved distributional clusters' characterization. Finally, its simple formulation allows the derivation of efficient estimation algorithms to scale up its applicability to large multivariate datasets with thousands of observations and hundreds of groups.

The proposed work stimulates several questions that are worth investigating. 
The new insights into the interaction between common atoms and stick-breaking weights pave the way for additional research on other nested models with dependent DP~\citep{Quintana2022}. In particular, it brings attention back to the possibility of directly exploiting the stochastic ordering of the atoms for modeling purposes, as done, for example, by~\citet{Griffin2006} with the order-based dependent DP. 
This reasoning is not limited to nested mixtures: nonparametric priors based on common atoms have been used, for example, by~\citet{Chandra2022} in the context of regression in clinical trials.

\bedit
The advantages of using finite mixtures within the proposed nested setting could be further enhanced by employing more flexible distributions than the Dirichlet. \citet{Argiento2022}, for example, introduced a class of priors based on the normalization of a point process which includes the Dirichlet mixture model as a particular case, and that was recently extended to the hierarchical setting~\citep{Colombi2023}. 
Other finite-dimensional nonparametric priors were proposed by~\citet{rigon1} and~\citet{rigon2}, which introduced flexible and robust alternatives to the Dirichlet that reduce the influence of the mass parameter and allow for more control over the expected number of clusters and resulting partition. 
\eedit

\bedit
Another crucial aspect of the proposed framework is its applicability to data with large sample sizes and high dimensionality. We addressed this issue from a computational perspective, implementing an efficient posterior inference algorithm. 
Additionally, we focused on prior properties that are not affected by the particular choice of the likelihood and base measure, and thus apply also to the multivariate case. 
However, it could be interesting to study how the particular structure and covariance of the data affect the clustering results in such multivariate settings.
Moreover, instead of relying on fast computation, another possible approach could be to leverage models developed for high-dimensional frameworks and extend them to our nested setting. For example, one could induce a sparser partition of the data via repulsive mixtures~\citep[see, e.g.,][]{beraha2022,ghilotti2023}, or explicitly assume the existence of a set of low-dimensional latent variables that drive the clustering~\citep{chandra2023}.
\eedit

Finally, the proposed nested variational algorithm -- developed for partially exchangeable data -- can be easily extended to other similar frameworks. 
\bedit
Stochastic variational inference~\citep{hoffman13a} solutions might be investigated to grant the immediate applicability of these models to the original Spotify dataset, including hundreds of thousands of observations. 
\eedit
Additionally, one could devise a similar finite-infinite model to account for separable exchangeability. For example, our hybrid mixture weight specification, along with a VI approach, could be applied to the common atoms model proposed by~\cite{Lin2021}, granting an efficient and powerful nonparametric method for matrix biclustering.

%% file: 03_text_suppl.tex


{\hypersetup{linkcolor=black}\parskip=0.2em
\tableofcontents
}
\clearpage

\clearpage
\section{Background}
\subsection{Review and comparison of nDP, CAM, and fCAM}
In this Section, we provide additional details about the nested Dirichlet process~\citep[nDP,][]{Rodriguez2008}, the common atoms model~\citep[CAM,][]{Denti2021} and its finite version~\citep[fCAM,][]{DAngelo2022}.
We consider the setting outlined in Section~2 of the main paper, particularly in Equation~1. All these processes define a prior on the collection of dependent random distributions $G_1, \dots, G_J$ that induces a simultaneous clustering of the groups and observations. Here, we summarize their structure, highlighting the differences in the underlying distributional assumptions.
\paragraph{The nested Dirichlet process}
The nDP was introduced by~\citet{Rodriguez2008} and assumes that
\begin{gather*}
       G_1,\dots,G_J \mid Q \overset{iid}{\sim} Q, \quad \quad
    Q = \sum_{k= 1}^{\infty} \pi_k \delta_{G^*_k},\\
    G^*_k = \sum_{l= 1}^{\infty}\omega_{l,k}\delta_{\theta^*_{l,k}},\\
    \theta^*_{l,k}\overset{iid}{\sim}H \quad \text{for }\,  l,k\geq 1,
\end{gather*}
with $H$ a non-atomic distribution defined on $(\Theta,\mathcal{X})$, $\{\pi_k\}_{k=1}^{\infty} \sim \mathrm{GEM}(\alpha)$ and $\{\omega_{l,k}\}_{l=1}^{\infty} \sim \mathrm{GEM}(\beta)$ for $k\geq 1$.
Such a construction implies that, \textit{a priori}, for two groups $j$ and $j'$, $\mathbb{P}[G_j=G_{j'}] = 1/(\alpha+1)>0$, hence inducing a clustering of the groups. Additionally,~\citet{Rodriguez2008} noticed how the discreteness of the $G^*_k$ simultaneously conveys a clustering of observations. Since each distributional atom $G^*_k$ is based on a different set of observational atoms $\{\theta^*_{l,k}\}_{l=1}^{\infty}$ sampled from a continuous distribution, two observations $\theta_{i,j}$ and $\theta_{i',j'}$ can be clustered together only if they belong to the same group (i.e., $j=j'$), or if their groups are assigned to the same distributional cluster (i.e., $G_j = G_{j'}$). 

However, ~\citet{Camerlenghi2019} noticed how this construction leads to an undesired side effect. Whenever two observations in different groups are assigned to the same observational atom ($\theta_{i,j} = \theta_{i',j'}$), the nDP automatically treats the two groups as fully exchangeable, forcing unwanted homogeneity in the two distributions. 

\paragraph{The common atoms model} To overcome this issue,~\citet{Denti2021} introduced the CAM, a modification of the nDP that enhances the flexibility of the dependence structure between groups. This improvement is achieved by forcing common observational atoms among all the $G_j$'s. Specifically, the CAM reformulates the distributional atoms $G^*_k$ as 
\begin{gather*}
    G^*_k = \sum_{l= 1}^{\infty}\omega_{l,k}\delta_{\theta^*_{l}} ,\\
    \theta^*_{l}\overset{iid}{\sim}H  \quad \text{for }\,  l\geq 1,
\end{gather*}
where, similarly to the nDP, $\{\pi_k\}_{k=1}^{\infty} \sim \mathrm{GEM}(\alpha)$ and $\{\omega_{l,k}\}_{l=1}^{\infty} \sim \mathrm{GEM}(\beta)$ for $k\geq 1$. Here, it is key that the set of observational atoms $\{\theta^*_{l}\}_{l=1}^{\infty}$ is shared by all distributional atoms. This modeling choice prevents the model from collapsing since now it is possible to allocate two observations in different groups to the same cluster, even if the two samples are modeled by different distributions (i.e., $\mathbb{P}[\theta_{i,j}=\theta_{i',j'}\mid G_j \neq G_{j'}]>0$ for $j\neq j'$).

In this construction the distributional atoms $G^*_k$ differentiate only through the weights $\{\omega_{l,k}\}_{l=1}^{\infty}$, which are sampled independently for each $k\geq 1$. However, the commonality of the atoms induces a strong similarity between the $G^*_k$'s. This similarity becomes apparent when considering the prior correlation $\rho_{j,j'}=\mathrm{Corr}(G_j(A), G_{j'}(A))$, which is restricted to the interval $(0.5,1)$.

\paragraph{The finite common atoms model} Leveraging on the recent developments in finite mixture models,~\citet{DAngelo2022} combined Dirichlet mixtures with the structure of the CAM, proposing the fCAM. This model reformulates both $Q$ and the $G^*_k$ as discrete finite distributions, i.e.,
\begin{gather*}
        Q = \sum_{k= 1}^{K} \pi_k \delta_{G^*_k},\quad \quad
    G^*_k = \sum_{l= 1}^{L}\omega_{l,k}\delta_{\theta^*_{l}} ,
\end{gather*}
where, similarly to the CAM, $\theta^*_{l}\overset{iid}{\sim}H$ for $l=1,\dots,L$. Here, however, the sequences of weights are finite, with 
\begin{gather*}
  \{\pi_k\}_{k=1}^{K}\mid K \sim \mathrm{Dirichlet}_K(\alpha/K,\dots,\alpha/K),\quad \quad K\sim p(K),\\ 
\{\omega_{l,k}\}_{l=1}^{L}\mid L \sim \mathrm{Dirichlet}_L(\beta/L,\dots,\beta/L) 
 \quad \text{for }\,  l\geq 1, \quad \quad L\sim p(L).  
\end{gather*}
To ensure flexibility, they further assumed beta-negative-binomial distributions on the dimensions $K$ and $L$, following the work of~\citet{fs2011}. As explained in our main paper, the fCAM showed improved performance and flexibility compared to CAM. Indeed, given the terminology we introduce, fCAM can be seen as a first proposal of a shared atom nested model.

The three processes thus differentiate through the structure of the distributional atoms and the distributions of the sequences of weights. Table~\ref{tab:models} summarizes these specifications for the three models. Although it might seem that the different models only present minor differences, the implications of these modifications on the prior behaviors are significant.
\clearpage

\begin{table}[ht]
    \centering
    \begin{tabular}{cccc}
    \toprule
         Model & Distributional atom  & Distribution of $\bpi$ & Distribution of $\bomega_k$  \\
         \midrule
        \midrule
         nDP   &    $\displaystyle G^*_k = \sum_{l= 1}^{\infty}\omega_{l,k}\delta_{\theta^*_{l,k}}$ & $\{\pi_k\}_{k=1}^{\infty} \sim \mathrm{GEM}(\alpha)$   & $\{\omega_{l,k}\}_{l=1}^{\infty} \sim \mathrm{GEM}(\beta)$ \\
        \midrule
         CAM   &    $ \displaystyle G^*_k = \sum_{l= 1}^{\infty}\omega_{l,k}\delta_{\theta^*_{l}}$ & $\{\pi_k\}_{k=1}^{\infty} \sim \mathrm{GEM}(\alpha)$     & $\{\omega_{l,k}\}_{l=1}^{\infty} \sim \mathrm{GEM}(\beta)$ \\
        \midrule
         fCAM   &    $ \displaystyle G^*_k = \sum_{l= 1}^{L}\omega_{l,k}\delta_{\theta^*_{l}}$ & $\{\pi_k\}_{k=1}^{K} \sim \mathrm{Dir}_K(\alpha/K)$ & $\{\omega_{l,k}\}_{l=1}^{L} \sim \mathrm{Dir}_L(\beta/L)$ \\
    \bottomrule
    \end{tabular}
    \caption{Summary of the nDP, CAM, and fCAM.}
    \label{tab:models}
\end{table}

\subsection{Interaction between the SB weights and the shared atoms in the CAM: visual intuition}\label{subsec::interaction_SB}
\begin{figure}[bh!]
    \centering
    \includegraphics[width=.9\linewidth]{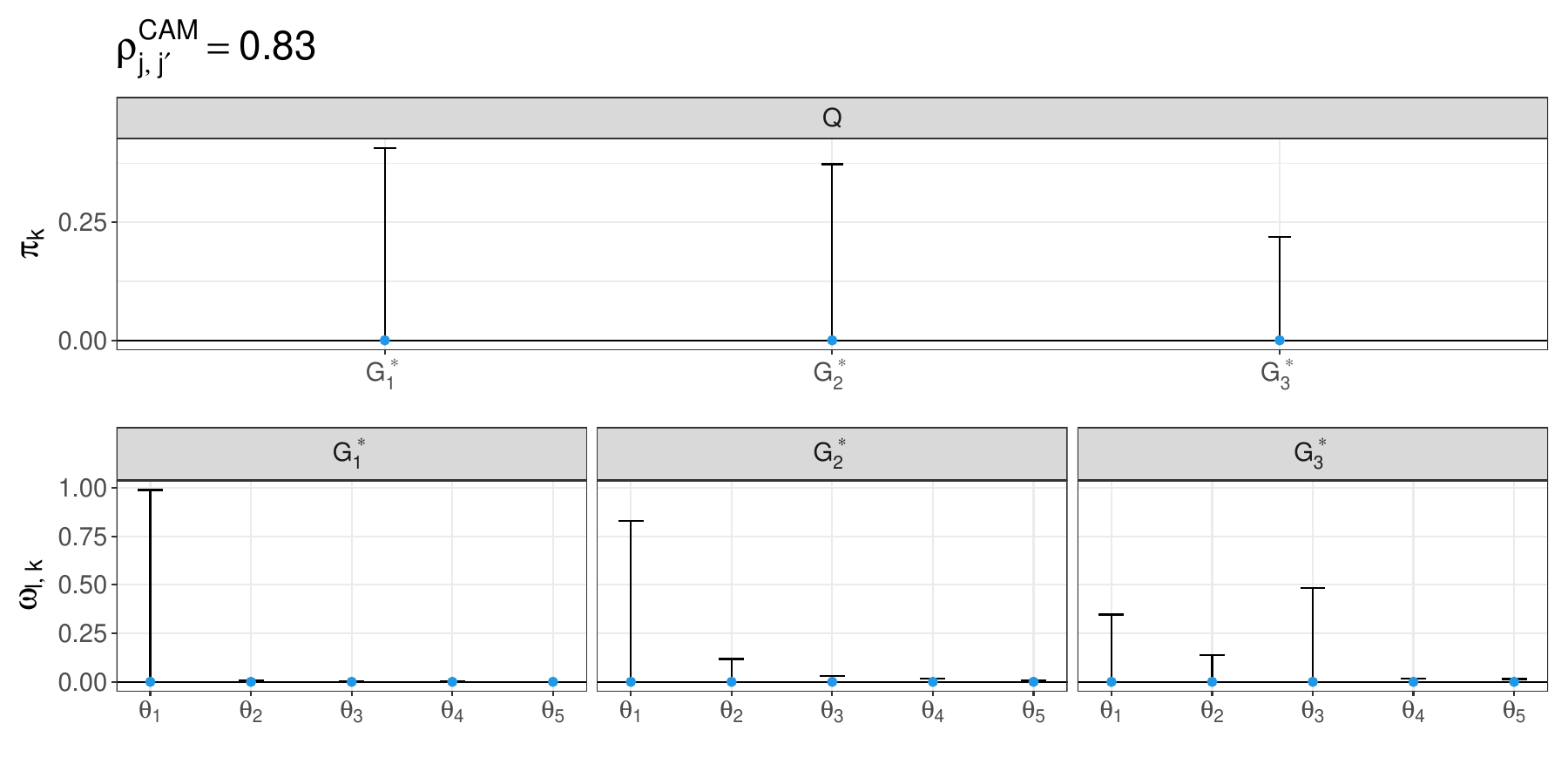}
    \caption{Realizations of stick-breaking weights of the CAM, sampled from processes with concentration parameters $\alpha=\beta=1$.}
    \label{suppl::fig::ordering1}
\end{figure}

Visual examples of the interaction between common atoms and stick-breaking weights using the nonparametric CAM are displayed in Figure~\ref{suppl::fig::ordering1}. The top two panels show an instance of how the mass can be placed on distributional and observational atoms, generated via a DP stick-breaking with concentration parameters $\alpha=\beta=1$. We note that the random measures $G_1^*$ and $G_2^*$ are similar even in such a common specification, placing most of the mass on the shared atom $\theta_1$, inducing very high correlation ($0.83$). This drawback is exacerbated in the bottom two rows, generated by considering the ``extreme'' setting $\alpha=\beta=0.001$.

\begin{figure}
    \centering
    \includegraphics[width=.9\linewidth]{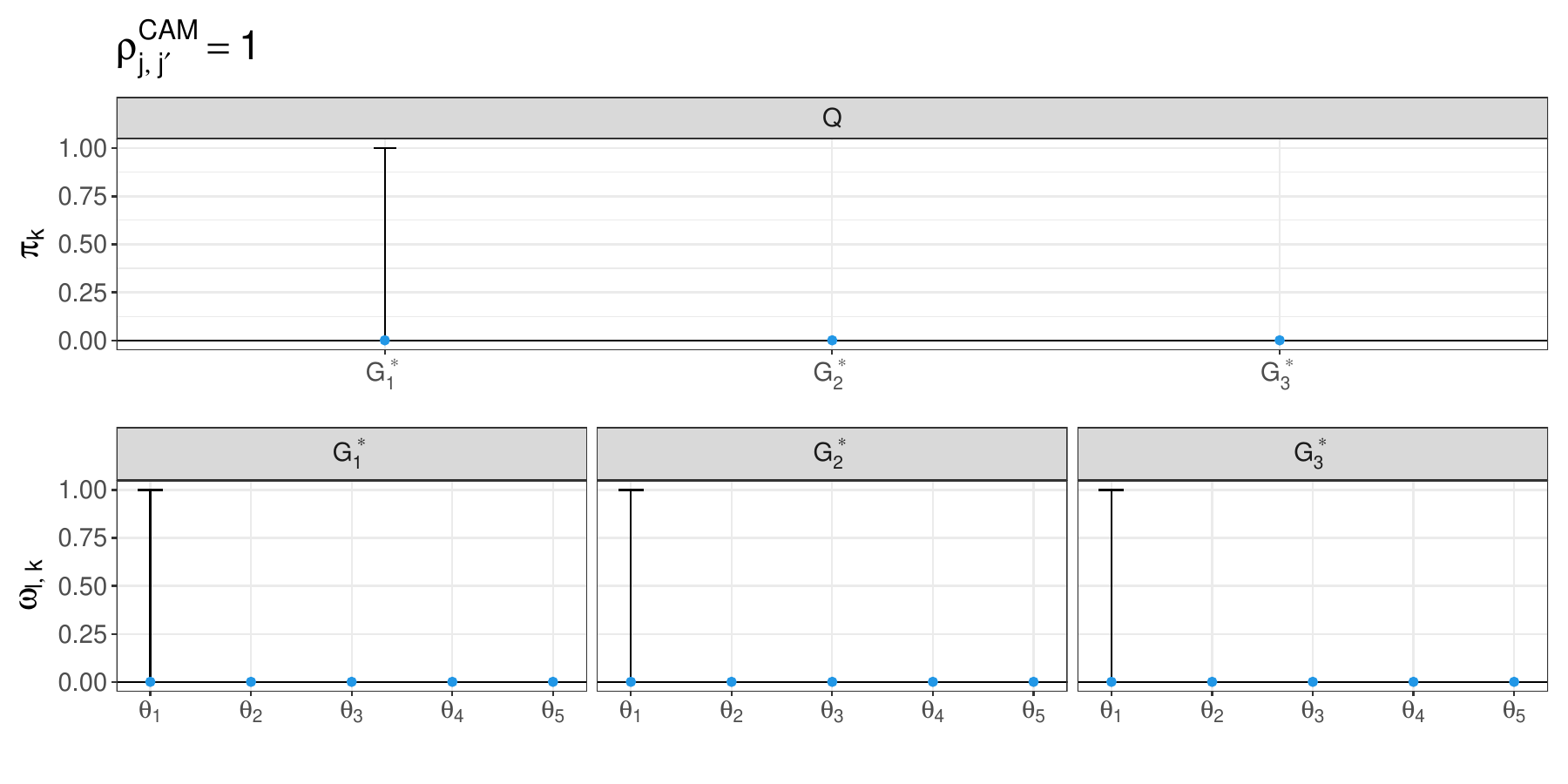}
    \caption{Realizations of stick-breaking weights of the CAM, sampled from processes with concentration parameters $\alpha=\beta=0.001$}
    \label{suppl::fig::ordering2}
\end{figure}

\section{Derivation of the prior properties} 
\label{suppl::sec::fSAN_proofs}

\subsection{Proofs for the finite SAN prior}
\subsubsection*{Co-clustering probability}
Consider the distributions $Q$, $G_j$, and $G_{j'}$ as defined in Section~2.1 of the paper. We first derive the distributional co-clustering probability, i.e., $\prob{G_j = G_{j'}}$:
\begin{align*}
    &\begin{aligned}
    \prob{G_j = G_{j'}\mid Q} &= \sum_{k=1}^K \prob{G_j = G_{j'} = G_k^*\mid Q}\\
    &= \sum_{k=1}^K \prob{G_j = G_k^*\mid Q} \prob{G_{j'} = G_k^*\mid Q} = \sum_{k=1}^K \pi_k^2,
    \end{aligned}\\
    &\begin{aligned}
    \prob{G_j = G_{j'}} &= \E{\prob{G_j = G_{j'}\mid Q} }\\
    &= \E{\sum_{k=1}^K \pi_k^2} = \sum_{k=1}^K \E{\pi_k^2} = \frac{1+a}{1+Ka},
    \end{aligned}
\end{align*}
where the last step follows from the assumption that $\bm{\pi} \sim \Dirichlet_K(a,\dots,a)$, for which is known that $\E{\pi_k^2} = (1+a)/(K(Ka+1))$. \\

\noindent We now turn to the observational co-clustering probability, i.e., $\prob{\theta_{i,j} = \theta_{i',j'}}$. Then,
\begin{align*}
    \prob{\theta_{i,j} = \theta_{i',j'}} 
    &= \E{\prob{\theta_{i,j} = \theta_{i',j'}\mid G_j,G_{j'}}} \\
    &= \E{\prob{\theta_{i,j} = \theta_{i',j'}\mid G_j=G_{j'}} \prob{G_j=G_{j'}} + \prob{\theta_{i,j} = \theta_{i',j'}\mid G_j\neq G_{j'}} \prob{G_j\neq G_{j'}} }\\
    &= \frac{1+a}{1+Ka}\, \E{\sum_{l=1}^L \omega_{l,j}^2} + \frac{a(K-1)}{1+Ka} \E{\sum_{l=1}^L \omega_{l,j} \,\omega_{l,j'}}\\
    & \overset{(\star)}{=} \frac{1}{(1+Ka)} \left[ \frac{(1+a)(1+b)}{(1+Lb)} + \frac{a(K-1)}{L}\right],
\end{align*}
where in step $(\star)$ we used
\begin{equation*}
    \E{\sum_{l=1}^L \omega_{l,j}^2} = \sum_{l=1}^L \E{ \omega_{l,j}^2} = \sum_{l=1}^L \frac{(1+b)}{L(1+Lb)} = \frac{1+b}{1+Lb},
\end{equation*}
and
\begin{equation*}
    \E{\sum_{l=1}^L \omega_{l,j} \, \omega_{l,j'}} = \sum_{l=1}^L \E{ \omega_{l,j} \, \omega_{l,j'} } \overset{\mathrm{ind.}}{=} \sum_{l=1}^L \E{ \omega_{l,j} }^2 = \frac{1}{L}.
\end{equation*}\\

\subsubsection*{Correlation}
To derive the correlation, the first step is to compute the covariance between random probability measures evaluated at two Borel sets $A$ and $B$, i.e., 
$$\text{Cov}(G_j(A),G_{j'}(B)) = \E{G_j(A)G_{j'}(B)} - \E{G_j(A)}\E{G_{j'}(B)}.$$
The first term is computed as
\begin{align*}
        \E{G_j(A)G_{j'}(B)} &= \E{\E{G_j(A)G_{j'}(B)\mid Q}} \\
        &= \E{ \sum_{k=1}^K \pi_k^2 G^*_k(A)G^*_k(B) + \sum_{k_1\neq k_2} \pi_{k_1}\pi_{k_2} G^*_{k_1}(A)G^*_{k_2}(B)}\\
        &= \sum_{k=1}^K \E{\pi_k^2} \E{ G^*_k(A)G^*_k(B)} + \E{\sum_{k_1\neq k_2} \pi_{k_1}\pi_{k_2}} \E{G^*_{k_1}(A)G^*_{k_2}(B)}\\
        &\overset{(\star\star)}{=} q_a^K \left( \frac{1+b}{1+Lb} H(A\cap B) + \frac{b(L-1)}{1+Lb} H(A) H(B) \right) +\\ &\qquad+ (1-q_a^K) \left(\frac{1}{L} H(A\cap B) + \frac{L-1}{L} H(A)H(B) \right)\\
        &= H(A\cap B) \left[ q_a^K \frac{1+b}{1+L b} + \frac{1-q}{L} \right] + H(A)H(B) \left[ q_a^K\frac{b(L-1)}{1+Lb} + (1-q_a^K) \frac{L-1}{L}\right],
\end{align*}
where we denoted with $q_a^K = \sum_{k=1}^K \E{\pi_k^2}$.
Step $(\star\star)$ follows from the fact that
\begin{align*}
  \E{ G^*_k(A)G^*_k(B)} &= \E{ \sum_{l=1}^L \omega_{l,k} \delta_{\theta^*_l}(A) \sum_{l=1}^L \omega_{l,k} \delta_{\theta^*_l}(B) } \\
    &= \E{ \sum_{l=1}^L \omega_{l,k}^2 \delta_{\theta^*_l}(A)  \delta_{\theta^*_l}(B) + \sum_{l \neq h} \omega_{l,k} \omega_{h,k}\delta_{\theta^*_l}(A)  \delta_{\theta^*_h}(B)}\\
    &= \sum_{l=1}^L \E{\omega_{l,k}^2 } H(A\cap B) + \left(1-\sum_{l=1}^L \E{\omega_{l,k}^2 }\right) H(A)H(B)\\
    &= \frac{1+b}{1+Lb} H(A\cap B) + \frac{b(L-1)}{1+Lb} H(A) H(B)  
\end{align*} and that
\begin{align*}
   \E{ G^*_{k_1}(A)\:G^*_{k_2}(B)} &= 
   \E{ \sum_{l=1}^L \omega_{l,k_1} \delta_{\theta^*_l}(A) \sum_{l=1}^L \omega_{l,k_2} \delta_{\theta^*_l}(B) } \\
    &= \E{ \sum_{l=1}^L \omega_{l,k_1}\omega_{l,k_2} \delta_{\theta^*_l}(A)  \delta_{\theta^*_l}(B) + \sum_{l \neq h} \omega_{l,k_1} \omega_{h,k_2}\delta_{\theta^*_l}(A)  \delta_{\theta^*_h}(B)}\\
    &\overset{i.i.d.}{=} \sum_{l=1}^L \E{\omega_{l,k}}^2 H(A\cap B) + \left( 1-\sum_{l=1}^L \E{\omega_{l,k}}^2 \right) H(A)H(B)\\
    &= \frac{1}{L} H(A\cap B) + \frac{L-1}{L} H(A)H(B).
\end{align*} 
The second term that appears in the covariance is simply $\E{G_j(A)}\E{G_{j'}(B)} = H(A)H(B)$, hence with some simple algebra we obtain
\begin{equation*}
    \begin{aligned}
    \text{Cov}(G_j(A), G_{j\prime}(B)) &= \frac{q_a^K(L-1)+Lb+1}{L(1+Lb)}\left(H(A\cap B)-H(A)H(B)\right).
    \end{aligned}
\end{equation*}

The correlation follows immediately by dividing by the variance $\mathrm{var}(G_j(A)) = \big(H(A)(1-H(A))\big) (1+b)/(1+Lb)$.

\subsection{Proofs for the finite-infinite SAN}
\label{suppl::sec::fiSAN_proofs}
The derivations of correlation between random measures and the co-clustering probability for the fiSAN model follow the ones just outlined for the finite case. The two models only differ in the specification of the distributional weights $\{\pi_k\}_{k=1}^K$, which are now assigned a Dirichlet process prior with concentration parameter $\alpha$ in place of the finite-dimensional Dirichlet. Hence we can exploit of the derivation of the original CAM model for the nonparametric part (see the Supplementary Material of~\citealp{Denti2021}). In particular, the distributional co-clustering probability is now
\begin{align*}
    \prob{G_j = G_{j'}} &= \E{\prob{G_j = G_{j'}\mid Q} }\\
    &= \E{\sum_{k=1}^{\infty} \pi_k^2} = \sum_{k=1}^{\infty} \E{\pi_k^2} = \frac{1}{\alpha + 1}.
\end{align*}
This result can be used in the previous computation to obtain the corresponding quantities.

\subsection{Derivation of the partially exchangeable partition probability functions}\label{suppl::subseq::peppf}
Recall the notation and setting described in Sec.~2.2 of the paper.
Our goal is to study the distribution of the partition of $(\boldsymbol{\theta}_1,\boldsymbol{\theta}_2)$ induced by $(\bM_1,\bM_2)$, which is equivalently expressed as a function of the frequencies $(\bm{n}_1,\bm{n}_2) = \big( (n_{1,1},\dots,n_{L,1}), (n_{1,2},\dots,n_{L,2}) \big)$.

First, we need to compute the marginal distribution of $(\bM_1,\bM_2\mid S_1,S_2)$, with $\bomega_k$, $k=1,2$, integrated out. In particular, we need to distinguish two cases: (i) $S_1 = S_2 = k$, (ii) $S_1 =k_1$ and $S_2=k_2$ with $k_1 \neq k_2$. From these expressions, we can derive a distribution on the partitions by integrating out the specific cluster labels, thus obtaining a function of the frequencies. The expression of the partially exchangeable partition probability function (pEPPF) is then obtained as
\begin{equation*}
\begin{aligned}
  \Pi_{N_1,N_2,s}(\bm{n}_1,\bm{n}_2)=&  \prob{\bm{n}_1,\bm{n}_2} \\
  =& \underbrace{\prob{\bm{n}_1,\bm{n}_2\mid S_1 = S_2} \prob{S_1 = S_2}}_{(i)} + \underbrace{\prob{\bm{n}_1,\bm{n}_2\mid S_1 \neq S_2} \prob{ S_1 \neq S_2}}_{(ii)}.
\end{aligned}
    \label{suppl::eq::peppf_generic}
\end{equation*}

We first investigate case (i). Whenever the two units are assigned to the same distributional cluster, we find ourselves in the fully exchangeable setting. Therefore, we can consider $\tilde{\boldsymbol{\theta}} =\{\boldsymbol{\theta}_1 \cup \boldsymbol{\theta}_2\}$ as a unique sample of size $N_1+N_2$. The membership labels for this sample are in expressed in $\tilde{\bM}$, obtained concatenating the vectors $\bM_1$ and $\bM_2$, each label being extracted from a categorical random variable with weights $\bomega_k$. The marginal distribution of $\tilde{\bM}$ is given by:
\begin{align*}
    \prob{\tilde{\bM} \mid S_1 = S_2 = k} 
    &= \int_{(0,1)^L} {\omega_{1,k}}^{n_{1,1}+n_{1,2}} \dots {\omega_{L,k}}^{n_{L,1}+n_{L,2}} p(\bomega_k) \mathrm{d}\bomega_k\\
    &=  \int_{(0,1)^L} \frac{\Gamma(Lb)}{ \Gamma(b)^L} \, {\omega_{1,k}}^{n_{1,1}+n_{1,2}+b-1} \dots {\omega_{L,k}}^{n_{L,1}+n_{L,2}+b-1} \mathrm{d}\bomega_k\\
    &= \frac{\Gamma(Lb) \, \Gamma(b)^{-L}}{ \Gamma(Lb+ N_1+N_2)} \prod_{l=1}^L \Gamma(b + n_{l,1} + n_{l,2})\\
    &= \frac{\Gamma(Lb) \, \Gamma(b)^{-(s_1+s_0+s_2)}}{ \Gamma(Lb+ N_1+N_2)} \prod_{l: n_{l,1} + n_{l,2} > 0} \Gamma(b + n_{l,1} + n_{l,2}).
\end{align*}
Following an argument similar to the proof presented in~\citet{miller2019}, we can obtain the desired distribution $\prob{(\bm{n}_1,\bm{n}_2)\mid S_1 = S_2 = k}$ by summing over all configurations $(\tilde{\bM}) \in \{1,\dots,K\}^{N_1+N_2}$ such that the induced partition $C_M$ is equal to the configuration~$(\bm{n}_1,\bm{n}_2)$:
\begin{align*}
    \prob{\bm{n}_1,\bm{n}_2 \mid S_1 = S_2 = k} &= \sum_{\tilde{\bM} \in \{1,\dots,K\}^{N_1+N_2}} \ind{C_M = (\bm{n}_1,\bm{n}_2)} \frac{\Gamma(Lb) \, \Gamma(b)^{-L}}{ \Gamma(Lb+ N_1+N_2)} \prod_{l=1}^L \Gamma(b + n_{l,1} + n_{l,2}) \\
    & = \binom{L}{s_0, s_1, s_2}\, s_0!\, s_1!\, s_2!\, \frac{\Gamma(Lb) \, \Gamma(b)^{-L}}{ \Gamma(Lb+ N_1+N_2)} \prod_{l=1}^L \Gamma(b + n_{l,1} + n_{l,2}) \\
    & = \frac{L!}{(L-s)!} \frac{\Gamma(Lb) \, \Gamma(b)^{-L}}{ \Gamma(Lb+ N_1+N_2)} \prod_{l=1}^L \Gamma(b + n_{l,1} + n_{l,2}). \\
    & = \Phi^{D_L}_{N_1+N_2,s}(\bm{n}_{1} + \bm{n}_{2}).\\
\end{align*}

Then, we consider case (ii). Here, the two samples are conditionally independent, although having support on the same set $\{1,2,\dots, L\}$. We obtain
\begin{align*}
    \mathbb{P}(\bM_1, &\bM_2  \mid S_1 = k_1, S_2 = k_2)=
    \prob{ \bM_1\mid S_1 = k_1 } \prob{ \bM_2 \mid S_2 = k_2} \\
    &= \int_{(0,1)^{L}} {\omega_{1,k_1}}^{n_{1,1}} \dots {\omega_{L,k_1}}^{n_{L,1}} p(\bomega_{k_1}) \mathrm{d}\bomega_{k_1}  \int_{(0,1)^{L}} {\omega_{1,k_2}}^{n_{1,2}} \dots {\omega_{L,k_2}}^{n_{L,2}} p(\bomega_{k_2}) \mathrm{d}\bomega_{k_2}\\
    &= \int_{(0,1)^L} \frac{\Gamma(Lb)}{ \Gamma(b)^L} \, {\omega_{1,k_1}}^{n_{1,1}+b-1} \dots {\omega_{L,k}}^{n_{L,1}+b-1} \mathrm{d}\bomega_{k_1}  \int_{(0,1)^L} \frac{\Gamma(Lb)}{ \Gamma(b)^L} \, {\omega_{1,k_2}}^{n_{1,2}+b-1} \dots {\omega_{L,k}}^{n_{L,2}+b-1} \mathrm{d}\bomega_{k_2}\\
    &= \frac{\Gamma(Lb) \Gamma(b)^{-L}}{\Gamma(Lb+N_1)} \prod_{l=1}^L \Gamma(b + n_{l,1})  \:\frac{\Gamma(Lb) \Gamma(b)^{-L}}{\Gamma(Lb+N_2)} \prod_{l=1}^L \Gamma(b + n_{l,2})\\
    &= \frac{\Gamma(Lb)^2 \Gamma(b)^{-2L}}{\Gamma(Lb+N_1)\Gamma(Lb+N_2)} \prod_{l=1}^L \Gamma(b + n_{l,1})\Gamma(b + n_{l,2})
\end{align*}
and, following a similar argument as before, the distribution of the partition is 
\begin{align*}
    \mathbb{P}\big( (\bm{n}_1,&\bm{n}_2)\mid S_1 = k_1, S_2 = k_2 \big) = \prob{\bm{n}_1\mid S_1 = k_1} \prob{\bm{n}_2 \mid \bm{n}_1, S_1 = k_1, S_2 = k_2}\\
    &= \binom{L}{s_0 \:s_1} s_0! s_1! \, \frac{\Gamma(Lb) \Gamma(b)^{-L}}{\Gamma(Lb+N_1)} \prod_{l=1}^L \Gamma(b + n_{l,1}) \cdot   
    \binom{L-s_0-s_1}{s_2} s_2!\,
    \frac{\Gamma(Lb) \Gamma(b)^{-L}}{\Gamma(Lb+N_2)} \prod_{l=1}^L \Gamma(b + n_{l,2})\\
    &= \frac{L!}{(L-s_0-s_1-s_2)!}
    \frac{\Gamma(Lb)^2 \Gamma(b)^{-2L}}{\Gamma(Lb+N_1)\Gamma(Lb+N_2)} \prod_{l=1}^L \Gamma(b + n_{l,1})\Gamma(b + n_{l,2})\\
    & = \frac{(L-s_0-s_1)!(L-s_0-s_2)!}{L!(L-s_0-s_1-s_2)!}
    \frac{L!}{(L-s_0-s_1)!}\frac{L!}{(L-s_0-s_2)!}\times\\&\quad\quad\times\frac{\Gamma(Lb)^2 \Gamma(b)^{-2L}}{\Gamma(Lb+N_1)\Gamma(Lb+N_2)} \prod_{l=1}^L \Gamma(b + n_{l,1})\Gamma(b + n_{l,2})\\
    & = \mathcal{C}^L_{s_0,s_1,s_2}
    \Phi^{D_L}_{N_1,s_0+s_1}( \bm{n}_{1})\Phi^{D_L}_{N_2,s_0+s_2}(\bm{n}_{2}).
\end{align*}

Finally, we only need to remove the dependence from the distributional cluster allocation. Since we assumed that all Dirichlet distributions have the same hyperparameter $a$, we do not need to keep track of the specific allocation $k$. We only need to know whether the two groups are assigned to the same distributional cluster or two different ones. This implies that the desired pEPPF is simply obtained as
\begin{equation}
\begin{aligned}
    \Pi_{N_1,N_2,s}(\bm{n}_1,\bm{n}_2)  = q_\alpha \Phi^{D_L}_{N_1+N_2,s}(\bm{n}_{1} + \bm{n}_{2})
     + (1-q_\alpha) \mathcal{C}^L_{s_0,s_1,s_2}
    \Phi^{D_L}_{N_1,s_0+s_1}( \bm{n}_{1})\Phi^{D_L}_{N_2,s_0+s_2}(\bm{n}_{2})
\end{aligned}
\label{suppl::eq::peppf_fiSAN}
\end{equation}
where $q_\alpha = q^K_a$ for the finite SAN, and $q_\alpha = (\alpha+1)^{-1}$ for the finite-infinite SAN.

\subsection{Limiting behavior of the pEPPF}
\label{subsec::relationships}
Once we obtain the pEPPF for the finite SAN, we can investigate the behavior of our shared atoms model at the limit. Specifically, we now consider $\alpha = a/K$, and $b=\beta/L$. The pEPPF of the fiSAN can indeed be derived also as  $\lim_{K\rightarrow +\infty}\Pi_{N_1,N_2,s}^{(\fSAN)}(\bm{n}_1,\bm{n}_2)$. However, we focus now on $\lim_{L\rightarrow +\infty}\Pi_{N_1,N_2,s}^{(\fiSAN)}(\bm{n}_1,\bm{n}_2)$, where the latter is defined in~\eqref{suppl::eq::peppf_fiSAN}.

Recall that, for all $k$, $\bomega_k \sim \Dirichlet_L\left({\beta}/L,\dots,{\beta}/{L}\right)$.
It is immediate to see that under this specification of the parameters, the first term of the above convex combination converges to the $\eppf$ of a Dirichlet process with concentration parameter $\beta$~\citep{Green2001}.
The same holds for the two independent single-sample EPPFs in the second term.

Thus, we only need to compute the limit of the correction constant $\mathcal{C}^L_{s_0,s_1,s_2}$. Recall that by Stirling's approximation, for a given integer $r$, it holds $L!/(L-r)! \asymp L^{r}$, for $L\rightarrow+\infty$. Therefore $\mathcal{C}^L_{s_0,s_1,s_2} \asymp L^{-s_0}$.
Clearly, if $s_0>0$, the second term of the pEPPF converges to zero, and we are left with the standard EPPF of a DP in a fully exchangeable setting. When the two groups are assigned to separate distributional clusters, the necessary condition to prevent the model from collapsing is $s_0 = 0$.
Thus the limit of the pEPPF of the fiSAN can be written as
\begin{align*}\label{eq:cam_eppf}
\lim_{L\rightarrow\infty} \Pi_{N,s}^{(\fiSAN)}(\boldsymbol{n}_1,\boldsymbol{n}_2) 
&=  \frac{1}{\alpha+1} \Phi^{(DP)}_{N_1+N_2,s}(\boldsymbol{n}_1+\boldsymbol{n}_2) + \frac{\alpha}{\alpha+1} \Phi^{(DP)}_{N_1,s}(\boldsymbol{n}_1) \Phi^{(DP)}_{N_2,s}(\boldsymbol{n}_2)\ind{s_0=0}\\
&= \Pi_{N,s}^{(\mathrm{nDP})}(\boldsymbol{n}_1,\boldsymbol{n}_2),
\end{align*}
where $\Pi_{N,s}^{(\mathrm{nDP})}(\boldsymbol{n}_1,\boldsymbol{n}_2)$ is the pEPPF of the nested DP.
This limiting pEPPF is a linear combination between the fully exchangeable and the unconditional independence cases, coherently with the results displayed in~\citet{Camerlenghi2019}. This equation provides additional evidence that, at the limit, it is not enough to define the model using a shared set of atoms to prevent the collapsing to the fully exchangeable case while retaining a shared observational cluster. 
This is clear if we notice that we are indirectly asking to \textit{independently} sample the same value twice from an infinite sequence of values, an event with almost surely null probability.

\begin{figure}[ht!]
    \centering
    \includegraphics[width=\linewidth]{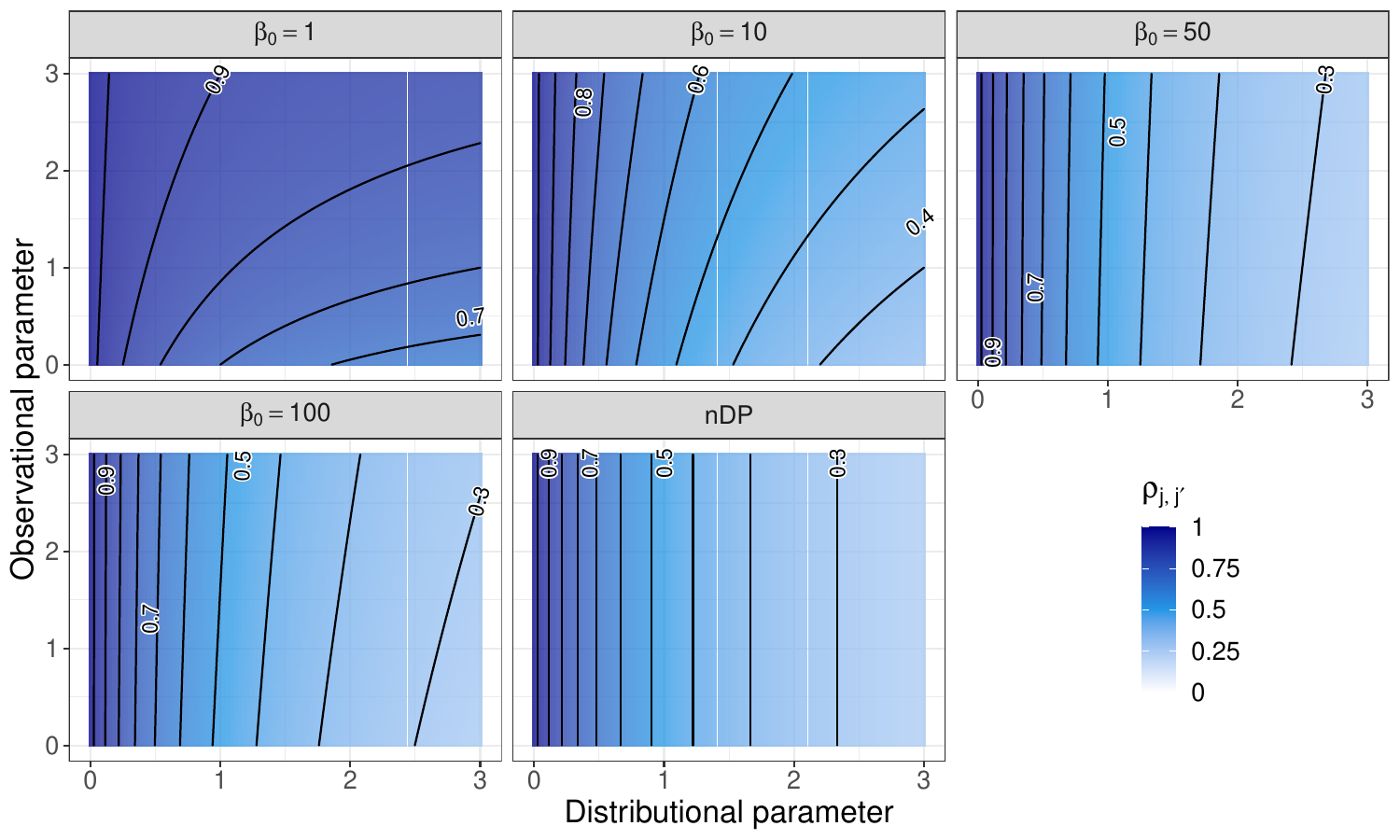}
    \caption{Heatmaps representing the correlation structures under the HHDP model for $\beta_0\in \{1,10,150,100\}$ and under the nDP model.}
    \label{fig:HHDPcorr}
\end{figure}

To actually avoid the issue, CAM exploits the combination of the order of the atoms and the stochastic order given by the stick-breaking construction. 
Similarly, the HHDP of~\cite{lijoi2023} prevents the collapsing by sampling the atoms of the different $G_k^*$'s from a discrete, albeit stochastic, distribution. This feature implies the unequal weighting the observational atoms across all the random measures (also inducing a sort of ``atom preference''). Once that preference is removed, the HHDP converges to the nDP as well, as $G_0$ converges to the non-atomic base measure $H$ (see also Figure~\ref{fig:HHDPcorr}). 

\begin{figure}
    \centering
\tikzstyle{block} = [rectangle, draw, text width=4em, text centered, rounded corners, minimum height=3em]
\tikzstyle{line} = [draw, -latex']
\begin{tikzpicture}[node distance=1cm and 3 cm]
    \node (init) {};
    \node [block] (B) {fnDP};
    \node [block, right=of B] (C) {nDP};
    \node [block, right=of C] (Q) {HHDP};
    \node [block, below=of C] (D) {fiSAN};
    \node [block, below=of B] (E) {fSAN};
    \node [block, below=of Q] (F) {CAM};

    \path [line] (B) -- node [midway,above] {$L,K \rightarrow +\infty$} (C);
    \path [line] (Q) -- node [midway,above] {$\beta_0 \rightarrow +\infty$} (C);
    \path [line] (E) -- node [midway,above] {$K \rightarrow +\infty$} (D);
    \path [line] (D) -- node [midway,right] {$L \rightarrow +\infty$} (C);
\end{tikzpicture}
    \caption{A diagram depicting the relations between the nested priors we discussed.}
    \label{fig:diagram}
\end{figure}
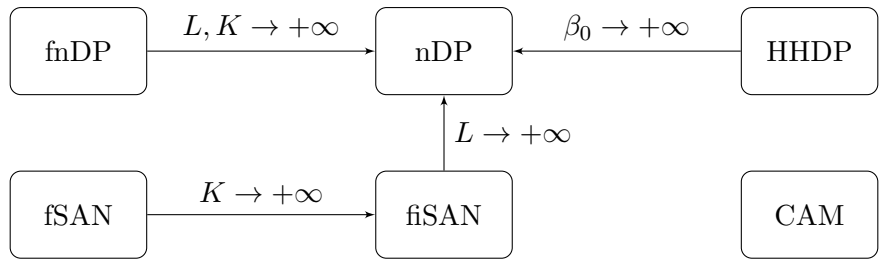

In Figure~\ref{fig:diagram}, we report a summarizing chart of the connections between the nested priors we discussed (fnDP indicates a ``finite nested DP'', i.e., a fSAN prior with non-shared atoms $\theta^*_{l,k}\sim H$). 


\clearpage
\section{Additional details on posterior inference}\label{suppl::sec::posterior_inference}
\subsection{Gibbs sampler algorithm}
\label{suppl::subsec::gibbs}
In this section, we derive the full conditional distributions to perform posterior inference using a Gibbs sampler algorithm.
To this end, we use the model formulation introduced in Sec.~3 of the paper based on the cluster allocation variables $\bM = \{ M_{i,j}, i=1,\dots,N_j, j=1,\dots, J\}$ and $\bS = \{S_j, j=1,\dots,J\}$.

For the observational mixtures, we consider $\bomega_k \sim \Dirichlet_{L}(b,\dots,b)$ for all $k$.
The algorithm for the finite and for the finite-infinite SAN models is distinguished only in the steps involving the update of the distributional weights and of the associated hyperparameters. 
In particular, for the fSAN, the distributional mixture is $\bpi \sim \Dirichlet_{K}(a,\dots, a)$.
Differently, for the fiSAN, the steps for updating the distributional mixture are based on the slice sampler of~\citet{Denti2021}. Following their suggestion, for the slice sampler we use a deterministic sequence $\bm{\xi} = \{\xi_k\}_{k\geq 1}$, with $\xi_k = 0.5^k$ for $k\geq 1$.
Hence in the following, we distinguish the two cases when necessary, while for the rest of the steps, we outline a unique procedure.

The steps of the algorithm are:
\begin{enumerate}
\item Update the distributional cluster probabilities $\bpi$:
\begin{itemize}
    \item[fiSAN] \begin{enumerate}
        \item For $j=1,\dots,J$, sample the uniform random variables $u_j\sim \mathrm{Unif}(0,\xi_{S_j})$; compute the threshold $K^*$ for the number of needed mixture components~\citep{Denti2021}.
        \item For $k=1,\dots,K^*$, sample the distributional stick-breaking proportions $v_k$ independently from $v_k \sim \mathrm{Beta}(\bar{a}_{k} , \bar{b}_{k})$, where $ak = 1 + \sum_{j=1}^J \ind{S_j=k}$ and $\bar{b}_{k} = \alpha + \sum_{j=1}^J \ind{S_j>k}$. 
        \item Compute the mixture weights $\pi_k = v_k \prod_{j=1}^{k-1}(1-v_k)$ for $k=1,\dots,K^*$.
    \end{enumerate}
    \item[fSAN] Sample the weights from a Dirichlet distribution of dimension $K$: 
    $$(\pi_1,\dots,\pi_{K})\mid \bS, a \sim \Dirichlet_{K}\bigg(a + \sum_{j=1}^J \ind{S_j=1},\dots, a + \sum_{j=1}^J \ind{S_j=K}\bigg).$$
\end{itemize}
\item Update the observational cluster probabilities $\bomega_k$, for $k=1,\dots,K^*$ (fiSAN) or $k=1,\dots,K$ (fSAN),
\begin{equation*}
    (\omega_{1,k},\dots,\omega_{L^*,k}) \mid \bM, \bS, b \sim \Dirichlet_{L}\left(b + n_{1,k}, \dots, b + n_{L^*,k}\right)
\end{equation*}
where $n_{l,k} = \sum_{j=1}^J \sum_{i=1}^{N_j} \ind{ M_{i,j} = l \mid S_j=k }$.
\item Update the distributional cluster assignment $S_j$: to improve the mixing we use a collapsed Gibbs sampling step~\citep{Denti2021}. For $j=1,\dots,J$ sample a multinomial random variable
\begin{itemize}
    \item[fiSAN] $\pr(S_j = k \mid \bpi,\bomega, \bM) \propto \ind{u_j < \xi_k} \frac{\pi_k}{\xi_k}\: \prod_{i=1}^{N_j}\omega_{M_{i,j},k}, \quad k=1,\dots,K^*.$
    \item[fSAN] $\pr(S_j = k \mid \bpi,\bomega, \bM) \propto \pi_k \: \prod_{i=1}^{N_j}\omega_{M_{i,j},k}, \quad k=1,\dots,K.$
\end{itemize}
\item Update the observational cluster assignment $M_{i,j}$: again, it is multinomial with updated hyperparameters. For $j=1,\dots,J$, $i=1,\dots,N_j$, sample
\begin{equation*}
    \pr(M_{i,j}=l\mid \by, \bS,\bomega ) \propto \omega_{l,k}\: \phi_d(\by_{i,j} \mid \bmu_l,\bLambda^{-1}).
\end{equation*}
\item Update the distributional mixtures hyperparameters.
\begin{itemize}
    \item[fiSAN] Update the distributional DP hyper-parameter $\alpha$ as in Step 8 of Algorithm 1 of~\citet{Denti2021}.
\end{itemize}
\item Sample the model parameters $(\bmu_l,\bLambda_l)$ for $l = 1,\dots,L$ from a normal-Wishart distribution with updated hyperparameters (exploiting conjugacy).
\end{enumerate}

\subsection{Variational inference for the fSAN}\label{subsec::VI_fsan}
The structure of the VI algorithm for the fSAN model is analogous to that outlined in Section~3 for the fiSAN model, hence here we will only discuss the main differences. 
Clearly, being all quantities finite, here we do not need to define a truncation level $T$ for the variational family, hence $T=K$ in this framework.

The fully factorized family of distributions that we assume can be written as
\begin{equation*}
\begin{aligned}
    q_{\boldsymbol{\lambda}}(\bM, \bS, & \{\boldsymbol{\omega}_k\}_{k=1}^K, \bv, \{\bmu_l,\bLambda_l\}_{l=1}^L) =
     \prod_{j=1}^J q(S_j; \{\rho_{j,k}\}_{k=1}^K )\:
     \prod_{j=1}^J \prod_{i=1}^{N_j} q(M_{i,j}; \{ \xi_{i,j,l} \}_{l=1}^{L} ) \times\\      & 
     \times q(\bpi ; \{\tilde{p}_k\}_{k=1}^K)\:
     \prod_{k=1}^{K} q(\bomega_k ; \{p_{l,k}\}_{l=1}^L)  
     \prod_{l=1}^L q(\bmu_l,\bLambda_l ; \bm{m}_l, t_l, c_l, D_l),
\end{aligned}
\end{equation*}
where $q(S_j; \{\rho_{j,k}\}_{k=1}^K )$ and $q(M_{i,j}; \{ \xi_{i,j,l} \}_{l=1}^{L} )$ are multinomial distributions; $q(\bpi ; \{\tilde{p}_k\}_{k=1}^K)$ and $q(\bomega_k ; \{p_{l,k}\}_{l=1}^L)$ are Dirichlet distributions; and $q(\bmu_l,\bLambda ; \bm{m}_l, t_l, c_l, \boldsymbol{D}_l)$ are normal-Wishart distributions. 
Under this representation, the set of latent variables is $\boldsymbol{\Theta} = \left(\bS, \bM, \bpi, \{\boldsymbol{\omega}_k\}_{k=1}^K, \{\bmu_l,\bLambda\}_{l=1}^L \right)$ and the set of variational parameters is $\boldsymbol{\lambda} =\left(\boldsymbol{\rho}, \boldsymbol{\xi}, \tilde{\boldsymbol{p}}, \boldsymbol{p},\boldsymbol{m},\boldsymbol{t},\boldsymbol{c},\boldsymbol{D}\right)$.

\begin{algorithm}
\SetAlgoLined
{\small
\hspace*{0em}\textbf{Input:} $t \gets 0$. Randomly initialize $\boldsymbol{\lambda}^{(0)}$.\\
\hspace*{0em}\phantom{\textbf{Input:}  } Define the error bound $\epsilon$ and randomly set $\Delta>\epsilon$. 

\While{$\Delta(t-1,t) > \varepsilon$}{
    Let $\boldsymbol{\lambda} = \boldsymbol{\lambda}^{(t)};$ \\
    Set $t = t+1$ ; \\
    
    Update the variational parameters according to the following CAVI steps:
    \begin{enumerate}
    \item For $j=1,\dots,J$, $q^\star(S_{j})$ is a $K$-dimensional multinomial, with $q^\star(S_{j}=k)=\rho_{j,k}$.\\ Thus, for $k=1,\dots,K$,
      \begin{align*}
       \log\rho_{j,k} = 
            h_k(\tilde{\boldsymbol{p}})+ \sum_{l=1}^L \left(\sum_{i=1}^{N_j} \xi_{i,j,l}\right) h_l(\boldsymbol{p}_k),
    \end{align*}
    where $h_k(\boldsymbol{x}) = \psi(x_{k}) - \psi(\sum_{k=1}^K x_{k})$, with $\psi$ the digamma function.
    \item For $j=1,\dots,J$ and $i=1,\dots,N_j$, $q^\star(M_{i,j})$ is a $L$-dimensional multinomial, with $q^\star(M_{i,j}=l)=\xi_{i,j,l}$ for $l=1,\dots,L$,
    \begin{align*}
       \log\xi_{i,j,l} = \frac12\ell^{(1)}_{l} +\frac12\ell^{(2)}_{i,j,l} + \sum_{k=1}^T \rho_{j,k} h_{l}(\boldsymbol{p}_k).
    \end{align*}
    where $\ell^{(1)}_{l} =  \sum_{x=1}^d\psi\left( (c_l-x+1)/2\right) +d\log2 + \log|\boldsymbol{D}_l|$ and $\ell^{(2)}_{i,j,l}= - d/t_l- c_l  (\by_{i,j}-\bm{m}_l)^T \boldsymbol{D}_l(\by_{i,j}-\bm{m}_l)$
    \item For $k=1,\dots,K$, {$q^\star(\boldsymbol{\omega}_k)$} is $\Dirichlet_L(\boldsymbol{p}_k)$ with
    $ p_{l,k} = b + \sum_{j=1}^J \sum_{i=1}^{N_j}\xi_{i,j,l}\rho_{j,k}.$
    \item $q^\star(\bpi)$ is $\Dirichlet_K(\boldsymbol{s})$ with $\tilde{p}_k = a + \sum_{j=1}^J \rho_{j,k}$ for $k=1,\dots,K$.
    \item  For $l=1,\dots,L$ {$q^\star(\theta_l)$} is a $\mathrm{NW}(\bm{m}_l,t_l,c_l,\boldsymbol{D}_l)$ distribution with parameters
    \begin{align*}
        \bm{m}_l &= t_l^{-1}(\kappa_0\:\boldsymbol{\mu}_0+N_{\boldsymbol{\cdot} l}\bar{\by}_l),  \quad 
        t_l = \kappa_0 + N_{\boldsymbol{\cdot} l}, \quad 
        c_l = \tau_0 + N_{\boldsymbol{\cdot} l},\\
        \boldsymbol{D}_l^{-1} &= \boldsymbol{\Gamma}_0^{-1} +\frac{\kappa_0 N_{\boldsymbol{\cdot} l}}{\kappa_0 + N_{\boldsymbol{\cdot} l}} \left(\bar{\by}_l-\bmu_0\right) \left(\bar{\by}_l-\bmu_0\right)^T + \mathbcal{S}_{\boldsymbol{\cdot} l},
    \end{align*}
    where $N_{\boldsymbol{\cdot} l} = \sum_{j=1}^J\sum_{i=1}^{N_j} \xi_{i,j,l}$, $\bar{\by}_l = 
N_{\boldsymbol{\cdot} l}^{-1} (\sum_{j=1}^J\sum_{i=1}^{N_j} \xi_{i,j,l}\by_{i,j})$, and $\mathbcal{S}_{\boldsymbol{\cdot} l}=\sum_{j=1}^J\sum_{i=1}^{N_j} \xi_{i,j,l}\left(\bm{y}_{i,j}-\bar{\by}_l\right)\left(\bm{y}_{i,j}-\bar{\by}_l\right)^T$.     \end{enumerate}
    Store the updated parameters in $\boldsymbol{\lambda}$ and let $\boldsymbol{\lambda}^{(t)}=\boldsymbol{\lambda}$. \\
    
    Compute $\Delta(t-1,t) = \mathrm{ELBO}(\boldsymbol{\lambda}^{(t)})- \mathrm{ELBO}(\boldsymbol{\lambda}^{(t-1)})$.\\ 
 }
 \Return{ $\boldsymbol{\lambda}^\star$, containing the optimized variational parameters. }
 }
 \caption{CAVI updates for the fSAN model}
\label{algo::CAVIfSAN}
\end{algorithm}

\subsection{Computation of the ELBO in the variational inference approach}\label{suppl::subsec::elbo}
Here we outline the ELBO evaluation for both SAN specifications since most steps are unchanged in the two procedures. We only split the steps when needed.
For the fSAN, the vector of latent variables and the set of variational parameters are introduced in the previous paragraph; for the fiSAN, we use the notation introduced in Section~3 of the paper.
Moreover, recall that $h_k(\boldsymbol{x}) = \psi(x_{k}) - \psi(\sum_{k=1}^K x_{k})$ and $g(x,y) = \psi(x) - \psi(x+y)$. The minimization of the Kullback-Leibler divergence between the posterior and the variational distributions is equivalent to the maximization of the ELBO, expressed as
$$ELBO(q)=\mathbb{E}_q\left[\log p(\by,\boldsymbol{\Theta})\right] - \mathbb{E}_q\left[\log q_{\boldsymbol{\lambda}}(\boldsymbol{\Theta})\right].$$

The first term, $\mathbb{E}_q\left[\log p(\by,\boldsymbol{\Theta})\right] $, can be decomposed into the following components (note that some parts are different under the fSAN and fiSAN models):


\begin{enumerate}
    \item 
   $ \mathbb{E}[\log p(\by\mid \bM,\{\bmu_l,\bLambda_l\}_{l=1}^L)] = \frac12\sum_{l=1}^L N_{\boldsymbol{\cdot} l}\big[\ell^{(1)}_{l}-d/t_l -c_l \mathbcal{T}( \mathbcal{S}_{\boldsymbol{\cdot} l}/N_{\boldsymbol{\cdot} l}\boldsymbol{D}_l) - D\log(2\pi) 
    \linebreak -c_l\left(\bar{\by}_l-\bm{m}_l\right)^T\boldsymbol{D}_l \left(\bar{\by}_l-\bm{m}_l\right)  \big]$
     where $\mathbcal{T}(\cdot)$ is the trace operator;

    \item $\mathbb{E}\left[\log p(\bM \mid \bS, \bomega)\right] =
    \sum_{j=1}^J \sum_{i=1}^{N_j} \sum_{k=1}^T \sum_{l=1}^L   \: \xi_{i,j,l} \: \rho_{j,k} \: h_l\left(\boldsymbol{p}_k\right); $
    \item fSAN ~~$
    \mathbb{E}\left[\log p(\bS \mid \boldsymbol{\pi})\right] = \sum_{j=1}^J \sum_{k=1}^K \rho_{jk}\:
    h_k(\tilde{\boldsymbol{p}});  $ \\
    fiSAN ~$\,\mathbb{E}\left[\log p(\bS \mid \bv)\right] =
    \sum_{j=1}^J \sum_{k=1}^T  \rho_{j,k}\left(
    g(\bar{a}_{k},\bar{b}_{k}) + \sum_{q<k} g(\bar{b}_{q},\bar{a}_{q}))
    \right);  $
    \item fSAN ~~$\mathbb{E}\left[\log p(\bpi)\right] =
    \log(\mathcal{C}_{\boldsymbol{\pi}}(\boldsymbol{a}))+
    \sum_{k=1}^K(a-1) h_k(\tilde{\boldsymbol{p}}),$ where $\mathcal{C}_{\boldsymbol{\pi}}(\cdot)$ is the normalizing constant of a $\Dirichlet_K$ distribution; \\
    fiSAN ~$\,\mathbb{E}\left[\log p(\bv)\right] = 
    (T-1)\left(\psi(s_1)-\log(s_2)\right) + \left(s_1/s_2-1\right)
    \sum_{k=1}^{T-1} g(\bar{b}_{k},\bar{a}_{k}); $ 
    \item $\mathbb{E}\left[\log p(\boldsymbol{\omega})\right] =
    T \log(\mathcal{C}_{\boldsymbol{\omega}}(\boldsymbol{b}))+
    \sum_{k=1}^T \sum_{l=1}^L (b-1) h_l(\boldsymbol{p}_k),$
    where $\mathcal{C}_{\boldsymbol{\omega}}(\cdot)$ is the normalizing constant of a $\Dirichlet_L$ distribution;
%
       \item $\mathbb{E}\left[\log p(\bmu,\bLambda)\right] =
    L\log \mathbcal{B}\left(\boldsymbol{\Gamma}_0,\tau_0\right)+
    0.5(\tau_0-d-1)\sum_{l=1}^L \ell^{(1)}_{l} - 0.5\sum_{l=1}^L c_l \mathbcal{T}(\boldsymbol{\Gamma}_0^{-1}\boldsymbol{D}_l) \linebreak +0.5 \sum_{l=1}^L
    \big[ d\log(\kappa_0/2\pi) +\ell^{(1)}_{l}- d\kappa_0/t_l- \kappa_0c_l(\bm{m}_l-\boldsymbol{\mu}_0)^T\boldsymbol{D}_l(\bm{m}_l-\boldsymbol{\mu}_0)\big]$,        
    where $\mathbcal{B}\left(\boldsymbol{\Gamma}_0,\tau_0\right)$ is the inverse of the normalizing constant of a Wishart distribution~\citep[see, for more details, Appendix B of][]{Bishop2006};
    \item fiSAN ~~$\mathbb{E}\left[\log p(\alpha)\right] =  \log(\mathcal{C}_{\alpha}(a_\alpha,b_\alpha)) + (a_\alpha-1)(\psi(s_1)-\log(s_2)) - b_\alpha s_1/s_2$, where $\mathcal{C}_{\alpha}(\cdot)$ is the normalizing constant of a Gamma distribution.
\end{enumerate}

The second term is decomposed into the following six components:

\begin{enumerate}
    \item $\mathbb{E}\left[\log q(\bM)\right] = \sum_{j=1}^J \sum_{i=1}^{N_j} \sum_{l=1}^L \xi_{i,j,l}\log(\xi_{i,j,l})$
    
    \item $\mathbb{E}\left[\log q(\bS)\right] = \sum_{j=1}^J \sum_{k=1}^T  \rho_{j,k}\log(\rho_{j,k})$    
    
    \item fSAN~~$\mathbb{E}\left[\log q(\bpi)\right] = 
    \log(\mathcal{C}_{\boldsymbol{\pi}}(\tilde{\boldsymbol{p}}))+
    \sum_{k=1}^K(\tilde{p}_k-1) h_k(\tilde{\boldsymbol{p}}),$ \\
   fiSAN~~$\mathbb{E}\left[\log q(\bv)\right] = \sum_{k=1}^T  \{\log(\mathcal{C}_{\bv}(\bar{a}_{k},\bar{b}_{k}))+
    (\bar{a}_{k}-1)g(\bar{a}_{k},\bar{b}_{k})+
   (\bar{b}_{k}-1)g(\bar{b}_{k},\bar{a}_{k})\}$, where $\mathcal{C}_{\bv}(\cdot)$ is the normalizing constant of a Beta distribution.    
    
    \item $\mathbb{E}\left[\log q(\bomega)\right] = \sum_{k=1}^T \log(\mathcal{C}_{\bomega}(\boldsymbol{p}_k))+  \sum_{k=1}^T \sum_{l=1}^L (p_{l,k}-1) h_l(\boldsymbol{p}_k)$.
    

       \item $\mathbb{E}\left[\log q(\bmu,\bLambda)\right] = 
           \sum_{l=1}^L \big[0.5\ell^{(1)}_{l} + 0.5d(\log(t_l/2\pi)-1) - \mathbcal{H}\left(q(\bLambda_l)\right)\big], $
    where $\mathbcal{H}\left(q(\bLambda_l)\right)$ is the entropy of a Wishart distribution.

    \item fiSAN~~$\mathbb{E}\left[\log(q(\alpha))\right] = \log(\mathcal{C}_{\alpha}(s_1,s_2)) + (s_1-1)(\psi(s_1)-\log(s_2)) - s_1$, where $\mathcal{C}_{\alpha}(\cdot)$ is the normalizing constant of a Gamma distribution.
\end{enumerate}

\clearpage

\section{Additional details on the simulation studies}

\subsection{Label-switching}\label{sec::suppl::labelsw}
\begin{figure}[h!]
    \centering
    \includegraphics[width=\linewidth]{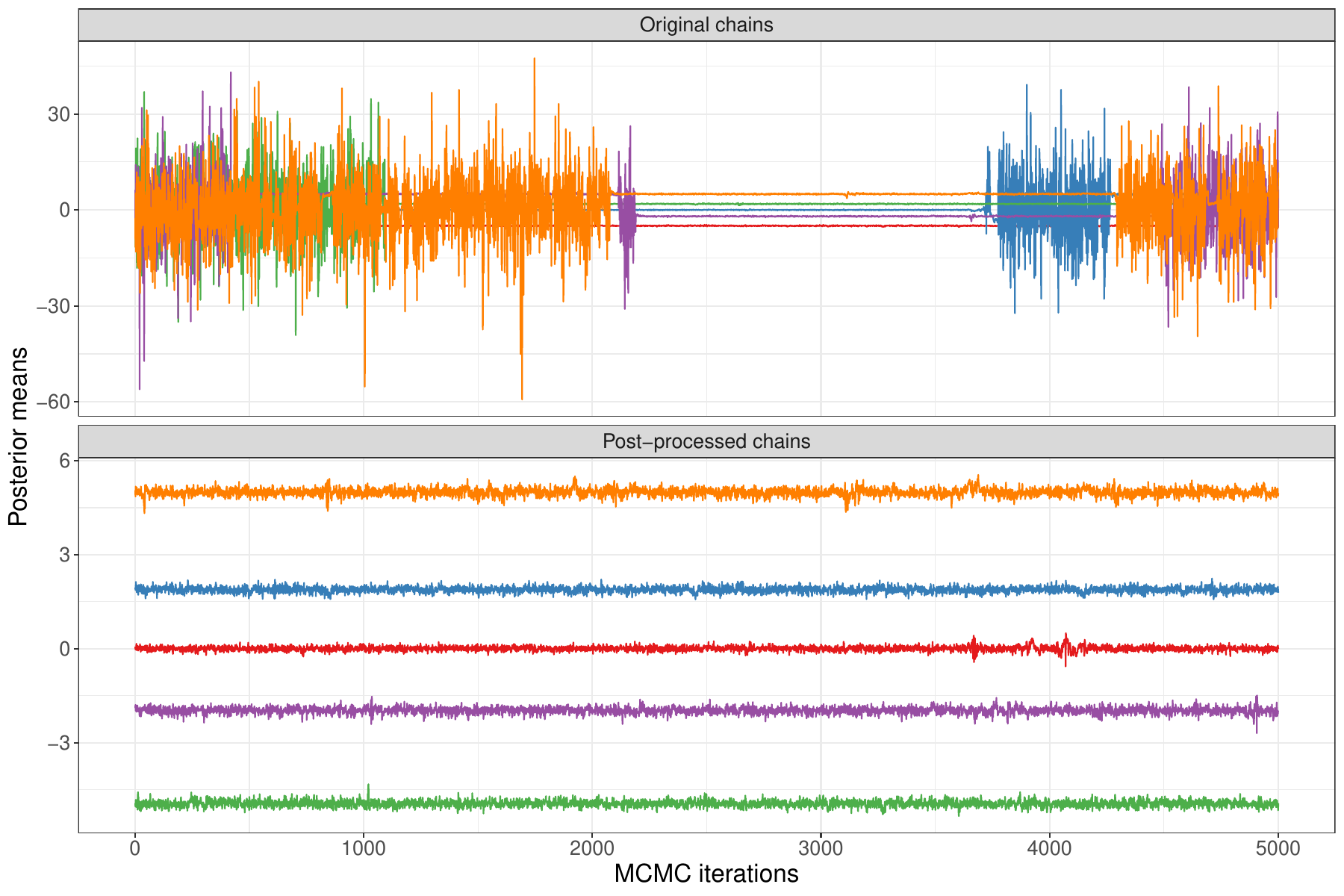}
    \caption{Top plot: five of the original chains, affected by label-switching. Note that all chains are processed with the ECR algorithm. Bottom plot: the five post-processed chains characterized by the lowest variability, disentangled from label-switching.}
    \label{fig:labelSW}
\end{figure}

Working with overfitted sparse mixtures, we noticed a sensible presence of label-switching in the chains of the component-specific parameters $(\mu_l,\sigma^2_l)$ in the univariate case and $(\bmu_l,\bLambda_l)$ in the multivariate case. Although it does not impact the convergence of the algorithms, it poses substantial challenges in drawing posterior inference about parameters that are component-specific.
In order to obtain ``workable'' chains, in the univariate case, we relied on the post-processing relabeling Equivalence Classes Representatives (ECR) algorithm~\citep{Rodriguezwalker2014} implemented in the R package \texttt{label.switching}~\citep{labswitchR}.
This post-processing phase to ``disentangle'' the chains turned out to be crucial to obtaining some of the main results. Figure~\ref{fig:labelSW} shows an example of this post-processing, presenting the chains of $(\mu_l,\sigma^2_l)$ in a single run before and after the application of the algorithm.

\clearpage
\subsection{Comparisons between MCMC and Variational Inference results}
\begin{figure}[th]
    \centering
    \includegraphics[width=\linewidth]{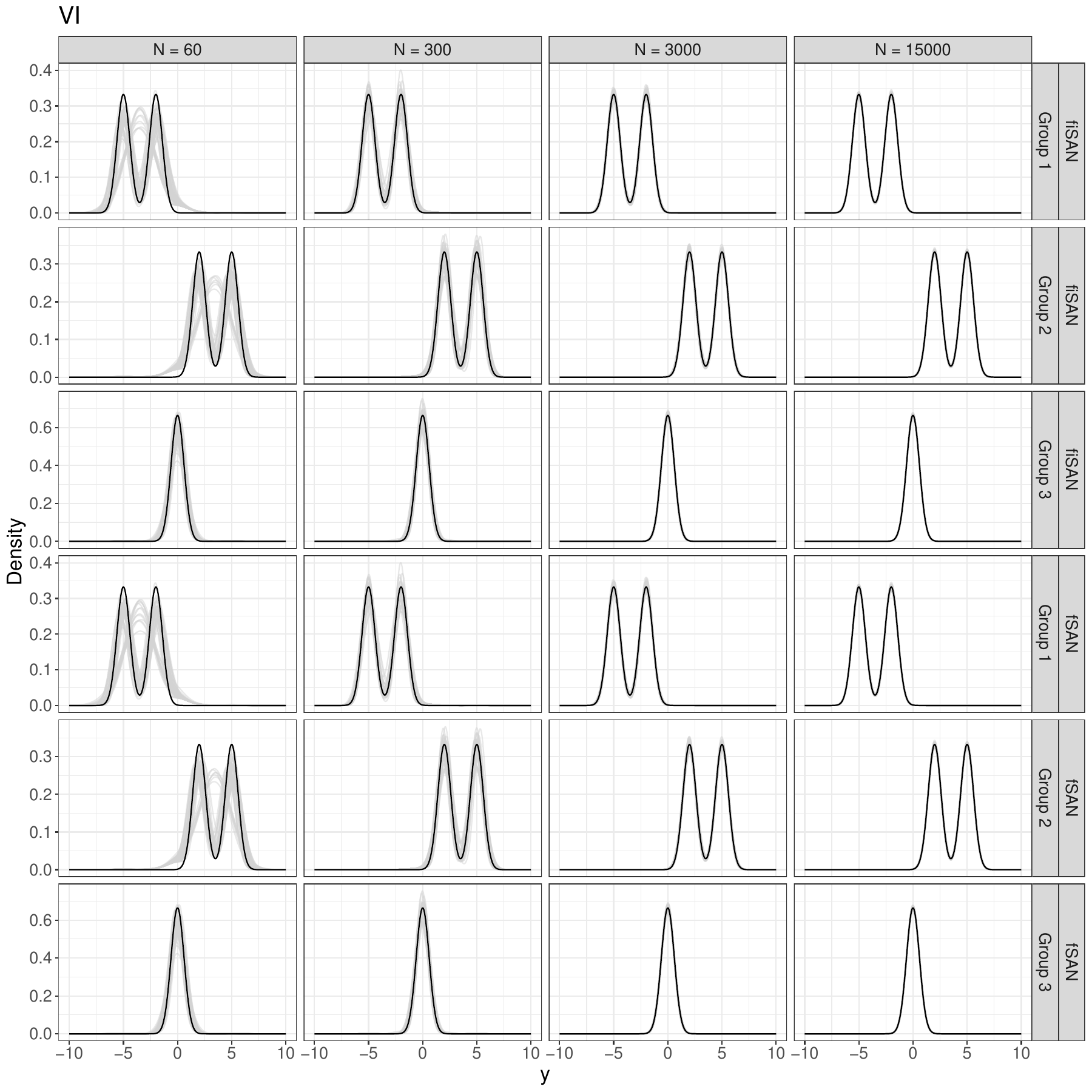}
    \caption{Posterior density estimates obtained with the fSAN and fiSAN using the CAVI algorithm for groups 1, 2, and 3, under all configurations. Each panel shows the true density (black line) and the posterior density estimates (grey lines) obtained over the 50 replications. }
    \label{fig:density_VB_fisan}
\end{figure}

Similarly to Figure~4 in the main text, which shows the posterior density estimates obtained using the MCMC approach, Figure~\ref{fig:density_VB_fisan} displays the estimates obtained using the CAVI algorithm for the fSAN and fiSAN model.

\begin{figure}[h!]
    \centering
    \begin{subfigure}[b]{0.32\textwidth}
    \centering
    \includegraphics[width=\linewidth]{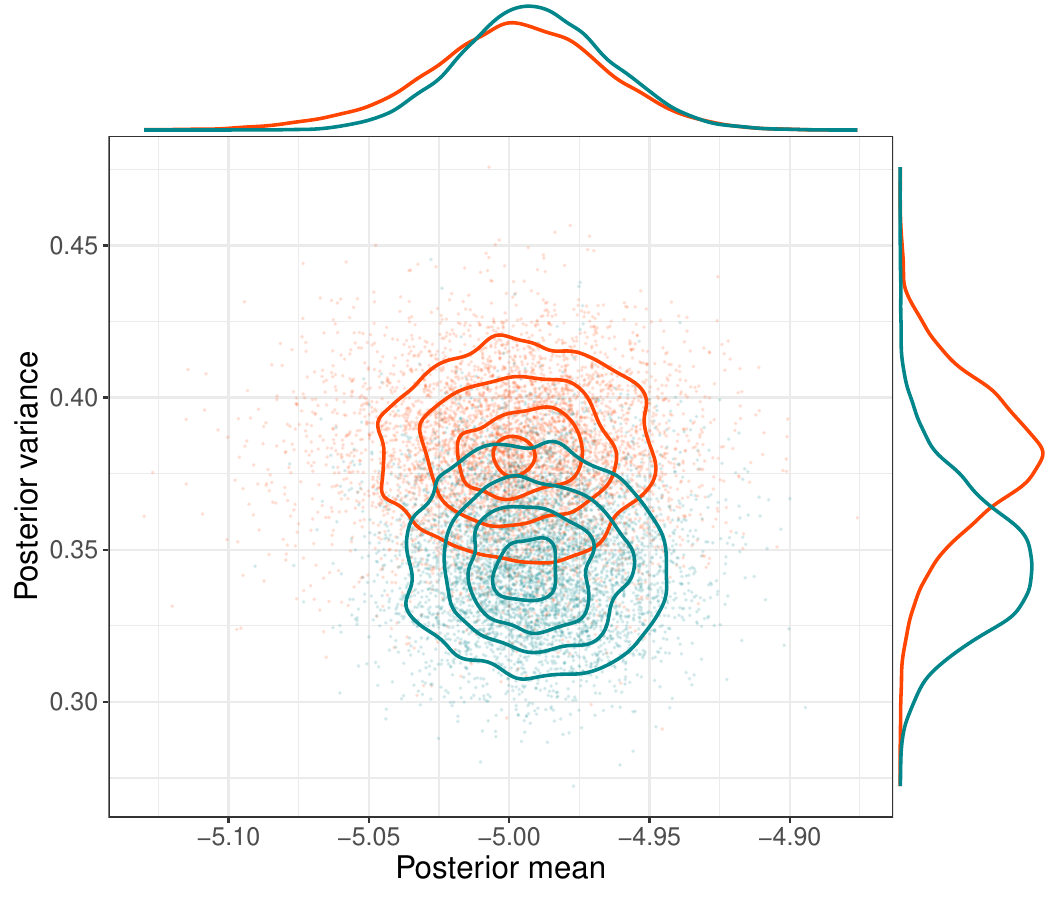} 
    \caption*{$l=1$}
    \end{subfigure}
    \begin{subfigure}[b]{0.32\textwidth}
    \centering
    \includegraphics[width=\linewidth]{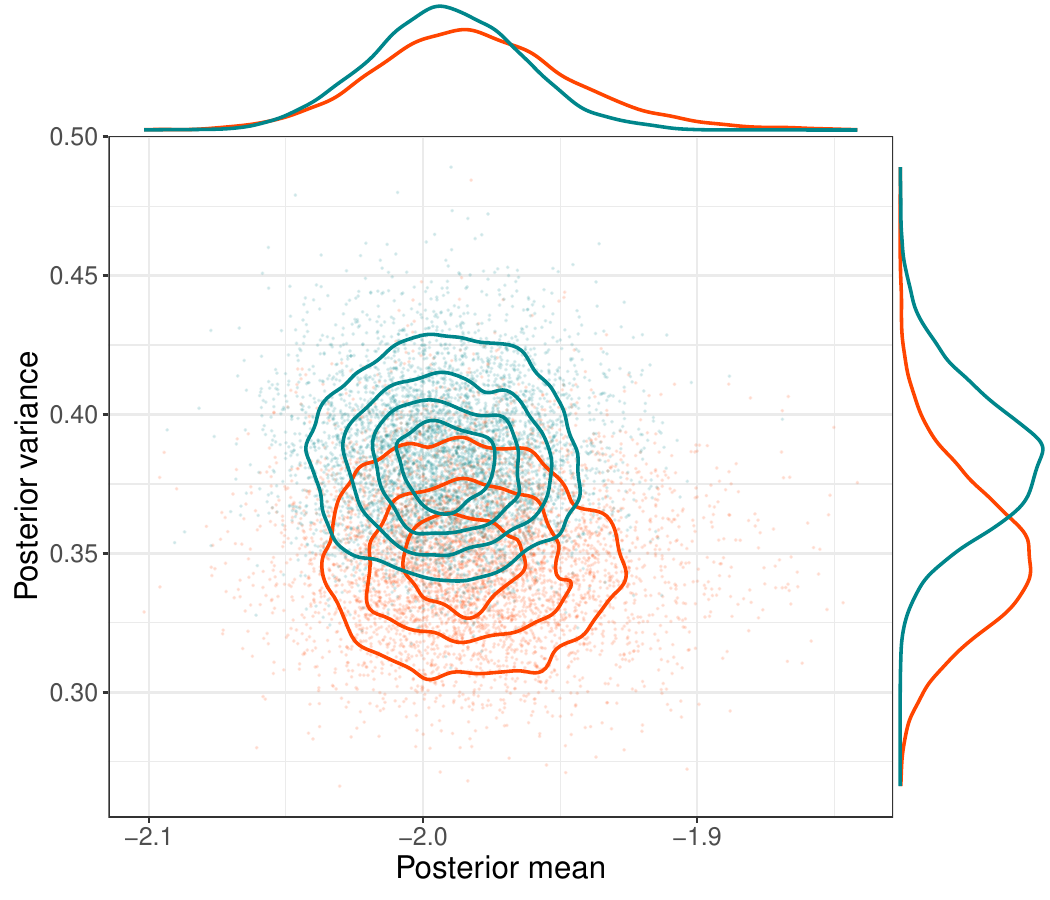}    \caption*{$l=2$}
    \end{subfigure}
    \begin{subfigure}[b]{0.32\textwidth}
    \centering
    \includegraphics[width=\linewidth]{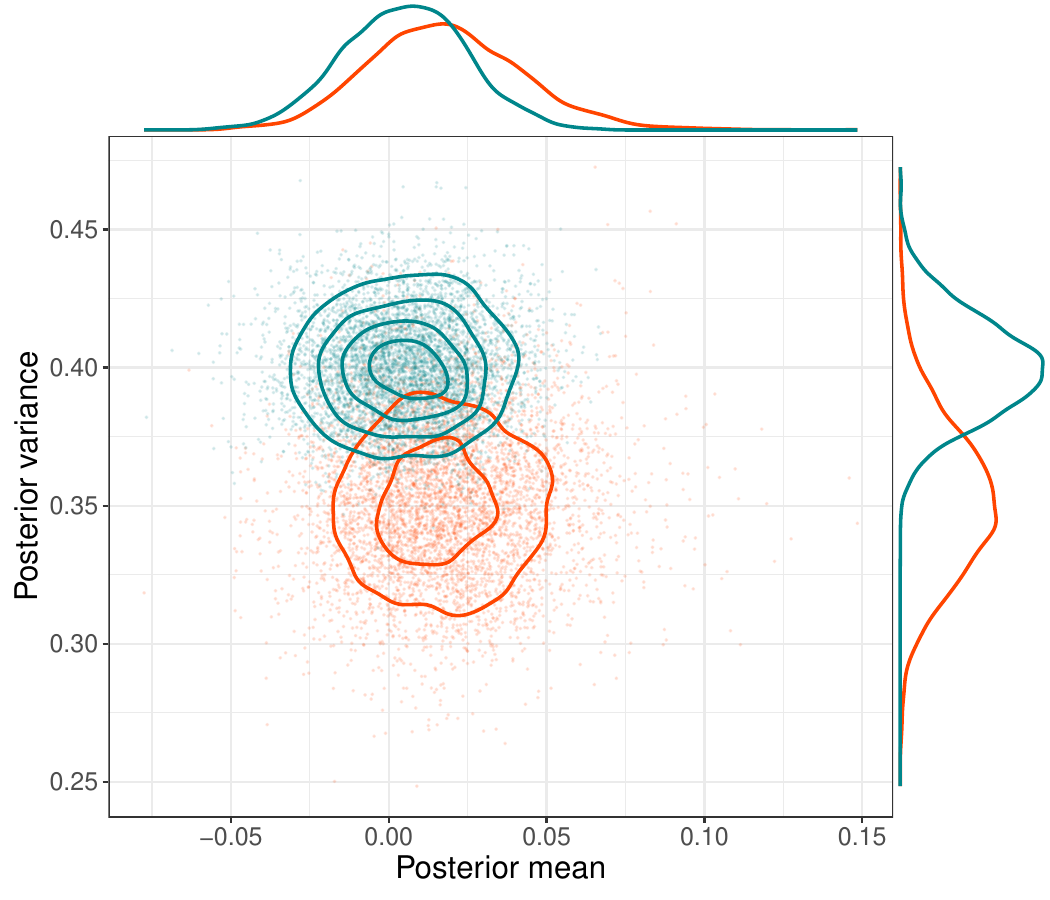}    \caption*{$l=3$}
    \end{subfigure} \\
    \begin{subfigure}[b]{0.32\textwidth}
    \centering
    \includegraphics[width=\linewidth]{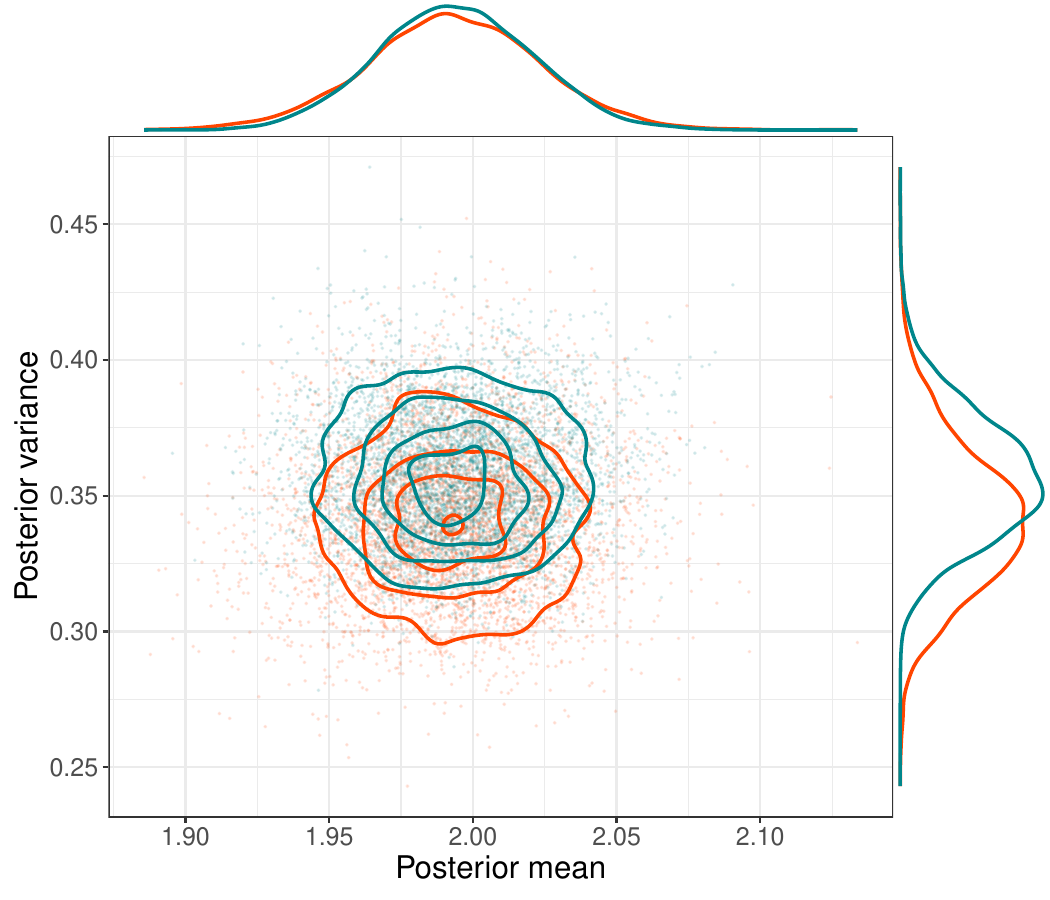}    \caption*{$l=4$}
    \end{subfigure}
    \begin{subfigure}[b]{0.32\textwidth}
    \centering
    \includegraphics[width=\linewidth]{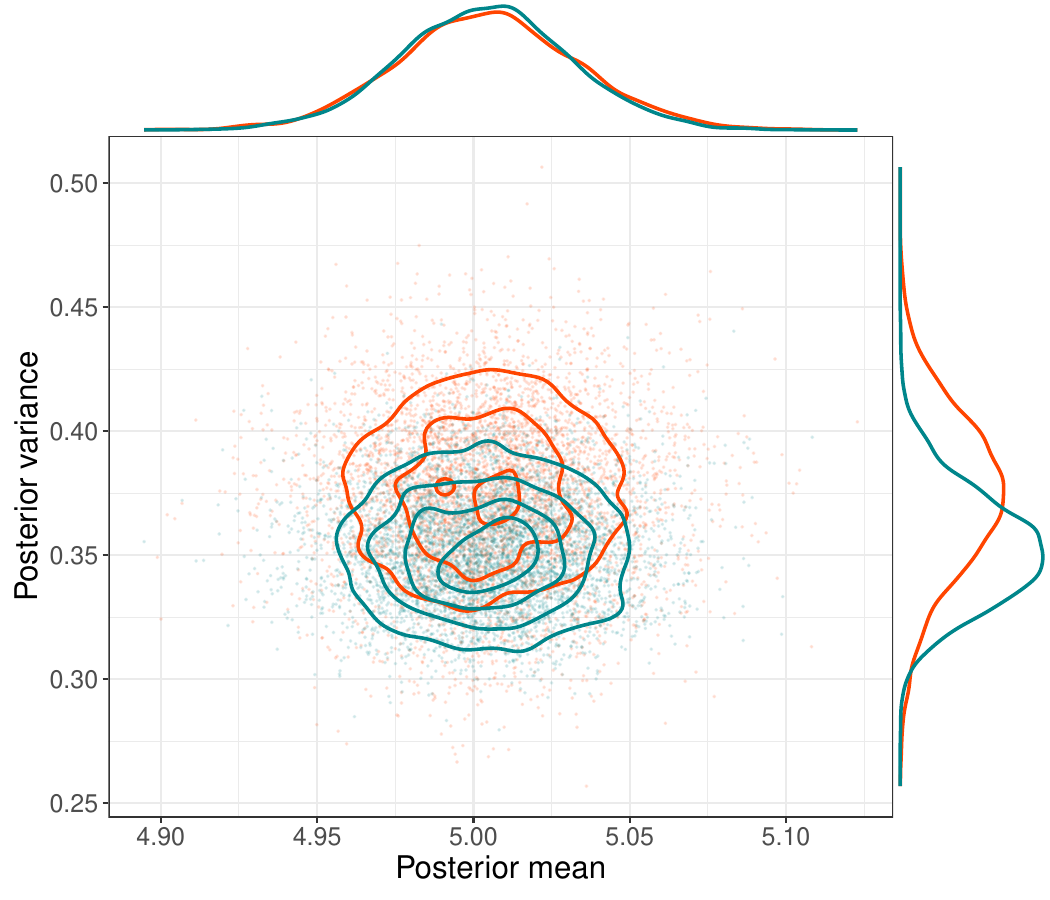}    \caption*{$l=5$}
    \end{subfigure}
    \caption{fiSAN prior - Configuration 3: posterior density estimate of $(\mu_l,\sigma^2_l)$, for $l=1,2,3,4,5$, obtained using a Gibbs sampler (orange line) and a CAVI algorithm (green line). Each panel shows the contour plot of the joint density, together with the two marginal densities.}
    \label{fig:scenario1:vi_vs_mcmc1_fiSAN_conf3}
\end{figure}

\begin{figure}[h]
    \centering
    \begin{subfigure}[b]{0.32\textwidth}
    \centering
    \includegraphics[width=\linewidth]{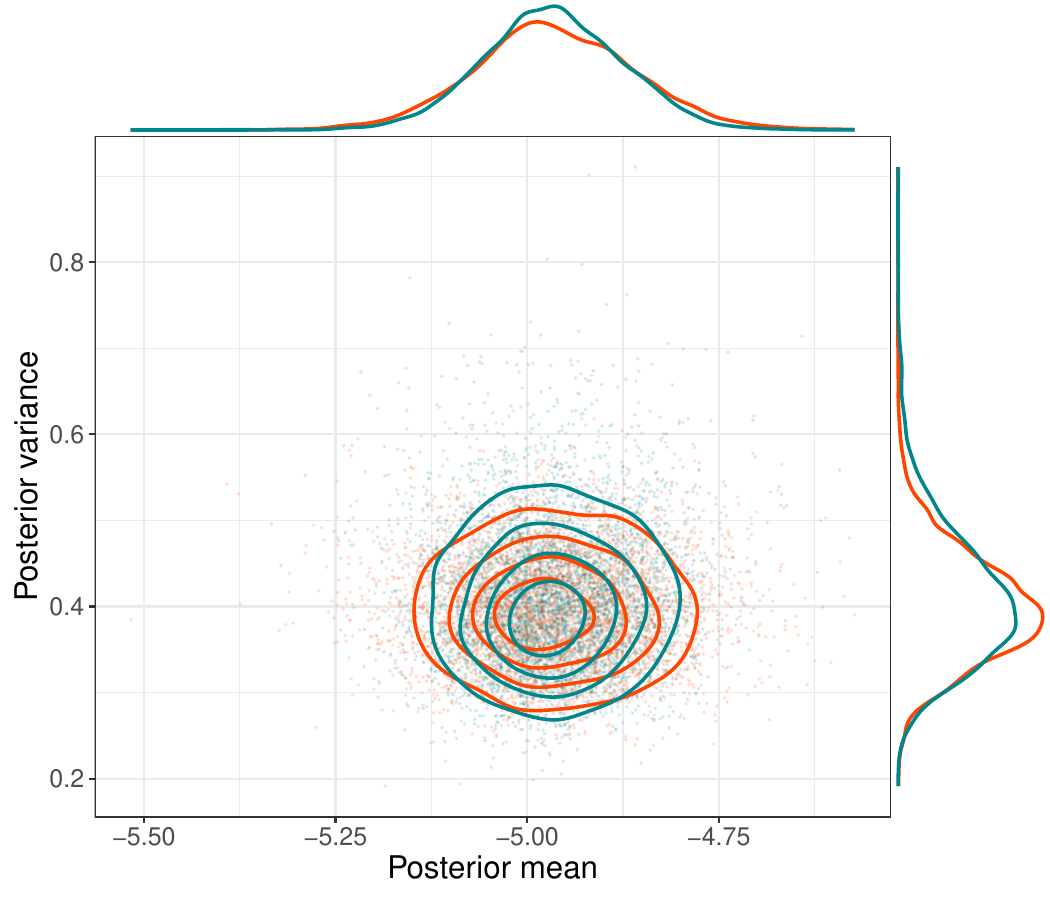}    \caption*{$l=1$}
    \end{subfigure}
    \begin{subfigure}[b]{0.32\textwidth}
    \centering
    \includegraphics[width=\linewidth]{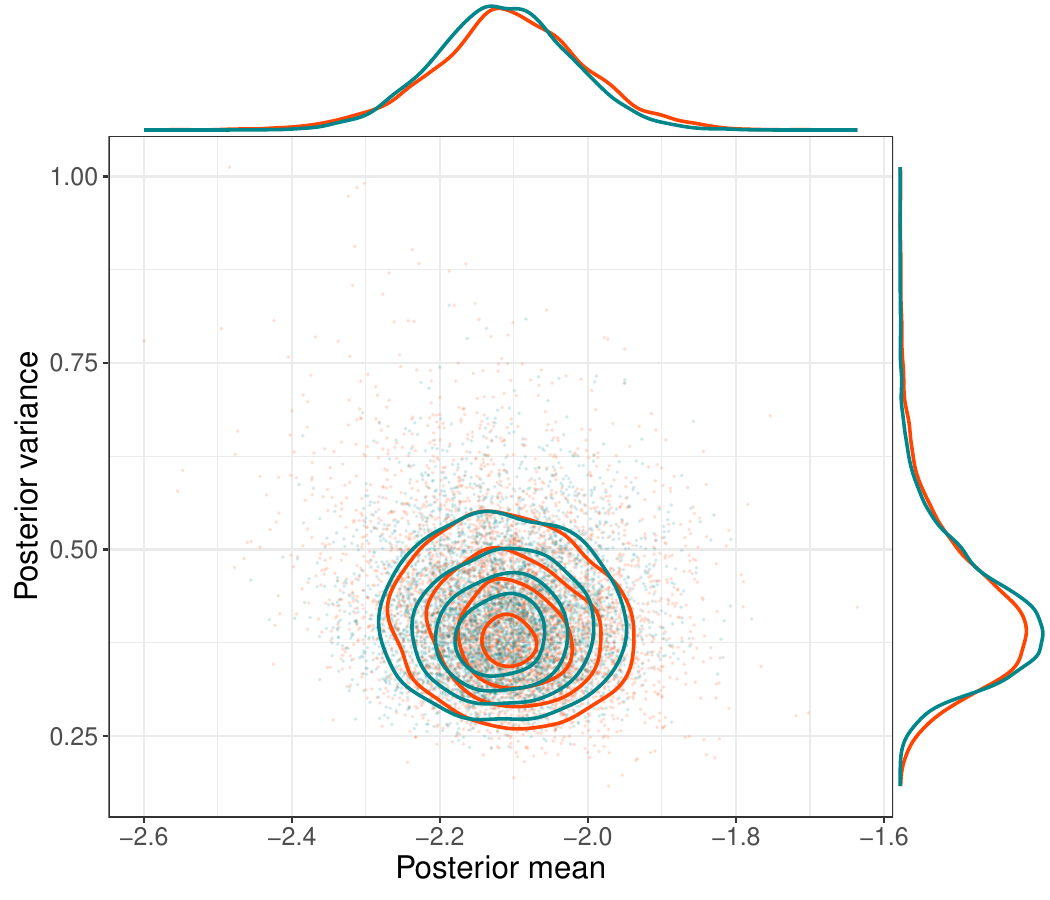}      \caption*{$l=2$}
    \end{subfigure}
    \begin{subfigure}[b]{0.32\textwidth}
    \centering
    \includegraphics[width=\linewidth]{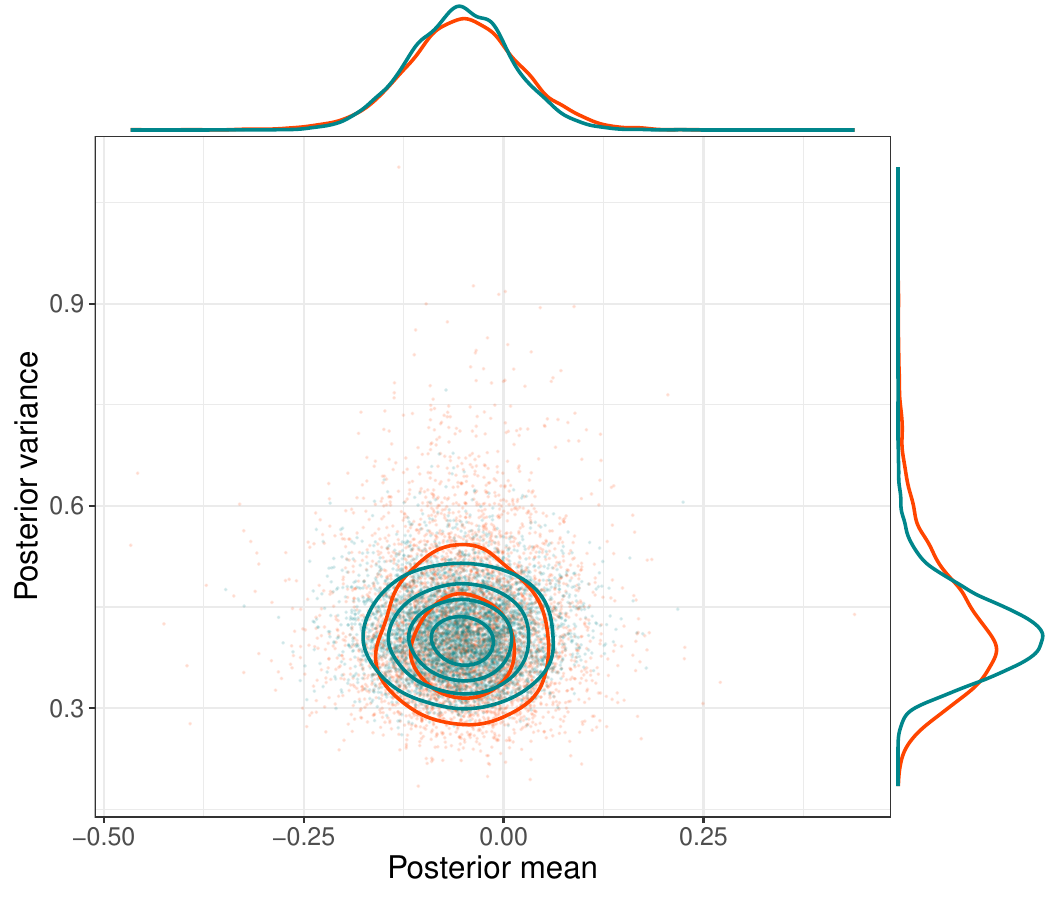}      \caption*{$l=3$}
    \end{subfigure} \\
    \begin{subfigure}[b]{0.32\textwidth}
    \centering
    \includegraphics[width=\linewidth]{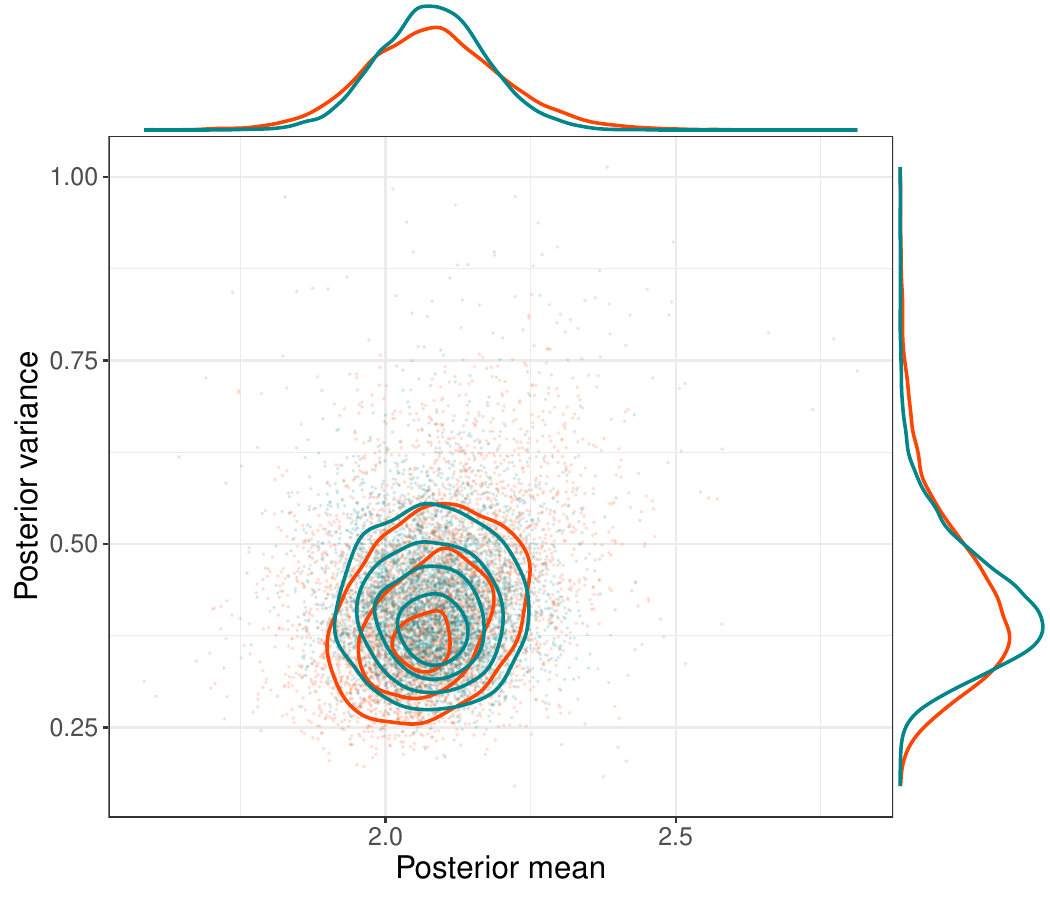}      \caption*{$l=4$}
    \end{subfigure}
    \begin{subfigure}[b]{0.32\textwidth}
    \centering
    \includegraphics[width=\linewidth]{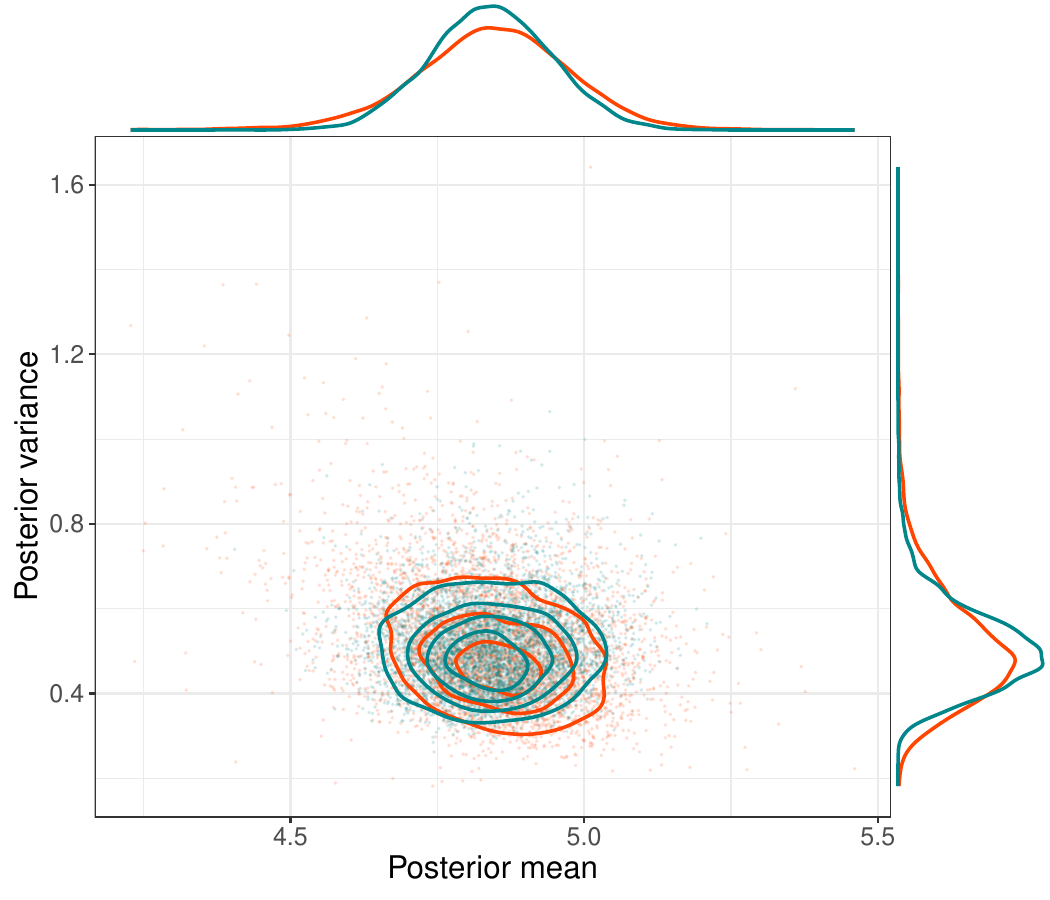}      \caption*{$l=5$}
    \end{subfigure}
    \caption{fSAN prior - Configuration 2: posterior density estimate of $(\mu_l,\sigma^2_l)$, for $l=1,2,3,4,5$, obtained using a Gibbs sampler (orange line) and a CAVI algorithm (green line). Each panel shows the contour plot of the joint density, together with the two marginal densities.}
    \label{fig:scenario1:vi_vs_mcmc1_fSAN_conf2}
\end{figure}

\begin{figure}[t]
    \centering
    \begin{subfigure}[b]{0.32\textwidth}
    \centering
    \includegraphics[width=\linewidth]{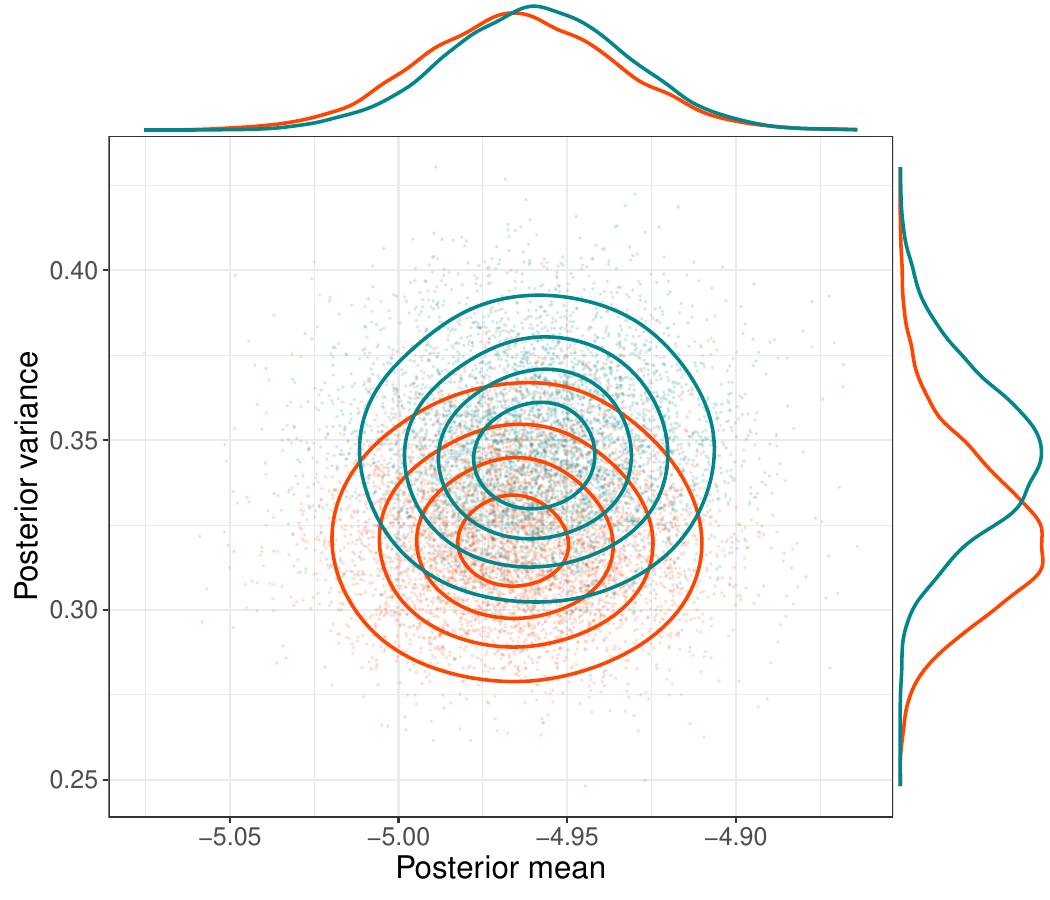}      \caption*{$l=1$}
    \end{subfigure}
    \begin{subfigure}[b]{0.32\textwidth}
    \centering
    \includegraphics[width=\linewidth]{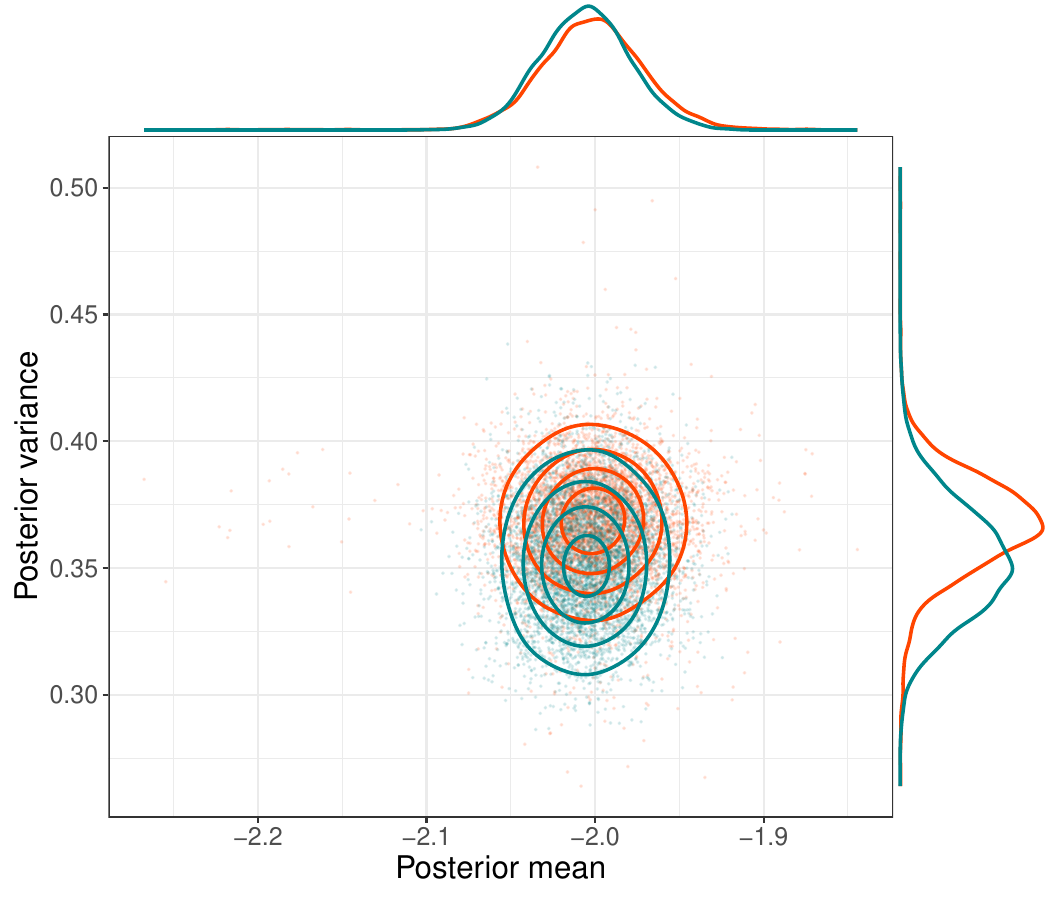}      \caption*{$l=2$}
    \end{subfigure}
    \begin{subfigure}[b]{0.32\textwidth}
    \centering
    \includegraphics[width=\linewidth]{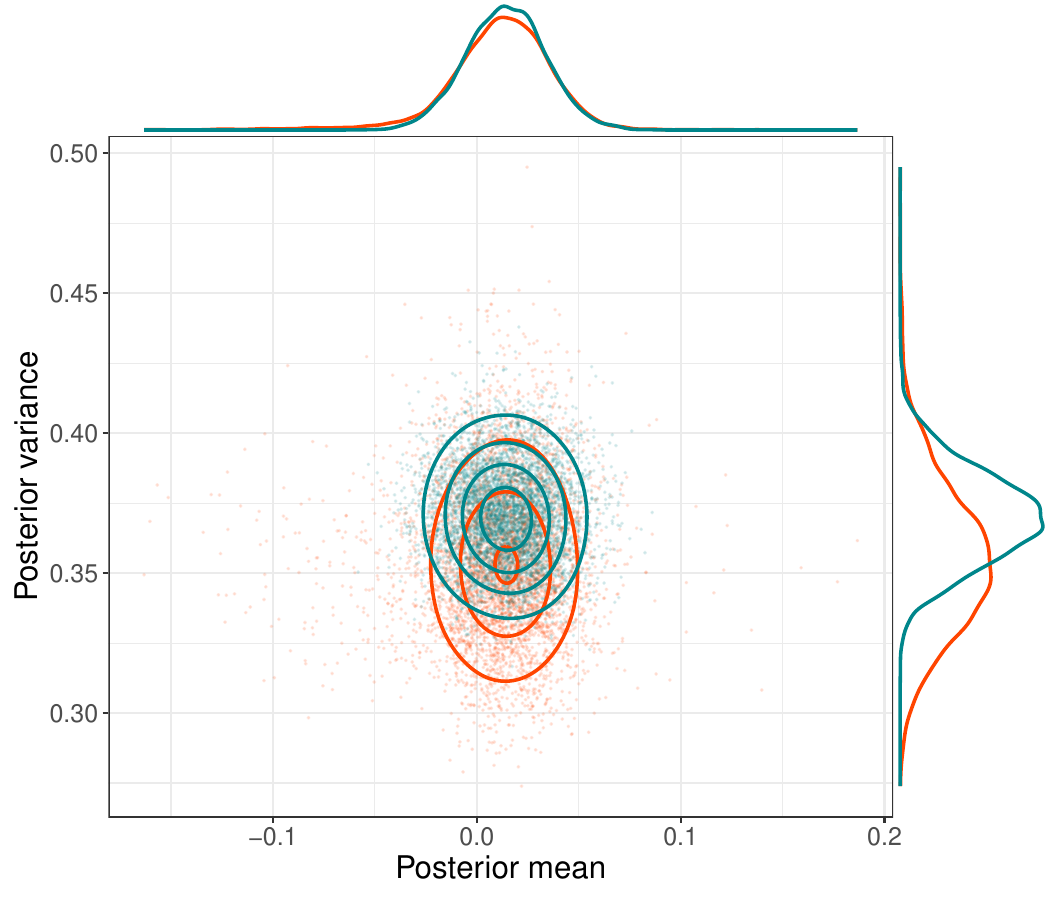}      \caption*{$l=3$}
    \end{subfigure} \\
    \begin{subfigure}[b]{0.32\textwidth}
    \centering
    \includegraphics[width=\linewidth]{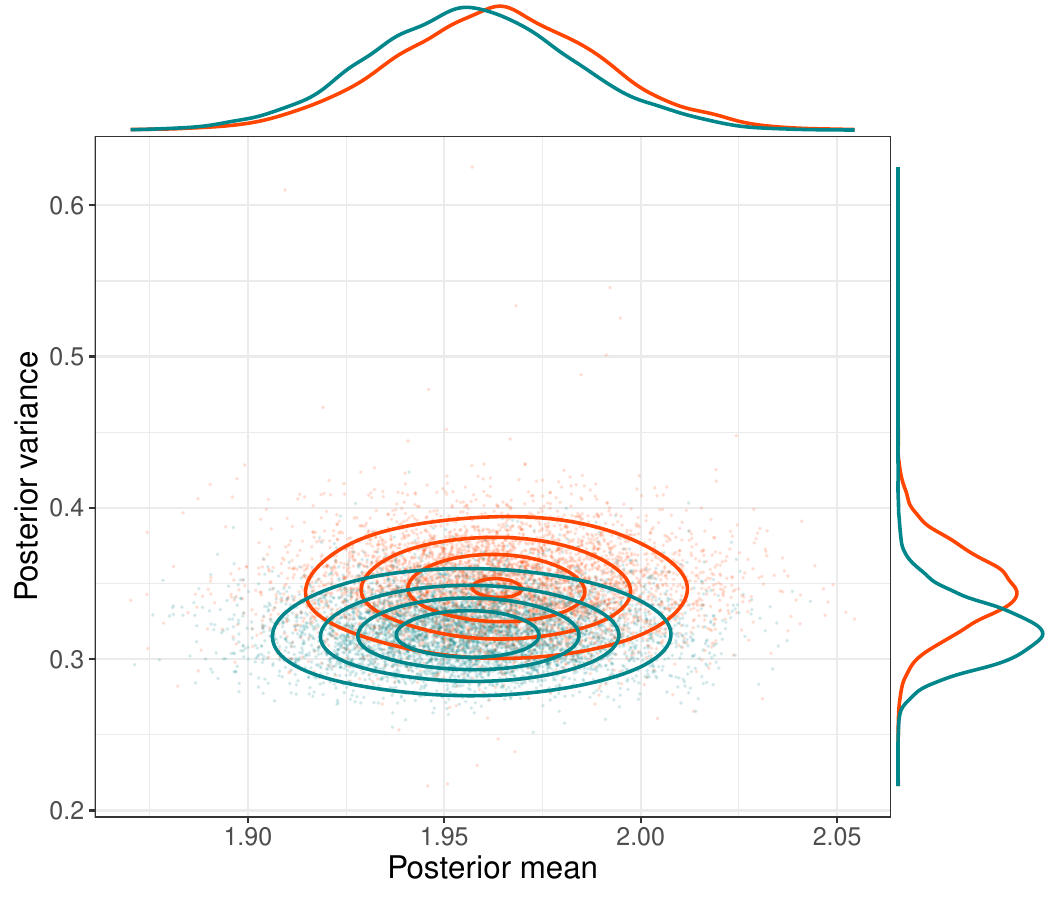}      \caption*{$l=4$}
    \end{subfigure}
    \begin{subfigure}[b]{0.32\textwidth}
    \centering
    \includegraphics[width=\linewidth]{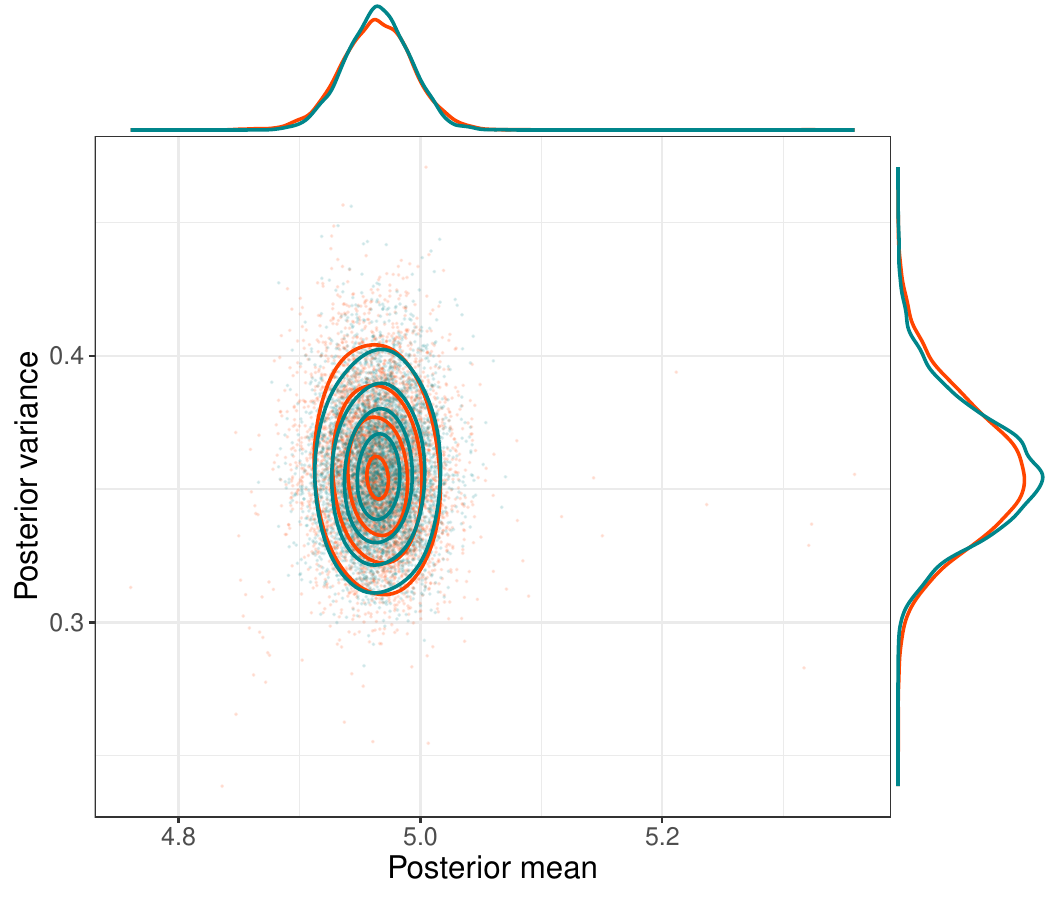}      \caption*{$l=5$}
    \end{subfigure}
    \caption{fSAN prior - Configuration 3: posterior density estimate of $(\mu_l,\sigma^2_l)$, for $l=1,2,3,4,5$, obtained using a Gibbs sampler (orange line) and a CAVI algorithm (green line). Each panel shows the contour plot of the joint density, together with the two marginal densities.}
    \label{fig:scenario1:vi_vs_mcmc1_fSAN_conf3}
\end{figure}

Similarly to Figure~5 in the main paper, Figures~\ref{fig:scenario1:vi_vs_mcmc1_fiSAN_conf3},~\ref{fig:scenario1:vi_vs_mcmc1_fSAN_conf2} and~\ref{fig:scenario1:vi_vs_mcmc1_fSAN_conf3} show the estimated posterior density of $(\mu_l,\sigma^2_l)$, for $l=1,2,3,4,5$ estimated via MCMC and VI. Each panel shows the contour plot of the joint density, together with the marginal densities obtained using the two algorithms. 

These figures are obtained based on a ``matching'' procedure between the estimates computed under the MCMC and VI approaches. To obtain results that are comparable between fundamentally different estimating procedures, we proceed as follows.

Regarding the results of the MCMC, the treatment of the label-switching described in Section~\ref{sec::suppl::labelsw} produces very ``clean'' chains, as shown in Figure~\ref{fig:labelSW}. The processed chains are characterized by a stable mean centered around the true values $\mu_l$ or zero (obtained from the priors). 
Moreover, the variance of chains for $\mu_l$ corresponding to active clusters is much smaller if compared to the chains associated with values sampled from the prior (notice that, by assuming a normal-inverse gamma prior of parameters $(\mu_0,\kappa_0,\tau_0,\Gamma_0) = (0,0.01,3,2)$, we have $\mathrm{var}(\mu_l) = 100$). 
Therefore, we selected the five chains with the smallest variability and matched them with the true atoms. Then, the chains of the cluster-specific variances $\sigma^2_l$ are recovered based on the correspondence of the index $l$.  

When dealing with the results of the CAVI algorithm, the matching procedure is even more straightforward, although the underlying idea is analogous. Similarly, in this case, it is crucial to distinguish which of the $L$ components have actually been employed (i.e., non-empty clusters). The variational parameters in this approach were initialized according to the hyperprior specification. Then, at each iteration, only the ``active'' parameters are updated as a function of observations with a large probability of being allocated to that atom. Thus, at the end of the optimization, the variational parameters clearly show what atoms have been updated (``activated''). It was then simple to match this set with the corresponding ground truth (and with the results of the MCMC). 

Finally, one last remark concerns the simulations where the observational partition was not \emph{perfectly} estimated. 
These cases were characterized by the presence of small, spurious clusters. To deal with this issue, both in the MCMC and the VI approach, we selected the clusters with the largest number of allocated observations.
Fortunately, these cases were rare and did not severely affect the results, as we can see from the overall good clustering performance.
The simplicity of the data-generating mechanism was indeed crucial to aid this analysis: it involves five cluster-specific parameters $(\mu_l,\sigma^2_l)$ where the means $\mu_l\in\{-5,-2,0,2,5\}$ are fairly distinct and hence easily distinguishable.

\FloatBarrier

\clearpage
\FloatBarrier

\subsection{Sensitivity analysis}

To evaluate the impact of the DP truncation parameter $T$ and the observational Dirichlet dimension $L$ on posterior inference, we performed a sensitivity analysis using the VI approach to fit the univariate fiSAN model.
To be as complete as possible, we considered varying $T$, $L$, and $b$ simultaneously - three crucial parameters in defining the mixture properties. We considered the 50 replications of the data of the simulation study in Section~4.1 and a sample size $N_j=50$.
In particular, we considered $T\in\{10,20,30\}$, $L\in\{15,25,35\}$, and $b\in\{0.001,0.05,1\}$. Notice that the common choice $b=1$, which assigns uniform mass to all points in the simplex, does not satisfy the condition for inducing sparsity (which requires $b$ strictly smaller than 1~\citep{Rousseau2011}). These experiments were conducted on a personal computer running an Intel(R) Core(TM) i7-9750H CPU @ 2.60GHz with 17.2 GB RAM.

\begin{figure}[ht]
    \centering
    \includegraphics[width=\linewidth]{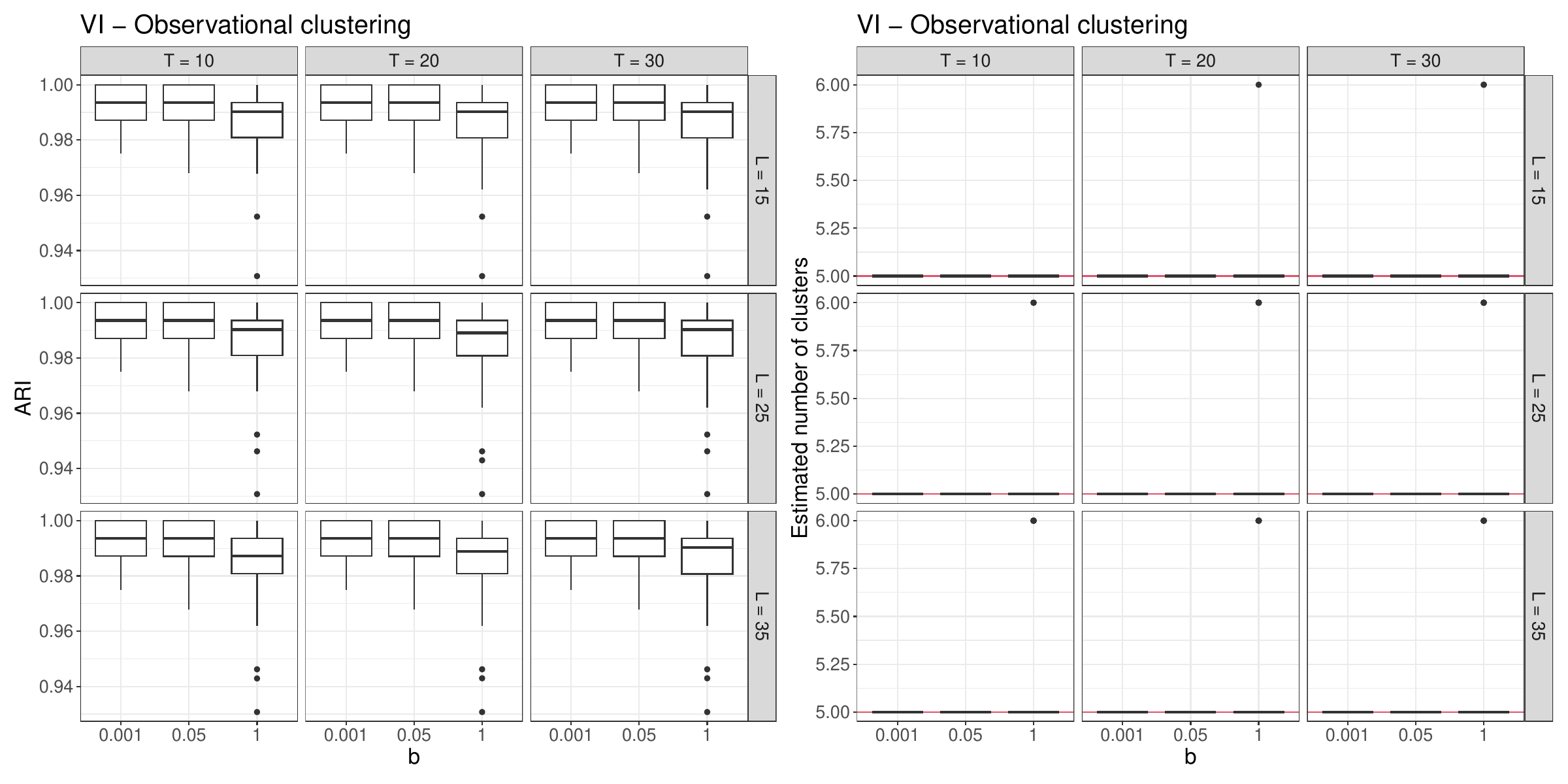}
    \caption{Left: observational Adjusted Rand Index. Right: estimated number of clusters (truth = 5). Each panel presents a combination of $T$ (columns) and $L$ (rows). Boxplots in each panel correspond to different values of $b$. \label{fig:VI-T-L-b}}
\end{figure}

We only report the results on the observational clustering since the distributional partition was correctly identified for all combinations. Figure~\ref{fig:VI-T-L-b} shows the observational ARI and the estimated number of clusters for each combination of the parameters.

\begin{figure}[th]
    \centering
    \includegraphics[width=.48\linewidth]{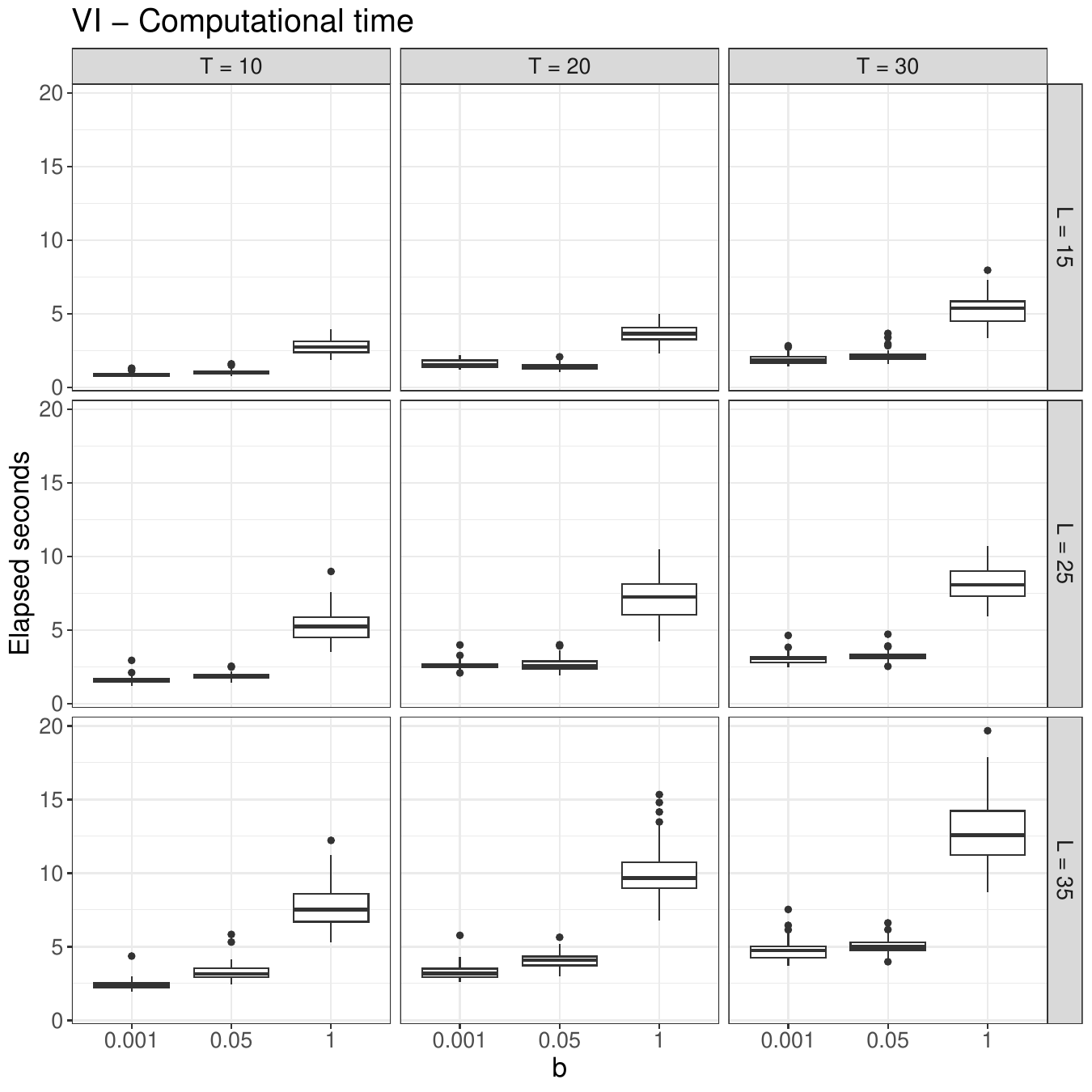}\hfill
    \includegraphics[width=.48\linewidth]{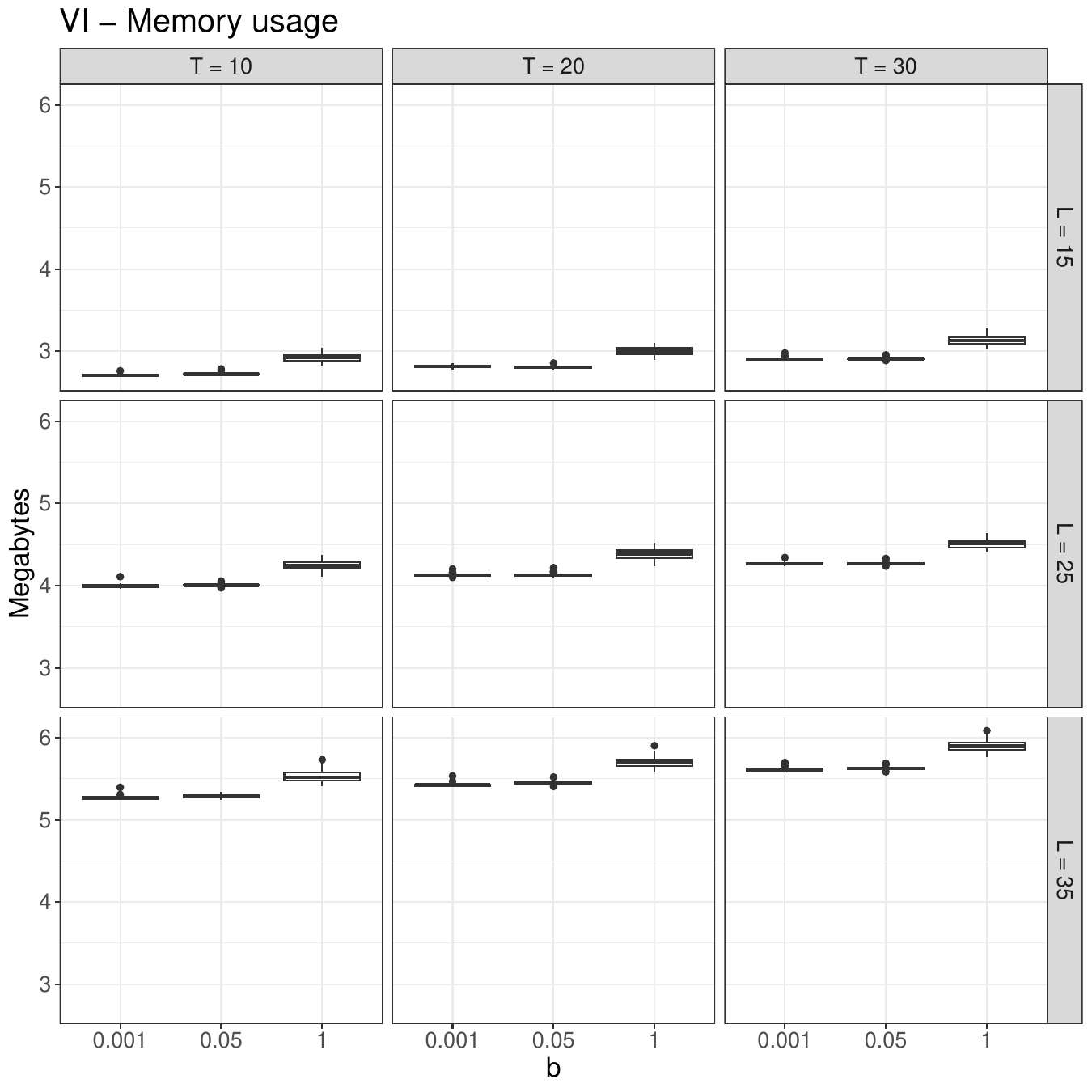}
    \caption{Left: elapsed computational time in seconds. Right: memory usage in megabytes. Each panel presents a combination of $T$ (columns) and $L$ (rows). Boxplots in each panel correspond to different values of $b$. \label{fig:VI-T-L-b2}}
\end{figure}

These experiments show that the clustering performance is not affected by the truncation as long as it is chosen to be large enough. Indeed, it becomes problematic only when set too close to the number of clusters estimated in the data. In this case, the algorithm cannot explore the entire space of reasonable partitions. In this regard, the truncation level $T$ is similar in spirit to the length of the allocated vectors when estimating DP mixtures via MCMC using a conditional sampler. 

Furthermore, we can see in Figure~\ref{fig:VI-T-L-b2} that this parameter does not particularly affect runtimes. It is $b$ that, when set equal to 1, leads to (slightly) higher computational cost, as the lack of explicit sparsity implies the necessity of more iterations to reach a stable value of the ELBO.

The main computational difference lies in the memory usage. However, the amount of used megabytes is mainly driven by $L$ rather than $T$. The fact that the memory usage increases with $T$ and $L$ is due to the larger matrix allocations done at the beginning of the algorithms. However, a great advantage of VI compared to MCMC is precisely the parsimony in the memory allocation. While MCMC algorithms require storing several iterations (hence, large \textit{arrays}), in VI, one is only interested in the final optimized parameters.

Finally, we highlight how the performances are not strongly affected by the value of $b$. However, with $b=1$, the performances are a little less precise, and the mixture once introduces an additional cluster, as evident from Figure~\ref{fig:VI-T-L-b}.

\clearpage
\section{Additional details on the Spotify data analysis}
\subsection{Preprocessing}\label{sec::preprocessing}
Here, we describe the preprocessing we performed to prepare the data for the analysis. The procedure we adopted is an extension of the one presented in \cite{Denti2022_sis}.
First, we noticed that 20\% of the songs in the dataset were authored by more than a single artist/band. To simplify the analysis, we assigned each of these songs to a single \emph{representative artist} that we identified as the first singer in the list of co-authors. From a preliminary exploratory analysis, we also noticed that most artists have authored a very limited number of songs: indeed, more than 90\% of the artists have less than 20 songs. 
The limited number of authored songs for a specific artist could make the estimation of the corresponding multivariate distributions challenging. To address this issue, we focused on the most productive artists: we included only the ones who authored at least 100 tracks in the analysis. We also filtered out the authors with more than 200 songs (0.3\% of the artists) to prevent them from dominating the analysis and simultaneously remove some ``author outliers''. For example, the profile named ``Ernest Hemingway'' (recorded in Cyrillic) is associated with more than one thousand tracks, likely representing audiobooks split into short parts. Another example of an outlier, this time in terms of duration, is provided by a single entry that lasts 45 minutes, comprising an entire concert. 

Recall that for this analysis, we focus on three song features: duration, energy, and speechiness. The energy and speechiness scores are measured over the $(0,1)$ interval, while duration is measured in milliseconds. Therefore, we map duration over the unit interval for homogeneity. Then, we filter out features identically equal to 0 and 1. For example, such extreme cases in terms of energy correspond to observations representing silent tracks or pure applause in live tracks. Finally, we mapped the three indexes from $(0,1)$ to the real line via a probit transform.

\subsection{Results}\label{sec::suppl_results}
The complete list of distributional clusters is reported in Table~\ref{table:fullDC}.

\begin{table}[h]
\footnotesize
\centering
\begin{tabular}{ll}
\toprule
DC & Artists\\
\midrule
\midrule
1 & 2Pac - Beastie Boys - Beyoncé - JAY-Z - Kanye West - Lil Uzi Vert - Lil Wayne - Mac Miller\\
& Sublime - The Notorious B.I.G.\\
\midrule
2 & Amirbai Karnataki - Asha Bhosle - Carmen Miranda - Dizzy Gillespie - Geeta Dutt \\& Mohammed Rafi - Suraiya\\
\midrule
3 & Antonio Aguilar - Dolly Parton - Juan Gabriel - Markos Vamvakaris - Merle Haggard \\& Otis Redding - Rita Ampatzi - Roza Eskenazi\\
\midrule
4 & Joan Sebastian\\
\midrule
5 & Lana Del Rey - The Moody Blues - The Velvet Underground - Van Morrison - Wings \\& Yusuf / Cat Stevens\\
\midrule
6 & Depeche Mode - Jimi Hendrix - Madonna - Michael Jackson - R.E.M. - Santana \\& Talking Heads - The Band - The Clash - The Cure - Tom Petty and the Heartbreakers - ZZ Top\\
\midrule
7 & B.B. King - Frank Zappa - James Brown - Los Tigres Del Norte - Louis Prima - The Weeknd\\
\midrule
8 & Andrew Lloyd Webber - Harry Belafonte - Tom Waits\\
\midrule
9 & Aaron Copland - Alexander Scriabin - Bill Evans - Bill Evans Trio - Charles Mingus \\& Chet Baker - Erik Satie - Felix Mendelssohn - Franz Joseph Haydn - Franz Liszt - Franz Schubert \\& George Frideric Handel - Gustav Mahler - Maurice Ravel - Pyotr Ilyich Tchaikovsky \\& Richard Strauss - Sergei Rachmaninoff - Stan Getz\\
\midrule
10 & Eric Clapton - Prince - The Doors - The Isley Brothers\\
\midrule
11 & Benny Goodman - Crosby - Sonny Rollins\\
\midrule
12 & AC/DC - Aerosmith - blink-182 - Bob Seger - BTS - Def Leppard - Green Day - Iron Maiden \\& Journey - Judas Priest - KISS - Linkin Park - Nirvana - Ramones - Rush - The Smiths - Van Halen\\
\midrule
13 & ABBA - Billy Joel - Black Sabbath - Bruce Springsteen - Chicago - Creedence Clearwater Revival \\& Eagles - Genesis - Kenny Chesney - Phil Collins - Stevie Ray Vaughan \\& The Allman Brothers Band - Yes\\
\midrule
14 & Bing Crosby - Charlie Chaplin - Judy Garland - Julie London - Leonard Bernstein \\& MGM Studio Orchestra - Peggy Lee - Raffi - Richard Rodgers - Sarah Vaughan\\
\midrule
15 & Dale Carnegie - Ernst H. Gombrich\\
\midrule
16 & Alabama - Jimmy Buffett - Neil Diamond - Neil Young - Paul Simon\\
\midrule
17 & Alan Jackson - Bee Gees - Chuck Berry - Daryl Hall \& John Oates - Electric Light Orchestra \\& John Mayer - Los Bukis - Luis Miguel - Mariah Carey - Steely Dan - The Byrds \\& The Monkees - The Ventures\\
\midrule
18 & George Jones - Henry Mancini - James Taylor - John Prine - Marty Robbins - Simon \& Garfunkel\\
\midrule
19 & José Alfredo Jimenez\\
\midrule
20 & Barbra Streisand - Dinah Washington - Ennio Morricone - Jackie Gleason - John Denver \\& Joni Mitchell - Lefty Frizzell - Nat King Cole - Wes Montgomery\\

\bottomrule
\end{tabular}

\caption{Artists organized by estimated DC.} \label{table:fullDC}
\end{table}

\clearpage
